\documentclass[twocolumn]{aastex631}

\usepackage{verbatim}
\usepackage{graphicx}
\usepackage{appendix}
\usepackage{hyperref}
 
\newcommand{\lsun}{L$_{\odot}$}
\newcommand{\msun}{M$_{\odot}$}

\newcommand{\mjyb}{mJy~beam$^{-1}$}

\newcommand{\kms}{km~s$^{-1}$}  
\newcommand{\mum}{$\mu$m}

\newcommand{\vlsr}{V$_{LSR}$}
\newcommand{\mstar}{M$_{\star}$}
\newcommand{\lstar}{L$_{\star}$}
\newcommand{\mearth}{M$_{\oplus}$}

\begin{document}

\title{Dynamical Masses for 23 Pre-Main Sequence Stars in Upper Scorpius: A Critical Test of Stellar Evolutionary Models}

\author{A.~P.~M. Towner}
\affiliation{University of Arizona Department of Astronomy and Steward Observatory, 933 North Cherry Ave., Tucson, AZ 85721, USA}

\author{J.~A. Eisner}
\affiliation{University of Arizona Department of Astronomy and Steward Observatory, 933 North Cherry Ave., Tucson, AZ 85721, USA}

\author{P.~D. Sheehan}
\affiliation{National Radio Astronomy Observatory, 520 Edgemont Rd., Charlottesville, VA 22903, USA}

\author{L.~A. Hillenbrand}
\affiliation{Department of Astronomy, MC 249-17, California Institute of Technology, Pasadena, CA 91125, USA}

\author{Y.-L. Wu}
\affiliation{Department of Physics, National Taiwan Normal University, Taipei 116, Taiwan}
\affiliation{Center of Astronomy and Gravitation, National Taiwan Normal University, Taipei 116, Taiwan}

\begin{abstract}
We present dynamical masses for 23 pre-main sequence K- and M-type stars in the Upper Scorpius star-forming region.
These masses were derived from the Keplerian rotation of CO disk gas using the MCMC radiative-transfer package {\tt pdspy} and a flared-disk model with 15 free parameters.
We compare our dynamical masses to those derived from five pre-main sequence evolutionary models, and find that most models consistently underestimate stellar mass by $\gtrsim$25\%.
Models with updated treatment of stellar magnetic fields are a notable exception $-$ they consistently return stellar masses in good agreement with the dynamical results. 
We find that the magnetic models' performance is valid even at low masses, in contrast with some literature results suggesting they may overestimate stellar mass for \mstar\/ $\lesssim$ 0.6~\msun.
Our results are consistent with dynamical versus isochronal evaluations for younger samples (e.g. Taurus, 1-3~Myr), and extend the systematic evaluation of stellar evolutionary models up to stars $\sim$11~Myr in age.
Finally, we derive disk dust masses to evaluate whether using dynamical masses versus isochronal masses changes the slope of the log(M$_{dust}$)$-$log(\mstar) relation.
We derive a slightly shallower relation using dynamical masses versus isochronal masses, but the slopes of these relations agree within uncertainties. 
In all cases, we derive a steeper-than-linear relation for log(M$_{dust}$)$-$log(\mstar), consistent with previous literature results for Upper Sco.
\end{abstract}

\section{Introduction}
The most common method of deriving the masses of pre-main sequence (PMS) stars is to use stellar temperature and luminosity in combination with models of PMS stellar evolution \citep{Cohen1979,Baraffe1998,Siess2000,Bressan2012,Soderblom2014,Baraffe2015,Feiden2016,Manara2023_review,Ratzenbock2023a}.
This method has enabled mass derivations for thousands of PMS stars in our galaxy, as it is easily scaled to large numbers of sources and does not require multiple epochs of observation \citep[e.g.][]{Luhman2020,Luhman2022,Carpenter2025}.
It does require precise and accurate measurements of T$_{eff}$ and \lstar, which can be confounded by line-of-sight extinction from protostellar disks (in edge-on systems) or intervening clouds. 

Stellar properties derived from isochrones are highly model-dependent, with some models producing preferentially higher or lower masses when given the same stellar data \citep[e.g.][]{Andrews2013,Pascucci2016,Manara2023_review}.
Additionally, some models preferentially over- or under-estimate stellar parameters in ways that are correlated with stellar mass or age \citep{Hillenbrand2004,Herczeg2015,Rizzuto2016} and 
are not always linear \citep{David2019}. 
All these factors together make systematic corrections for known model biases difficult to constrain. 

It is possible to measure stellar mass directly via dynamical interactions, e.g. Keplerian rotation in a disk or orbital motion in a binary system \citep[see e.g.][and references therein]{Rizzuto2016,Sheehan2019,Pegues2021,Manara2023_review}.
Once a mass has been obtained, a star's T$_{eff}$ and \lstar\/ can be used in combination with the preferred isochrones to derive a stellar age.
However, modeling a binary system or dynamical interaction typically requires multiple epochs of observations to obtain stellar proper motions, and is necessarily limited to a small subset of known PMS stars.
Keplerian rotation requires only one epoch of observation, but gas can be difficult to detect in older, smaller, or less gas-abundant disks \citep[][and references therein]{Manara2023_review,Carpenter2025}.
Consequently, dynamical modeling of such older, smaller, and fainter disks has historically been limited to a small number of targets.
It has only been with ALMA that we can achieve line sensitivities sufficient to detect these faint disks with finite integration times. 

It is, in theory, possible to calibrate (indirect) isochronal masses with (direct) dynamical masses.
This would both reveal any systematic bias in the isochronal masses and, if the dynamical mass is sufficiently precise, place tighter constraints on the inferred stellar age. 
Some comparisons between isochronal and dynamical results have been done, but most sample sizes have been quite small \citep[e.g. 7 and 9 binary systems, respectively, in][]{Rizzuto2016,David2019}. 
Crucially, the precision of Keplerian modeling is fundamentally limited by each source's distance uncertainty \citep[see e.g.][]{Czekala2015}, which has historically been 10\% or larger.

Since the advent of the {\it Gaia} mission, many nearby PMS stars have had their distances measured with 2\% precision or better \citep{Gaia_i,Gaia_DR2}.
Combined with the powerful capabilities of ALMA, precise measurements of dynamical mass can now be obtained for a much broader range of disk-bearing pre-main sequence stars. 
This both improves \mstar\/ measurements for individual sources and makes 
a self-consistent, systematic comparison between dynamical and isochronal masses possible.

\begin{deluxetable*}{lcccccc}
\tablecaption{Source Properties}
\tablecolumns{7}
\tablewidth{\textwidth}
\tablehead{\colhead{Source} & \colhead{RA} & \colhead{Dec} & \colhead{Distance$^a$} & \colhead{log(T$_{eff}$)$^b$} & \colhead{L$_{\star}^c$} & \colhead{Spectral$^b$} \\
& \colhead{($h$ $m$ $s$)} & \colhead{($^{\circ}$ $\arcmin$ $\arcsec$)} & \colhead{(pc)} & \colhead{(K)} & \colhead{(L$_{\odot}$)} & \colhead{Type} 
}
\startdata
J15521088-2125372 & 15:52:10.88 & -21:25:37.20 & 167.70$_{-7.19}^{+7.91}$ & 3.51 (0.02) & 0.010$^{+0.005}_{-0.005}$ & M4 \\
J15530132-2114135 & 15:53:01.32 & -21:14:13.5 & 146.27$_{-2.41}^{+2.62}$ & 3.51 (0.02) & 0.04$^{+0.02}_{-0.02}$ & M4\\ 
J15534211-2049282$^d$ & 15:53:42.11 & -20:49:28.2 & 135.64$_{-3.32}^{+3.38}$ & 3.52 (0.02) & 0.07$^{+0.03}_{-0.03}$ & M3.5\\
J15562477-2225552 & 15:56:24.77 & -22:25:55.2 & 141.24$_{-2.14}^{+2.13}$ & 3.51 (0.02) & 0.05$^{+0.02}_{-0.02}$ & M4 \\
J16001844-2230114$^d$ & 16:00:18.44 & -22:30:11.4 & 137.71$_{-7.64}^{+8.67}$ & 3.50 (0.02) & 0.05$^{+0.02}_{-0.02}$ & M4.5 \\
J16014086-2258103$^d$ & 16:01:40.85 & -22:58:11.3 & 124.26$_{-1.57}^{+1.56}$ & 3.51 (0.02) & 0.07$^{+0.03}_{-0.03}$ & M4 \\
J16020757-2257467 & 16:02:07.57 & -22:57:46.7 & 140.27$_{-1.22}^{+1.26}$ & 3.54 (0.02) & 0.15$^{+0.06}_{-0.07}$ & M2.5 \\
J16035767-2031055 & 16:03:57.67 & -20:31:05.5 & 142.59$_{-0.79}^{+0.77}$ & 3.64 (0.01) & 0.6$^{+0.3}_{-0.3}$ & K5 \\
J16035793-1942108 & 16:03:57.93 & -19:42:10.8 & 157.94$_{-1.97}^{+2.08}$ & 3.55 (0.02) & 0.13$^{+0.05}_{-0.06}$ & M2 \\
J16062277-2011243 & 16:06:22.77 & -20:11:24.3 & 151.14$_{-2.37}^{+2.41}$ & 3.49 (0.02) & 0.05$^{+0.02}_{-0.02}$ & M5 \\
J16075796-2040087$^d$ & 16:07:57.96 & -20:40:08.7 & 158.63$_{-6.02}^{+6.61}$ & 3.57 (0.02) & 0.09$^{+0.04}_{-0.05}$ & M1 \\
J16081566-2222199 & 16:08:15.66 & -22:22:19.9 & 140.17$_{-1.49}^{+1.61}$ & 3.53 (0.02) & 0.14$^{+0.06}_{-0.06}$ & M3.25 \\
J16082324-1930009 & 16:08:23.24 & -19:30:00.9 & 137.99$_{-1.07}^{+1.12}$ & 3.59 (0.01) & 0.2$^{+0.1}_{-0.1}$ & K9 \\
J16090075-1908526 & 16:09:00.75 & -19:08:52.6 & 137.63$_{-1.47}^{+1.46}$ & 3.59 (0.01) & 0.3$^{+0.1}_{-0.1}$ & K9 \\
J16095933-1800090 & 16:09:59.33 & -18:00:09.0 & 136.24$_{-2.23}^{+2.24}$ & 3.51 (0.02) & 0.07$^{+0.04}_{-0.03}$ & M4 \\
J16104636-1840598 & 16:10:46.36 & -18:40:59.8 & 143.04$_{-2.71}^{+2.83}$ & 3.50 (0.02) & 0.03$^{+0.01}_{-0.02}$ & M4.5 \\
J16113134-1838259 A & 16:11:31.36 & -18:38:26.3 & 127.91$_{-1.54}^{+1.63}$ & 3.64 (0.01) & 1.3$^{+0.5}_{-0.4}$ & K5 \\
J16113134-1838259 B$^{d,e}$ & 16:11:31.31 & -18:38:27.7 & 156.78$_{-4.44}^{+4.69}$ & 3.60 (0.01) & 0.4$^{+0.2}_{-0.1}$ & K7 \\
J16115091-2012098 & 16:11:50.91 & -20:12:09.8 & 152.17$_{-3.58}^{+3.78}$ & 3.52 (0.02) & 0.10$^{+0.04}_{-0.04}$ & M3.5 \\
J16123916-1859284 & 16:12:39.16 & -18:59:28.4 & 139.15$_{-1.63}^{+1.64}$ & 3.58 (0.01) & 0.3$^{+0.1}_{-0.1}$ & M0.5 \\
J16142029-1906481 & 16:14:20.29 & -19:06:48.1 & 142.99$_{-2.47}^{+2.51}$ & 3.59 (0.01) & 0.05$^{+0.02}_{-0.02}$ & M0 \\
J16143367-1900133 & 16:14:33.61 & -19:00:14.8 & 142.01$_{-2.19}^{+2.25}$ & 3.53 (0.02) & 0.10$^{+0.04}_{-0.04}$ & M3 \\
J16154416-1921171 & 16:15:44.16 & -19:21:17.1 & 131.76$_{-2.1}^{+2.13}$ & 3.64 (0.01) & 0.03$^{+0.01}_{-0.01}$ & K5 \\
J16163345-2521505 & 16:16:33.45 & -25:21:50.5 & 162.65$_{-1.45}^{+1.47}$ & 3.58 (0.01) & 0.17$^{+0.08}_{-0.07}$ & M0.5 \\
J16181904-2028479 & 16:18:19.04 & -20:28:47.9 & 137.94$_{-2.37}^{+2.45}$ & 3.5 (0.02) & 0.03$^{+0.02}_{-0.02}$ & M4.75 \\
\enddata
\tablenotetext{a}{Distances and their uncertainties come from {\it Gaia} parallax measurements.}
\tablenotetext{b}{Taken from \citet{Barenfeld2016}, except for J16113134-1838259 A \& B which are taken from \citet{Eisner2005}.}
\tablenotetext{c}{Derived using OIR photometry and BT Settl stellar atmospheric models, except for J16113134-1838259 A \& B which are taken from \citet{Eisner2005}.}
\tablenotetext{d}{These sources are either known binaries or have a candidate companion identified in \citet{Barenfeld2017_disksize}. In all cases, the candidate companion lies within the 1.3~mm continuum disk emission observed by \citet{Barenfeld2016}, making it possible that some or all of these disks are circumbinary.}
\tablenotetext{e}{Source J16113134-1838259 B is a spectroscopic binary. The luminosity, temperature, and spectral-type data reported in the body of the table are for the higher-mass member. The lower-mass companion is an M0 star with log(T$_{eff}$) $=$ 3.58$\pm$0.01 and log(L$_{\star}$) $=$ $-$0.59$\pm$0.15 \citep{Eisner2005}.}
\label{source_properties}
\end{deluxetable*}

In this paper, we present dynamical masses for 23 pre-main sequence stars in the Upper Scorpius OB association, and perform a systematic evaluation of five commonly-used PMS stellar evolutionary models. 
Our targets are all low-mass (K, M) stars with CO-detected disks, and 
we derive stellar masses by modeling each disk's Keplerian rotation. 
The structure of the paper is as follows: in Section~\ref{obs_section}, we discuss the sample and describe our observations.
We discuss our Keplerian modeling procedure in \S~\ref{diskmodel}, including the details of the model itself, the fixed vs. free parameters and their priors, and our criteria for evaluating convergence of the MCMC fit. 
In \S~\ref{results}, we report the results of our MCMC fits.
In \S~\ref{analysis}, we evaluate our dynamical results against five sets of PMS evolutionary tracks, 
derive a log(M$_{disk}$)$-$log(\mstar) relation using our dynamical masses, and briefly describe potential sources of bias in our sample statistics. 
We summarize our key findings in \S~\ref{summary}.
We show additional figures, including examples from online figure sets, in Appendix~\ref{additional_figs}, give additional notes on selected sources in Appendix~\ref{individual_notes}, discuss the impact of mass range, binarity, and sample bias in Appendices~\ref{full_sample_discussion} and \ref{uncertainty_appendix}, and summarize the key components of the stellar evolutionary models in Appendix~\ref{track_descriptions}.

\section{Sample \& Observations}
\label{obs_section}

\subsection{The Upper Scorpius OB Association}
The Upper Scorpius OB association (Upper Sco) is the nearest massive star-forming region to Earth \citep[$\sim$145~pc,][]{Preibisch2008}.
It is part of the larger Scorpius-Centaurus OB association \citep{Blaauw1946,Blaauw1964}, which also includes the Upper Centaurus-Lupus (UCL) and Lower Centaurus-Crux (LCC) regions \citep{Preibisch2008}.
Recent (sub)millimeter, targeted surveys have shown that Upper Sco contains both a large population of low-mass PMS stars overall \citep[e.g.][]{Luhman2022}, and a large population ($\gtrsim$250) of G-, K-, and M-type stars with disks detectable in the continuum \citep{Carpenter2025}.
Unlike most of the other nearby, canonical star-forming regions \citep[e.g. Taurus, Lupus, 1-3~Myr; see][and references therein]{Pascucci2016}, Upper Sco is generally considered to be older - 5-11~Myr, depending on the method used to determine age \citep{Preibisch2008,Pecaut2012}.
This makes it an excellent laboratory for studying how low-mass PMS stars and their disks might evolve with time.

\subsection{The Sample}
\label{sample_description}
\citet{Barenfeld2016} searched for protoplanetary disks around 106 G-, K-, and M-type pre-main sequence stars in Upper Sco.
They detected 58 targets in continuum (3$\sigma$ detection threshold, $\sim$0.48\mjyb) and 26 in CO (5$\sigma$ detection threshold, $\sim$70~\mjyb).
Of those 26 disks, \citet{Barenfeld2016} detected 24 in both CO and continuum, and 2 in CO only (J15521088$-$2125372 and J15562477$-$2225552).

\begin{figure*}
    \centering
    \includegraphics[width=\textwidth]{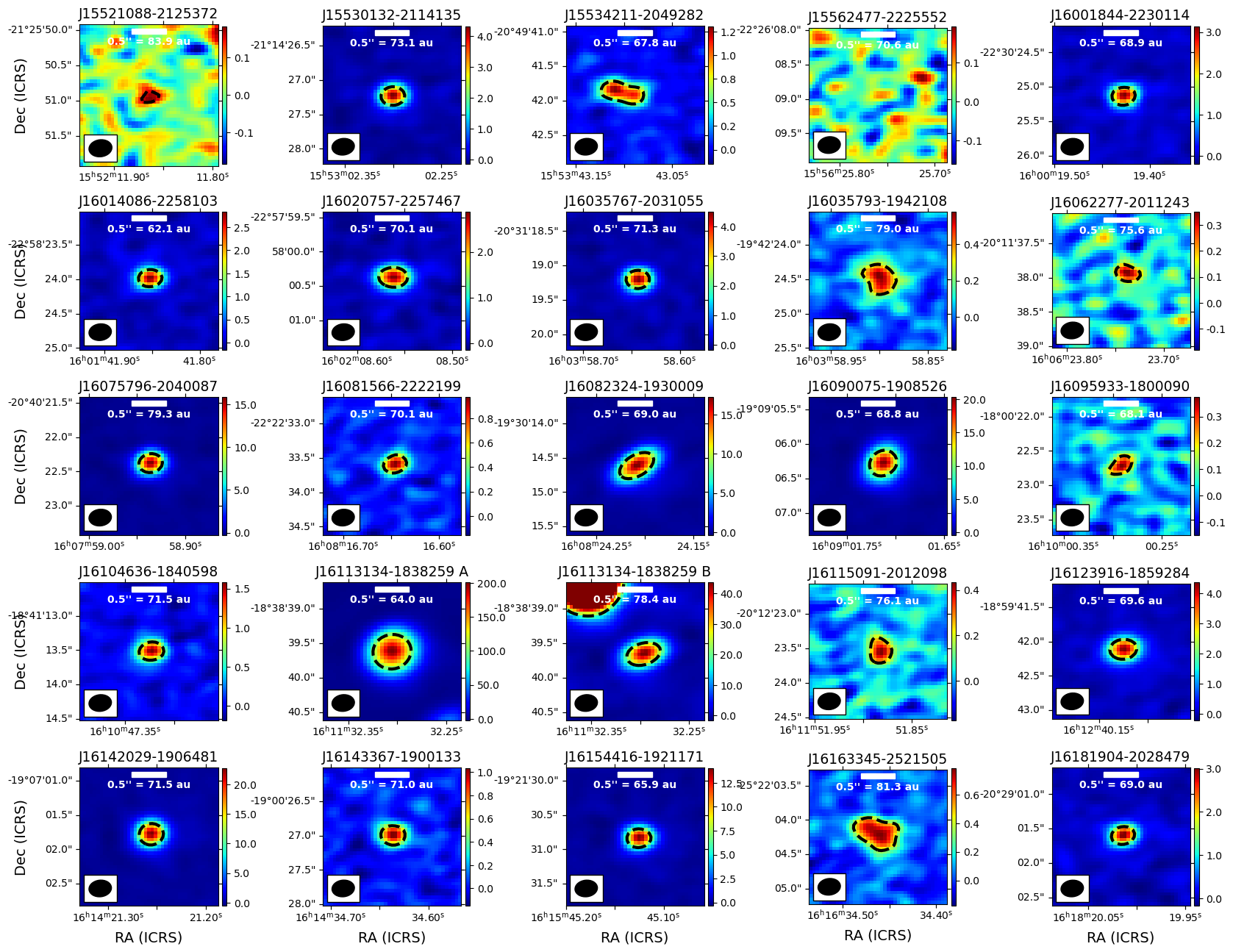}
    \caption{345~GHz continuum images for all 25 targets in the sample. All panels are in ICRS coordinates, with a field of view of 2$\arcsec$. All colorbars are in units of \mjyb. The 0.5$\times$I$_{peak}$ contour for each source is shown as a black dashed line. The continuum beam for each field is shown in the bottom left corner of each image, and an angular and physical scalebar in the top center. Source J15562477-2225552 is a non-detection at 3$\sigma$.}
    \label{cont_all}
\end{figure*}

We conducted follow-up observations in ALMA Cycle 7 of the CO-detected \citet{Barenfeld2016} targets with spectral type K5 or later. 
We also observe an additional target (16113134$-$1838259 B) that is in the same field as one of the \citet{Barenfeld2016} targets but not reported in that work.

The basic properties of each source are listed in Table~\ref{source_properties}.
The source distances and uncertainties are derived from {\it Gaia} parallax measurements.
The spectral types and T$_{eff}$ come from \citet{Barenfeld2016}, and references therein.
\citet{Barenfeld2016} use the spectral types of \citet{Luhman2012}, and adopt an uncertainty of $\pm$1 spectral subclass.
\citet{Barenfeld2016} then convert spectral type into T$_{eff}$ using the temperature scales of \citet{Schmidt-Kaler1982}, \citet{Straizys1992}, and \citet{Luhman1999}. 
The T$_{eff}$ uncertainties reflect the $\pm$1 spectral-subclass uncertainty.

The \lstar\/ values are derived by fitting optical-infrared (OIR) photometry to BT Settl stellar atmospheric models \citep{Allard2011} for all sources except J16113134-1838259 A \& B. 
Sources J16113134-1838259 A \& B are not separable in most OIR images, and so could not be fit with this method; for these sources, luminosities come from \citet{Eisner2005}, who use OIR spectroscopy and SED modeling to derive \lstar.
Full details of our BT Settl fitting procedure, as well as plots showing OIR photometry and the best-fit BT Settl models, can be found in Appendix~\ref{additional_figs}. 

Since the acquisition of the ALMA data, group membership information for some of the sources has been revised. 
One of our original targets, J16113134-1838259 A\footnote{J16113134-1838259 is also known as the T-Tauri star AS 205. Our source J16113134-1838259 A corresponds to AS 205 N, and our source J16113134-1838259 B corresponds to AS 205 S.}, is now explicitly identified in as belonging to $\rho$ Ophiuchus and not Upper Sco \citep{Luhman2022,Carpenter2025}.
However, sources J16113134-1838259 A \& B are in the same field of view. 
For this reason, source J16113134-1838259 A is included in the tables and figures detailing our line and continuum detections (see \S~\ref{obs_subsection} below), 
but is excluded from our subsequent analysis.
This gives us a final sample size of 24 disks.

\subsection{Observing Details and Data Products}
\label{obs_subsection}
Our tuning covers $^{12}$CO J$=$3$-$2 (345.7959899~GHz), $^{13}$CO J$=$3$-$2 (330.5879653~GHz), and 870~\mum\/ (345~GHz) continuum emission (ALMA Project 2019.1.00493.S, PI P. Sheehan).
The $^{12}$CO J$=$3$-$2 line spectral window was centered on 345.795990~GHz with 468.75~MHz bandwidth (406.4~\kms) and 282~kHz (0.245~\kms) spectral resolution.
The $^{13}$CO J$=$3$-$2 line spectral window was centered on 330.587965~GHz with 468.75~MHz bandwidth (406.4~\kms) and 282.227~kHz (0.256~\kms) spectral resolution.
The continuum data were observed in two spectral windows centered on 332 and 344~GHz, respectively, each with 1.875~GHz bandwidth and 31.25~MHz spectral resolution.
We requested a target angular resolution of 0$\farcs$27 and Largest Angular Scale (LAS) of 2$\farcs$4.
Our integration time was 24 minutes per target, designed to achieve 5~\mjyb\/ sensitivity per $\sim$0.5~\kms\/ channel (assuming $\sim$2$\times$ channel binning) and 55~$\mu$Jy beam$^{-1}$ in continuum.

\begin{figure*}
    \centering
    \includegraphics[width=\textwidth]{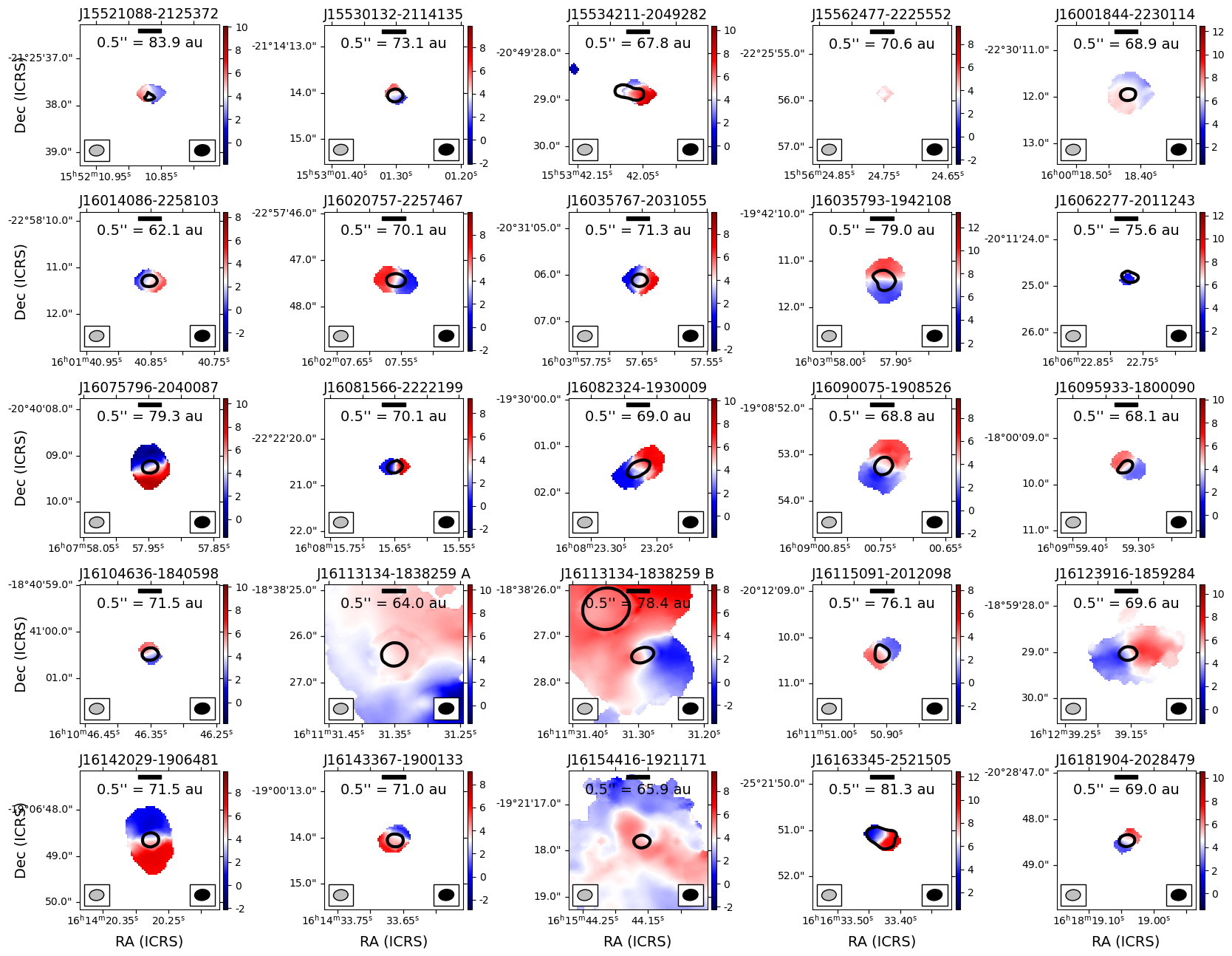}
    \caption{$^{12}$CO intensity-weighted velocity (moment 1) maps for all 25 targets. All panels are in ICRS coordinates, with a 4$\arcsec$ field of view. All colorbars are in \kms, and are centered on a \vlsr\/ determined from a by-eye examination of each target's spectrum. Each image is masked such that only pixels with a) S/N $>$ 2$\sigma$ in b) at least two consecutive channels and c) at least four pixels square are included in the moment maps. The 0.5$\times$I$_{peak}$ continuum contours are overlaid in black. 
    The $^{12}$CO beam is shown in the bottom left corner of each panel, the continuum beam in the bottom right, and an angular and physical scalebar in the top center.}
    \label{mom1_12co}
\end{figure*}

The data were taken in C-4 between 27 February 2020 and 20 May 2022.
All data calibration was performed by the ALMA pipeline in CASA \citep{CASA_2022,McMullin2007_casa}.
We use the ALMA pipeline image products for the continuum data.
For the line data, we resample the line cubes to 0.4~\kms\/ velocity resolution using the {\tt cvel2} task. 
The purpose of the resampling is to increase per-channel signal-to-noise in the image cubes for our initial search for line detections, and to reduce fitting time in our data modeling procedure (see \S~\ref{diskmodel}). 
We image the resampled data using {\tt tclean} with Briggs weighting, a robust value of 0.5, and no multiscale clean for most sources.
For sources with extended emission (e.g. J16113134-1838259, J16154416-1921171), we use multiscale clean and robust values between 0.0 and 1.0.

\begin{figure}
    \centering
    \includegraphics[width=0.48\textwidth]{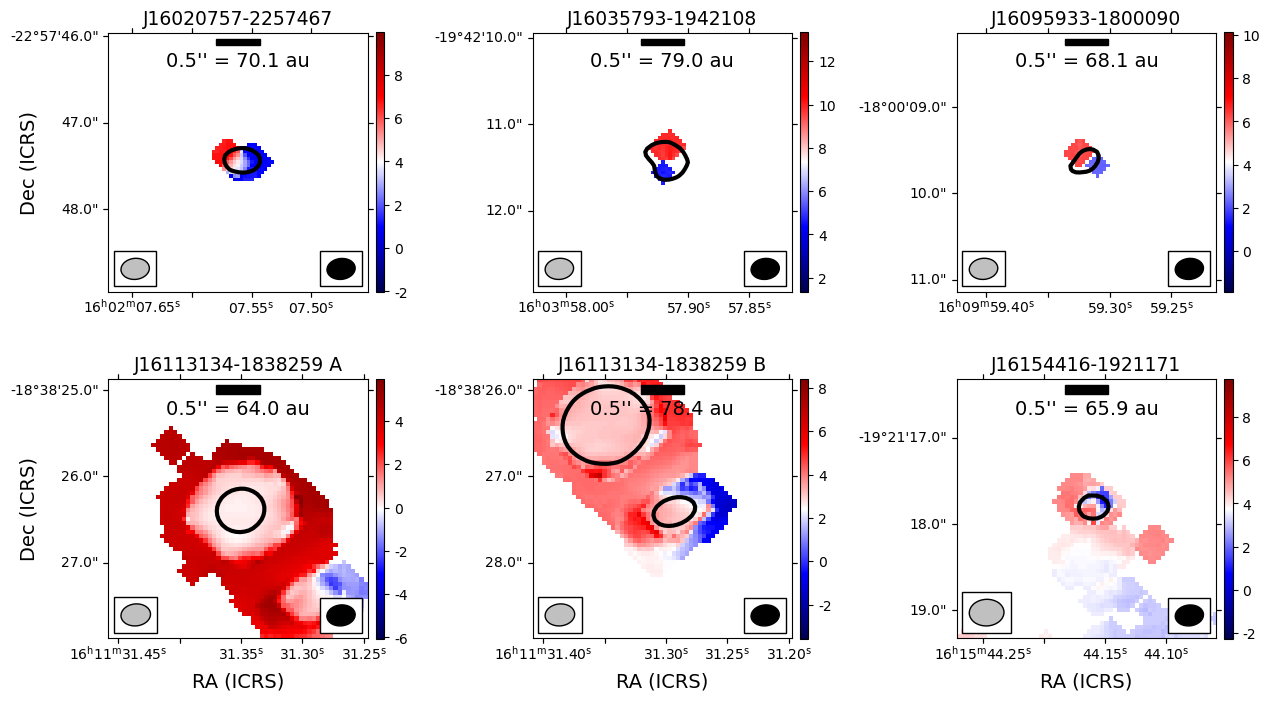}
    \caption{As in Figure~\ref{mom1_12co}, but for $^{13}$CO emission. Only sources with $^{13}$CO detections are shown.}
    \label{mom1_13co}
\end{figure}

For the CO line cubes, we define ``detection'' to mean emission $>$5$\sigma_{i}$ in at least two consecutive channels, where $\sigma_{i}$ is the rms in channel $i$.
We detect $^{12}$CO emission associated with all 25 targets, and $^{13}$CO emission associated with 6/25 targets.
We detect continuum emission associated with 24 out of 25 targets with a detection limit of 3$\sigma$ ($\sim$0.30-0.50~\mjyb\/ in most cases).
We use a lower detection threshold in continuum because the positions of all sources are already known from the $^{12}$CO detections; the statistical likelihood of multiple $>$3$\sigma$ detections at the same location in different images is much smaller than the likelihood of a 3$\sigma$
signal in a single image. 
The source that remains undetected in continuum is J15562477-2225552. 
This source has I$_{peak}$ $=$ 0.113~\mjyb\/ (1.9$\sigma$). 

\rotate
\begin{deluxetable*}{l|ccccc|cc|ccccc}
\tablecaption{Line Cube and Continuum Image Properties}
\tablecolumns{13}
\tablewidth{\textwidth}
\tablehead{
\colhead{Field} & \colhead{$\sigma_{^{12}CO}^a$} & \colhead{I$^a_{peak,\,^{12}CO}$} & \multicolumn{3}{c}{$^{12}$CO Moment 0 Gaussian Fit$^b$} & \colhead{$\sigma_{^{13}CO}^a$} & \colhead{I$^a_{peak,\,^{13}CO}$} & \colhead{$\sigma^c_{cont}$} & \colhead{I$^c_{peak,\,cont}$} & \multicolumn{3}{c}{Continuum Gaussian Fit$^d$}\\
 & & & \colhead{Major} & \colhead{Minor} & \colhead{P.A.} & & & & & \colhead{Major} & \colhead{Minor} & \colhead{P.A.}
}
\startdata
J15521088-2125372 & 5.4 & 42.8$\pm$5.6 & 0.3 (0.9) & 0.1 (0.06) & 80 (20) & 6.7 & $<$33.9 & 0.06 & 0.19$\pm$0.06 & 1.3 (0.9) & 0.1 (0.2) & 110 (20)\\
J15530132-2114135 & 5.4 & 42.8$\pm$5.5 & 0.7 (0.6) & 0.1 (0.3) & 40 (30) & 6.6 & $<$31.4 & 0.06 & 4.34$\pm$0.06 & 0.14 (0.02) & 0.11 (0.02) & 50 (30)\\
J15534211-2049282 & 5.5 & 52.5$\pm$5.6 & 0.6 (0.3) & 0.1 (0.1) & 70 (30) & 6.8 & $<$33.4 & 0.06 & 1.31$\pm$0.06 &  0.57 (0.05)$^{\dagger}$ & 0.25 (0.05)$^{\dagger}$ & 50 (20)$^{\dagger}$ \\ 
J15562477-2225552 & 5.5 & 30.5$\pm$5.3 & \nodata & \nodata & \nodata & 6.6 & $<$33.2 & 0.06 & $<$0.18 & \nodata & \nodata & \nodata \\
J16001844-2230114 & 5.5 & 185.9$\pm$5.4 & 0.41 (0.08) & 0.36 (0.07) & 110 (80) & 7.4 & $<$31.3 & 0.07 & 3.16$\pm$0.07 & 0.11 (0.06) & 0.08 (0.05) & 30 (90)\\
J16014086-2258103 & 5.5 & 75.7$\pm$5.5 & 0.1 (0.2) & 0.06 (0.07) & 100 (30) & 6.9 & $<$33.3 & 0.06 & 2.84$\pm$0.06 & 0.14 (0.03) & 0.09 (0.03) & 80 (30)\\
J16020757-2257467 & 5.5 & 75.7$\pm$5.4 & 0.4 (0.2) & 0.3 (0.3) & 200 (100) & 7.0 & 57.9$\pm$6.6 & 0.07 & 2.88$\pm$0.07 & 0.26 (0.02) & 0.15 (0.01) & 81 (6)\\
J16035767-2031055 & 5.5 & 67.8$\pm$5.4 & 0.4 (0.3) & 0.2 (0.1) & 110 (50) & 6.8 & $<$32.7 & 0.07 & 4.51$\pm$0.07 & 0.12 (0.01) & 0.08 (0.03) & 30 (20)\\ 
J16035793-1942108 & 5.5 & 111.7$\pm$5.3 & 0.5 (0.2) & 0.3 (0.3) & 170 (90) & 7.0 & 37.4$\pm$6.3 & 0.07 & 0.58$\pm$0.07 & 0.41 (0.09) & 0.3 (0.1) & 40 (30)\\
J16062277-2011243 & 5.5 & 29.7$\pm$5.7 & \nodata & \nodata & \nodata & 6.7 & $<$32.6 & 0.07 & 0.35$\pm$0.07 & \nodata & \nodata & \nodata \\
J16075796-2040087 & 5.5 & 153.3$\pm$5.6 & 0.3 (0.1) & 0.2 (0.2) & 20 (60) & 7.2 & $<$33.5 & 0.08 & 15.85$\pm$0.08 & 0.78 (0.05)$^{\dagger}$ & 0.64 (0.05)$^{\dagger}$ & 90 (20)$^{\dagger}$ \\ 
J16081566-2222199 & 5.5 & 41.2$\pm$5.3 & 0.4 (0.3) & 0.2 (0.2) & 70 (50) & 6.6 & $<$33.0 & 0.06 & 0.98$\pm$0.06 & \nodata & \nodata & \nodata \\
J16082324-1930009 & 5.5 & 93.0$\pm$5.6 & 0.6 (0.3) & 0.2 (0.2) & 150 (30) & 7.0 & $<$32.8 & 0.09 & 17.25$\pm$0.09 & 1.05 (0.05)$^{\dagger}$ & 0.70 (0.05)$^{\dagger}$ & 140 (20)$^{\dagger}$\\ 
J16090075-1908526 & 5.5 & 86.8$\pm$5.4 & 0.6 (0.2) & 0.4 (0.1) & 160 (30) & 6.8 & $<$34.0 & 0.10 & 20.3$\pm$0.1 & 0.90 (0.05)$^{\dagger}$ & 0.80 (0.05)$^{\dagger}$ & 140 (20)$^{\dagger}$\\ 
J16095933-1800090 & 5.5 & 59.8$\pm$5.6 & 0.3 (0.3) & 0.2 (0.1) & 70 (80) & 7.3 & 35.8$\pm$7.0 & 0.07 & 0.38$\pm$0.07 & \nodata & \nodata & \nodata \\
J16104636-1840598 & 5.6 & 46.4$\pm$5.5 & \nodata & \nodata & \nodata & 6.6 & $<$32.5 & 0.06 & 1.59$\pm$0.06 & 0.15 (0.04) & 0.14 (0.06) & 10 (70)\\
J16113134-1838259 A & 6.9 & 809.6$\pm$7.1 & 1.20 (0.09) & 0.91 (0.07) & 60 (10) & 7.9 & 221.4$\pm$8.2 & 0.87 & 201.2$\pm$0.9 & 0.462 (0.007) & 0.433 (0.007) & 120 (10)\\
J16113134-1838259 B & 6.9 & 421.2$\pm$6.5 & 1.06 (0.06) & 0.80 (0.04) & 107 (9) & 7.9 & 96.1$\pm$8.1 & 0.87 & 43.6$\pm$0.9 & 0.38 (0.01) & 0.20 (0.01) & 114 (3)\\
J16115091-2012098 & 5.5 & 65.6$\pm$5.5 & 0.4 (0.2) & 0.3 (0.1) & 100 (50) & 6.6 & $<$32.4 & 0.06 & 0.43$\pm$0.06 & \nodata & \nodata & \nodata \\
J16123916-1859284 & 5.5 & 82.9$\pm$5.4 & 1.1 (0.3) & 0.6 (0.1) & 110 (20) & 6.9 & $<$32.7 & 0.07 & 4.38$\pm$0.07 & 0.26 (0.01) & 0.183 (0.008) & 99 (6)\\
J16142029-1906481 & 5.6 & 158.5$\pm$5.6 & 0.6 (0.1) & 0.3 (0.1) & 10 (20) & 7.1 & $<$34.1 & 0.12 & 22.90$\pm$0.12 & 0.83 (0.05)$^{\dagger}$ & 0.76 (0.05)$^{\dagger}$ & 90 (20)$^{\dagger}$\\ 
J16143367-1900133 & 5.6 & 68.4$\pm$5.5 & 0.3 (0.1) & 0.2 (0.1) & 100 (90) & 7.7 & $<$33.5 & 0.07 & 1.04$\pm$0.07 & 0.19 (0.05) & 0.04 (0.09) & 50 (20)\\
J16154416-1921171 & 9.4 & 277.0$\pm$13.3 & 1.28 (0.08) & 0.79 (0.06) & 42 (6) & 11.7 & 78.5$\pm$12.1 & 0.13 & 14.1$\pm$0.1 & 0.14 (0.02) & 0.12 (0.02) & 50 (40)\\
J16163345-2521505 & 5.6 & 37.5$\pm$5.7 & 0.5 (0.3) & 0.3 (0.2) & 70 (60) & 6.9 & $<$34.2 & 0.06 & 0.79$\pm$0.06 & 0.50 (0.06) & 0.27 (0.05) & 70 (10)\\
J16181904-2028479 & 5.6 & 44.7$\pm$5.7 & 0.3 (0.2) & 0.2 (0.2) & 200 (200) & 7.0 & $<$34.3 & 0.06 & 3.02$\pm$0.06 & 0.13 (0.03) & 0.09 (0.04) & 120 (50)
\enddata 
\tablenotetext{a}{Units are \mjyb. The $^{12}$CO beam sizes range from 0$\farcs$31$\times$0$\farcs$23 to 0$\farcs$32$\times$0$\farcs$23. $^{13}$CO beam sizes range from 0$\farcs$34$\times$0$\farcs$26 to 0$\farcs$35$\times$0$\farcs$26. Both beams have a position angle of $-$83$^{\circ}$. Representative noise levels are the mean noise across all channels for a given cube, measured in an image region that is emission-free in all channels. $\sigma_{peak}$ is the rms value of the emission-free region in the peak channel. For sources with I$_{peak}$ $<$ 5$\sigma$, we report the 5$\sigma$ upper limit.}
\tablenotetext{b}{The FWHM major axis, minor axis, and position angle of a 2D Gaussian fit to the moment 0 maps for each source. The fits were performed with CASA's {\tt imfit} task. Major and minor axes are in units of arcsec, and position angles are in degrees. Entries of ``\nodata'' indicate either a point source or a source too faint to be fit.}
\tablenotetext{c}{Units are \mjyb. The continuum beam is 0$\farcs$33$\times$0$\farcs$24 for all fields, with a position angle of $-$83$^{\circ}$. Source J15562477-2225552 is a non-detection at 3$\sigma$ (I$_{peak}$$=$0.113~\mjyb\/ $=$ 1.9$\sigma$), and we report the 3$\sigma$ upper limit above.}
\tablenotetext{d}{The FWHM major axis, minor axis, and position angle of a 2D Gaussian fit to the 870~\mum\/ continuum emission for each source. The fits were performed with CASA's {\tt imfit} task. Major and minor axes are in units of arcsec, and position angles are in degrees. Entries of ``\nodata'' indicate either a point source or a source too faint to be fit. Sources marked with a ``$\dagger$'' could not be fit, and had their FWHM size and position angle measured by hand. The uncertainties on the major and minor axis FWHM (0$\farcs$05) correspond to the size of a single pixel. The uncertainty on position angle (20$^{\circ}$ in all cases) reflects the uncertainty in our by-eye assessment of P.A.}
\label{obstable}
\end{deluxetable*}

Continuum images for all sources are shown in Figure~\ref{cont_all}.
Half-peak contours (0.5$\times$I$_{peak,cont}$) are shown in black in all panels. 
Figure~\ref{mom1_12co} shows $^{12}$CO intensity-weighted velocity (moment 1) maps for all 25 sources, with the 0.5$\times$I$_{peak,cont}$ continuum contours overlaid.
Figure~\ref{mom1_13co} shows $^{13}$CO moment 1 maps for sources with I$_{peak,^{13}CO}$ $>$ 5$\sigma$, again with the 0.5$\times$I$_{peak,cont}$ continuum contours overlaid.
Continuum and median line noise (rms) values are listed in Table~\ref{obstable}, along with peak intensities and upper limits.
Beam sizes are nearly identical for all sources: (0$\farcs$31-0$\farcs$32)$\times$0$\farcs$23 for the $^{12}$CO cubes, (0$\farcs$34-0$\farcs$35)$\times$0$\farcs$26 for the $^{13}$CO cubes, and 0$\farcs$33$\times$0$\farcs$24 for the continuum images.
The beam size consistency is a result of our observing strategy; all fields were observed in all scheduling blocks.

\subsection{Disks vs Bipolar Outflows}
\label{disks_vs_outflows}
\begin{figure*}
    \centering
    \includegraphics[width=\textwidth]{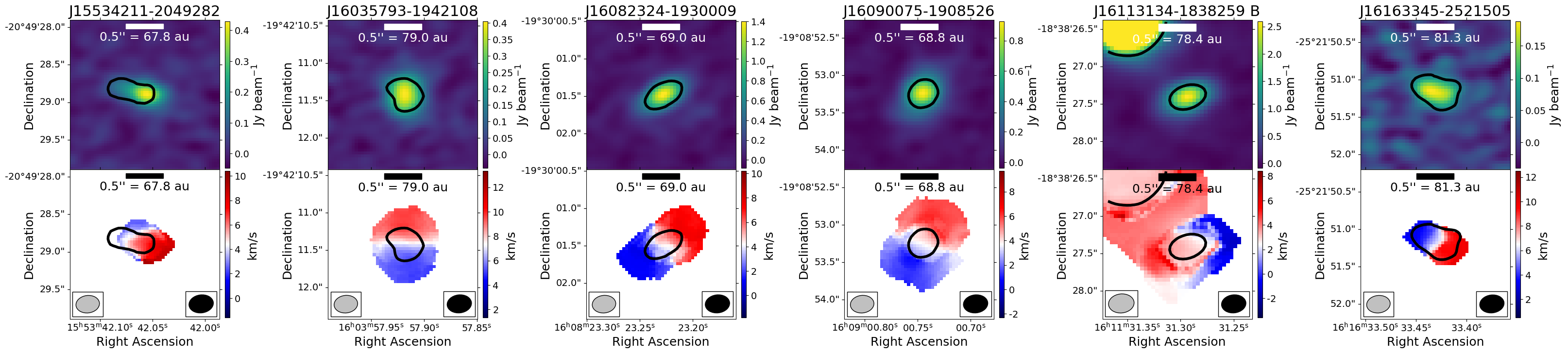}
    \caption{Moment 0 (top row) and moment 1 (bottom row) maps of the six sources whose continuum emission is reasonably well-resolved. The moment maps show $^{12}$CO emission in all cases except J16113134-1838259 B, which show $^{13}$CO. 870~\mum\/ continuum contours are overlaid in black; the contours show the 0.5$\times$I$_{peak}$ level. The $^{12}$CO beams are shown in the bottom left corner of the moment 1 images, and the continuum beams are shown in the bottom right corner.}
    \label{resolved_zooms}
\end{figure*}

Given the relationship between accretion and ejection \citep[e.g.][]{Frank2014}, it is reasonable to consider whether the CO emission in these sources is tracing disk winds/outflows in addition to (or instead of) a rotating disk.
The canonical morphological indicator of an outflow is a $\sim$90$^{\circ}$ misalignment between the observed disk major axes in continuum and line emission.
We examine our data for such a signature.
We use the {\tt imfit} task in CASA to fit 2-dimensional Gaussians to the $^{12}$CO moment 0 maps, and to the 870~\mum\/ continuum data.
We list the deconvolved {\tt imfit} sizes of all sources in Table~\ref{obstable}. 
Six sources appear to be reasonably well-resolved in the continuum: J15534211$-$2049282, J16035793$-$1942108, J16082324$-$1930009, J16090075$-$1908526, J16113134$-$ 1838259 B, and J16163345$-$2521505.
We compare the continuum and $^{12}$CO moment 0 position angles for these six sources.
In Figure~\ref{resolved_zooms}, we show close-in views of the moment 0 and moment 1 maps for these sources, with 870~\mum\/ continuum contours overlaid. 

The position angles of the continuum and moment-0 line emission agree within uncertainties for all six sources.
Furthermore, as seen in Figure~\ref{resolved_zooms}, the position angles of the continuum disks' major axes correspond to the direction of the velocity gradients seen in the moment 1 maps.
This is consistent with Keplerian rotation. 
These findings together are consistent with the interpretation that the CO is tracing disk gas and not outflows.
While we cannot rule out dust-gas disk misalignment for the remaining 18 sources, we find no evidence for this in our sample, and good evidence in favor of the assumption that the CO is tracing disk gas.

\section{Disk Model Fitting}
\label{diskmodel}
The Keplerian rotation of a protoplanetary disk is a direct probe of the enclosed mass.
However, because deriving source mass from Keplerian velocity requires knowing linear distances $r$ within the disk, and $r$ = $\theta$ $\times$ distance (where $\theta$ is the angular size), deriving high-accuracy stellar masses requires high-precision source distances.
{\it Gaia} has produced extremely high-precision parallactic distance measurements for many nearby pre-main sequence stars, including all of our targets in Upper Sco.

We use our high-sensitivity ALMA CO data, combined with the {\it Gaia} distances, to fit a Keplerian disk model to each of our targets and derive enclosed stellar mass.
We use the open-source, python-based {\tt pdspy} package to perform Markov Chain Monte Carlo (MCMC) fitting of disk models to our data in visibility space ($u,v$). 
We use {\tt radmc3d} \citep{Dullemond2012_radmc3d} to generate the disk models, {\tt galario} \citep{Tazzari2018_galario} to transform the model data from image space to Fourier space, and the {\tt emcee} package \citep{Foreman-Mackey2013_emcee} for the MCMC fitting. 
We describe our fitting procedure and convergence criteria in greater detail in the following sections.

\subsection{The Model}
In pure Keplerian rotation, the azimuthal velocity at any given stellocentric radius $r$ is given by

\begin{equation}
    v_{\phi}(r) = \sqrt{\frac{G\,M_{encl}}{r}},
\end{equation}

where $G$ is the gravitational constant and M$_{encl}$ is the total mass enclosed within the radius $r$. 
The corresponding observed velocity is 
\begin{equation}
    v_{obs}(r) = v_{\phi}(r) sin(i)
\end{equation}

where $i$ is the inclination angle of the disk ($i$ = 0 being face-on, and $i$=90 being edge-on).

In reality, the fluid dynamics and, especially, the pressure gradients in disks can result in sub-Keplerian velocities \citep[e.g.][]{Haworth2018,Sanna2019}.
Therefore, these effects must be accounted for as well.
We use an exponentially-tapered flared disk model based on the viscous-disk models of \citet{Lynden-Bell1974}.
The model has an azimuthally-symmetric radial structure with surface density given by 

\begin{equation}
    \Sigma(r) = \Sigma_0 \left( \frac{r}{r_c} \right)^{-\gamma} exp \left[ - \left( \frac{r}{r_c} \right)^{2-\gamma}\right].
\end{equation}

Here, $r_c$ is the radius beyond which the disk is exponentially tapered, $\gamma$ is the surface density power-law exponent, and $\Sigma_0$ is given by

\begin{equation}
    \Sigma_0 = \frac{(2-\gamma)M_d}{2\pi r_c^2},
\end{equation}

where $M_d$ is the total disk gas mass.
The disk flaring and scale-height parameters are calculated assuming hydrostatic equilibrium.
The disk scale height is given by

\begin{equation}
    h(r) = \left[ \frac{k_b\,r^3\,T_{gas}(r)}{G\,M_{\star}\,\mu\,m_H} \right],
\end{equation}

where $M_{\star}$ is the stellar mass and $\mu$ = 2.37 is the mean molecular weight of the gas assuming solar metallicity.
The model assumes the disk is vertically isothermal, with a radial temperature dependence given by

\begin{equation}
    T_{gas}(r) = T_0 \left( \frac{r}{1\,\,au} \right)^{-q},
\end{equation}

where $T_0$ is the temperature at 1 au and $q$ is the temperature power-law exponent.

These assumptions together give a 2-dimensional density structure of

\begin{equation}
    \rho(r,z) = \frac{\Sigma(r)}{\sqrt{2\pi}\,h(r)} exp \left[ -\frac{1}{2} \left( \frac{z}{h(r)} \right)^2 \right],
\end{equation}

where $z$ is the height above the disk midplane. 
Further details of this model and its implementation in {\tt pdspy} can be found in \citet{Sheehan2019}.

Following the method of \citet{Sheehan2019}, we use 15 free parameters:
the offset from the source sky center position (RA$_{off}$, Dec$_{off}$), distance ($d$), disk position angle (P.A.), inclination angle ($i$), and source velocity (\vlsr); stellar mass (\mstar), disk mass (M$_{d}$), disk inner radius (R$_{in}$) and outer radius (R$_{d}$); disk temperature at 1~au ($T_{0}$), disk temperature power-law exponent ($q$), stellar luminosity (\lstar), disk surface density power-law exponent ($\gamma$), and disk turbulence ($a_{turb}$). 
(We note that, in practice, \lstar\/ has very little impact on the fit results, as it is not directly incorporated into the disk model.)
The molecular-to-H$_2$ abundance ratios are fixed in all fits. 
We adopt a CO/H$_2$ abundance of 10$^{-4}$ \citep{Allen1973,Encrenaz1975,Dickman1978}.
We adopt a $^{13}$CO/H$_2$ abundance ratio of 10$^{-5.3}$, based on the assumption of a $^{12}$CO/$^{13}$CO ratio of 20:1 from literature observations of protoplanetary disks \citep{Bergin2024,Semenov2024}.

\subsection{Allowed Ranges for Free Parameters}
The free parameters and their allowed ranges are listed in Table~\ref{param_ranges}.
For ease of computation over multiple orders of magnitude, {\tt pdspy} calculates stellar mass, disk mass, disk inner and outer radius, disk inner edge temperature, turbulence, and luminosity in log space. 
Distance, \lstar, \vlsr, and position have ranges and initial values unique to each target.
The allowed ranges for most other free parameters were the same for all sources (see Table~\ref{param_ranges}), with the exception of position angle (P.A.) and temperature power-law slope ($q$). 
The conditions under which adjustments to these parameters were necessary are described in \S~\ref{pdspy_fitting}. 

\subsubsection{$D$ and \lstar}
For a Gaussian distribution, $\pm$2$\sigma$ will encompass $\pm$95\% of the distribution of the measured values.
We restrict $D$ to each target's {\it Gaia}-measured distance plus or minus twice its upper and lower limits, where the distance limits are calculated from the {\it Gaia} upper and lower limits on parallax.
We restrict \lstar\/ to the value reported in Table~\ref{source_properties} for each source, plus or minus twice its associated uncertainty. 

\subsubsection{\vlsr, RA$_{off}$, and Dec$_{off}$}
Our position and velocity initial values are estimated from the image cubes, and their allowed ranges reflect our uncertainty in this initial estimation. 
By default, we allow velocity to vary over \vlsr$\pm$1.2~\kms\/ (corresponding to $\pm$3 channels), where the \vlsr\/ is determined from a by-eye examination of the $^{12}$CO image cube and integrated spectrum. 
For some sources, especially those with low signal-to-noise ratios, our ability to estimate \vlsr\/ by eye is more limited, and we expand the parameter range to \vlsr$\pm$2.4~\kms\/ ($\pm$6 channels).
We estimate the center position of each disk by fitting a 2D gaussian in the image plane using the {\tt imfit} task in CASA. 
We calculate the distance from image center to disk center in RA and Dec, and use these as our initial values for RA$_{off}$ and Dec$_{off}$. 
We restrict the parameter range to this initial guess $\pm$0.25 arcsec, or about $\pm$6 pixels ($\sim$0$\farcs$04 each).
This range is $\sim$2-3$\times$ the typical uncertainty on the {\tt imfit} results. 

\begin{deluxetable}{lcc}
\tablecaption{{\tt pdspy} Parameter Ranges}
\tablecolumns{3}
\tablewidth{\textwidth}
\tablehead{
\colhead{Parameter$^a$} & \colhead{Units} & \colhead{Default Range}
}
\startdata
Distance ($D$) & pc & $D_{Gaia}-2(\sigma_{low}$/$\sigma_{upp}$)\\
Velocity (\vlsr) & \kms\/ & v$_{cen}$ $\pm$ 1.2 \\
RA offset (RA$_{off}$) & \arcsec & RA $\pm$ 0.25 \\
Dec offset (Dec$_{off}$) & \arcsec & Dec $\pm$ 0.25 \\
Inclination ($i$) & $^{\circ}$ & [0, 180] \\
Position Angle (P.A.) & $^{\circ}$ & [0, 360] \\
Stellar Mass (M$_{\star}$) & log(\msun) & [$-$1.5, 1.5] \\
Disk Mass (M$_d$) & log(\mearth) & [$-$10, $-$2.5] \\
Outer Radius (R$_d$) & log(AU) & [0, 4] \\
Inner Radius (R$_{in}$) & log(AU) & [$-$1, R$_d$) \\
Temp at 1 au (T$_0$) & log(K) &  [0, 3.5] \\
T(r) exponent ($q$) & \nodata & [0, 1] \\
Turbulence ($a_{turb}$) & log(\kms) & [$-$1.5, 1] \\
Luminosity ($L_{\star}$) & log(\lsun) & \lstar\/ $\pm$ 2$\sigma_{L_{\star}}$ \\
$\Sigma$ exponent ($\gamma$) & \nodata & [$-$0.5, 2)
\enddata
\tablenotetext{a}{Allowed parameter ranges for the 15 free parameters in our {\tt pdspy} fitting routine. Distance and distance uncertainties come from the {\it Gaia} parallax results and are listed in Table~\ref{source_properties}. \vlsr\/ comes from a by-eye examination of our data cubes; uncertainties are $\pm$3 channels (1.2~\kms) for sources with comparatively strong emission, and $\pm$6 channels (2.4~\kms) for sources with weaker emission. The RA and Dec for which the position offsets are calculated are listed in Table~\ref{source_properties}. Source luminosity and luminosity uncertainty are listed in Table~\ref{source_properties}.}
\label{param_ranges}
\end{deluxetable}

\subsubsection{\mstar, M$_{disk}$, R$_{in}$, R$_{d}$, T$_0$, $\alpha_{turb}$, $i$, and $\gamma$}
The allowed range for stellar mass (log(\mstar) $=$ [-1.5,1.5], \mstar\/ $=$ [0.03,30] \msun) encompasses all expected stellar mass values for our K- and M-type targets. 
The disk mass, outer radius, temperature at 1 au, and turbulence ranges are chosen to encompass reasonable expected values based on literature results for protoplanetary disks and the viscous, hydrostatic disk model.
The lower limit for disk inner radius is chosen to account for the absolute minimum physical resolution we can reasonably expect to achieve with our data (0.1~au).
For the disk inner radius upper limit, the model requires R$_{in}$ $<$ R$_d$ at all times. 
The range for $i$ allows for the possibility of differentiating between momentum vectors of equal magnitude but opposite line-of-sight directions (i.e. clockwise versus counterclockwise rotation for a source at a given $i$).
The range for $\gamma$ (the surface-density power-law exponent) reflects both the disk model itself and our data.
A lower limit of $\gamma$ $=$ $-$0.5, representing a sharply tapered disk surface density, is the the lower limit of what we can reasonably expect to constrain with our data.

\subsection{Fitting Procedure}
\label{pdspy_fitting}
MCMC fitting involves using ``walkers,'' which are individual evaluations of the model for different values of the model parameters.
The $\chi^2$ value of each walker is compared to that of another random walker in the set, and then the parameters of the original walker are adjusted toward or away from the comparison walker by a semi-randomized amount. 
One round of comparison and adjustment for all walkers is a single ``step'' of the MCMC fit. 
The walkers' initial distribution in parameter space (the ``priors'') are set by the user.
A uniform prior means that walkers are distributed uniformly across the parameter range, while a Gaussian prior means that more walkers have initial values close to the center of the parameter space, etc.
Full details of the MCMC procedure in general can be found in \citet{Foreman-Mackey2013_emcee}, and details of its implementation in {\tt pdpsy} can be found in \citet{Sheehan2019}.

In this work, we use 200 walkers for all fits. 
We assume uniform priors for all parameters except source distance and stellar mass.
For distance, we use a Gaussian distribution in parallax, where the distribution is centered on the {\it Gaia} parallax value and has a width of 2$\log$(2)$\times\sigma_{Gaia}$. 
We then calculate source distances from this distribution in parallax. 
For stellar mass, the initial distribution of walkers is set using the Charbrier IMF. 

Most MCMC procedures use a ``burn-in'' interval, i.e. some initial number of steps which will not be considered when determining the final MCMC results.
The goal is to allow the walkers time to settle from their initial distributions into - ideally - higher-probability regions for each parameter. 
Burn-in is most important when the MCMC best-fit values are calculated using bulk statistics (median, mean, $\sigma$) for all walkers and steps, such as the median-value method used in \citet{Sheehan2019}.
We use a minimum burn-in interval of 1000 steps, though some sources required longer (e.g. 4500 steps for J16075796-2040087).

After burn-in, we check the distribution of walkers for each parameter and adjust parameter ranges as needed for individual sources.
If any parameter range changes, we restart the {\tt pdspy} fitting for the source in question from 0 steps. 
The circumstances in which this was necessary are described in \S~\ref{pa_q}, below.
At 2500 steps, we begin evaluating the log(M$_{\star}$) parameter for convergence. 
Our convergence criteria are discussed in \S~\ref{stopping}, below.
If the stellar mass has converged at 2500 steps, we stop the fit.
If the stellar mass has not converged, we continue the fit until the convergence criteria are met. 
If the walker distributions at 2500 steps indicate that further adjustments to the parameter ranges are needed, we adjust the ranges, restart the fit from 0 steps, and evaluate the new fits at 1000 and 2500 steps as previously described. 
For most sources, the walkers meet the convergence criteria within $\sim$6500 total steps. 

\subsubsection{Modifying Parameter Ranges and Restarting Fits}
\label{pa_q}
For most parameters, it was not necessary to make adjustments to the parameter ranges after 1000 steps. 
However, two parameters required adjustment for $>$25\% of the sample: position angle (P.A.) and temperature exponent ($q$).
In the case of position angle, all sources had a total range of 360$^{\circ}$, but some sources had limits of [$-$180$^{\circ}$,180$^{\circ}$] or [$-$90$^{\circ}$,270$^{\circ}$] instead of [0$^{\circ}$,360$^{\circ}$].
This is for purely practical purposes: sources with position angles of $\sim$270$^{\circ}$ $\leq$ $\theta$ $\leq$ 90$^{\circ}$ often had some a fraction of their walkers ``pile up'' near the edge of the parameter range, causing a false bimodal distribution.
Resetting the position-angle range resulted in single solutions for position angle in all cases. 

In the case of $q$, we initially used a range of [0,1], i.e. a flat to linearly-decreasing temperature with radius, for all sources.
In some cases, however, most or all walkers were clustered near the top of the $q$ $=$ [0,1] range, 
with solutions showing no significant improvement over at least  1000 post-burn-in steps \citep[this is the autocorrelation time for 200 walkers and 15 degrees of freedom for our disk model; see][]{Sheehan2019}.
For these sources, we expand the parameter range to $q$ $=$ [0,2], i.e., between flat and a quadratic decrease in temperature with radius.

\subsubsection{Stopping Criteria}
\label{stopping}
The point at which an MCMC fit has converged does not have a universally agreed-upon definition \citep[see][for a recent review]{Roy2020_MCMCstop}. 
The mathematical basis of MCMC fitting says that, in theory, all walkers will converge to the same value for a given parameter, if given infinite time.
This is, for obvious reasons, not a useful definition in practice.
As a practical matter, most MCMC users apply empirical stopping criteria instead, such as fixed-width or relative fixed-width stopping rules, variance-based estimates, and density-based estimates \citep[see e.g.][and references therein]{Roy2020_MCMCstop}.

For the purposes of this work, we adopt a relative variance-based stopping rule in which we define ``convergence'' to mean that the value of log(\mstar) has varied by less than 1\% over a given interval.
Our assessment procedure is as follows: at each step past the initial burn-in (see \S~\ref{pdspy_fitting}), we apply a $\chi^2$ cut to the data to remove any walkers outside the 99th percentile ($\sim$2.5$\sigma$ for a normal distribution). 
We require that at least 101 walkers (i.e. a majority) survive the $\chi^2$ cut. 

Once $\geq$101 walkers remain, we make trace plots for each source in order to evaluate convergence.
To construct the trace plots, we calculate the mean, standard deviation, median, and median absolute deviation (MAD) of log(\mstar) for each step. 
The standard deviation and scaled MAD\footnote{Assuming a normal distribution of values, the scaled MAD (1.4826$\times$MAD) represents a 1$\sigma$ uncertainty about the median.} serve as the uncertainties for the mean and median, respectively.
We calculate the slope of the mean and median versus step over the previous 1000 steps, with each data point weighted by its uncertainty.
We choose the last 1000 steps for this fit because this is the autocorrelation time for an MCMC fit with 200 walkers and 15 free parameters \citep[see e.g. ][]{Sheehan2019}. 
If the magnitude of the slope over the last 1000 steps is less than 1\% of the value of the parameter itself, we consider the slope to be converged.
We require that both median and mean be converged in order to stop the fitting.
Representative trace plots from source J16090075-1908526 are shown in Figure~\ref{traceplot_example}.
Trace plots of log(\mstar) versus step number for each source can be found in Appendix~\ref{additional_figs}.

\begin{figure}
    \centering
    \includegraphics[width=0.47\textwidth]{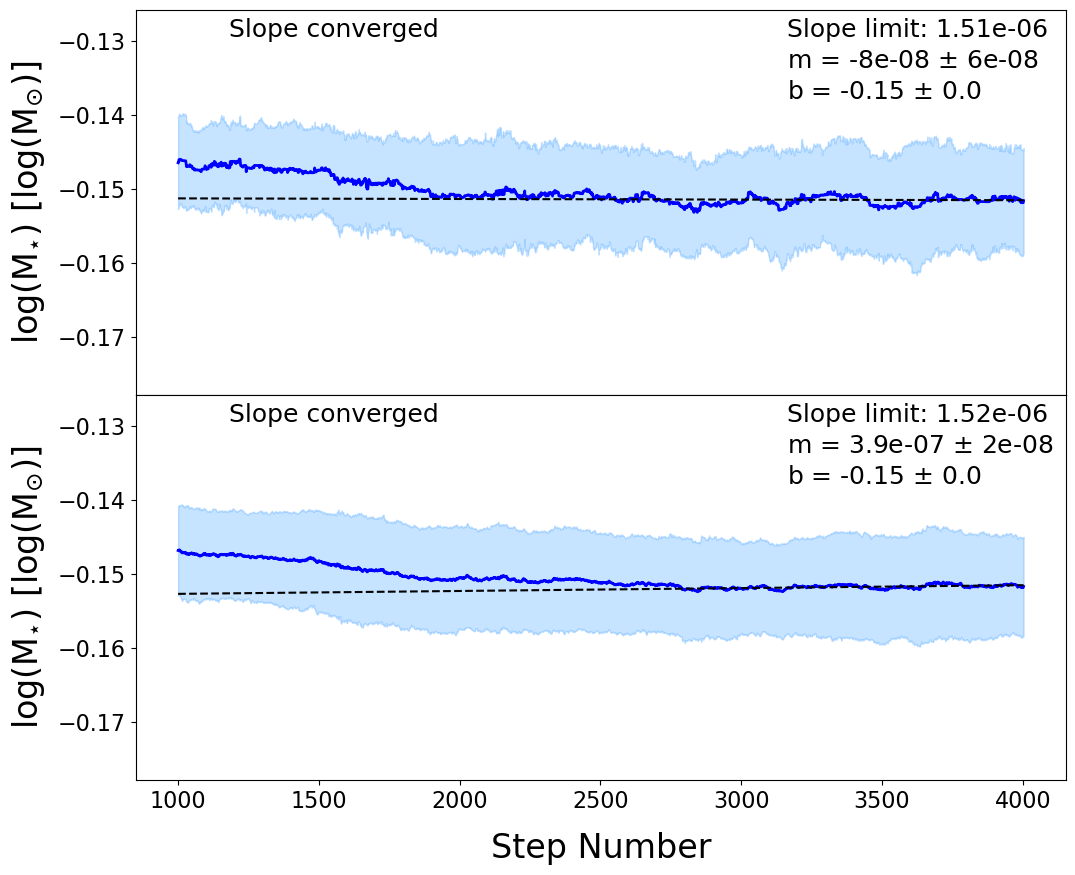}
    \caption{Representative example of the trace plots we used to determine fit convergence. The example shown is for source J16090075-1908526. In both panels, the first 1000 steps are excluded from plotting as they are considered burn-in steps. {\it Top:} The blue line is the median value of all walkers at each step. The blue shaded region is the scaled median absolute deviation from the median (scaled MAD) at each step, and is used to weight the linear fit. The black dotted line shows the best-fit line for the last 1000 steps. Properties of the best-fit line are listed in the upper right corner. {\it Bottom:} Same as for the top panel, except that the blue line is the mean value of all walkers and each step and the blue shaded region is the standard deviation at each step.}
    \label{traceplot_example}
\end{figure}

\subsection{Comparing Methods of Deriving Parameter Values and Uncertainties}
\label{chisq_uncertainties}
In addition to convergence criteria, choosing how to report the value and uncertainty for each parameter can also introduce some arbitrariness to the results. 
By default, {\tt pdspy} reports the median of all walker positions over all post burn-in steps as the parameter value, and the standard deviation of those values as the uncertainty on that value.
\citet{Sheehan2019} report the walkers' median as the parameter value and the $\pm$95\% percentile range as the parameter uncertainty.

We explore how the $\chi^{2}$ best-fit values and their associated uncertainties compare to the median and standard deviation of walker positions.
This exploration was motivated by the possibility that 1$\sigma$ values calculated from $
\Delta\chi^{2}$ might be larger than the walker standard deviations, which can occur when there is not a significant difference in $\chi^{2}$ between the best fit and the next-best fit.
For this comparison, we use only those walkers meeting the 99\% $\chi^2$ cut within the previous 1000 steps, in order to calculate our best-fit values over the same interval as we calculated the fit convergence. 

We calculate the median and standard deviation of the parameter values using the default {\tt pdspy} settings.
We also determine the fit with the lowest $\chi^2$ value, and extract the parameter values used for that fit.
We derive uncertainties for those best-$\chi^2$ parameters using the change in $\chi^2$. 
For a model with 15 free parameters, a value of $\Delta\chi^2$ $=$ 15.975 corresponds to a 1$\sigma$ uncertainty \citep{NumericalRecipesBook}.
We calculate the lower- and upper-bound uncertainties as the minimum and maximum parameter values across all fits with $\chi^2$ $\leq$ $\chi^2_{min}$ $+$ $\Delta \chi^2$ within the last 1000 steps.

We find that the minimum $\chi^{2}$- and median-derived parameter values typically agree within uncertainties, but can have up to a $\sim$1$\sigma$ disagreement (where $\sigma$ is derived using $\Delta\chi^2$).
More notably, we find that the standard deviation-based uncertainties are systematically lower than the $\Delta\chi^2$-based uncertainties by factors of $\sim$2-3.
The median standard deviation-based uncertainty in log(\mstar) is 9$\pm$7\%, whereas the median lower- and upper-bound $\Delta\chi^2$ uncertainties are 17$\pm$11\% and 30$\pm$25\%, respectively. 

\begin{figure*}
    \centering
    \includegraphics[width=\textwidth]{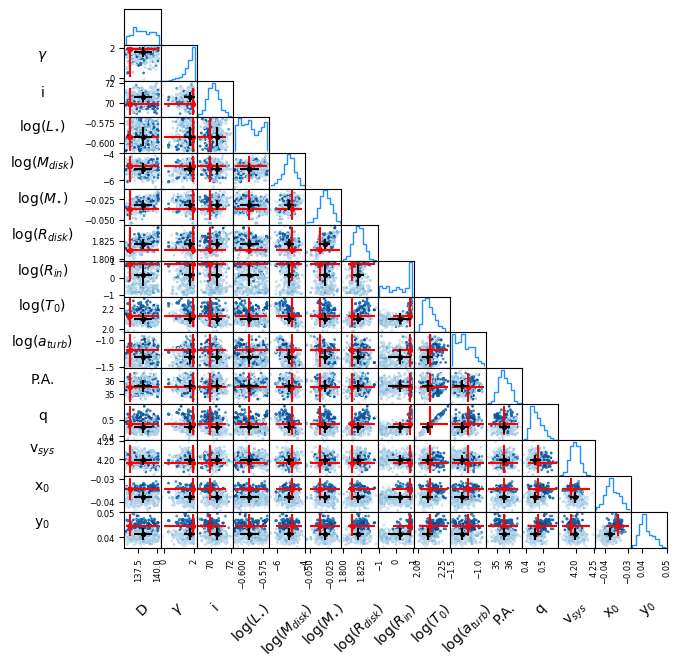}
    \caption{Corner plot showing the distribution of walker values for each parameter for the last 1000 steps. This example is for source J16082324-1930009. The ``x$_0$'' and ``y$_0$'' labels refer to the RA and Dec offsets, respectively. The data points are colored by their $\chi^2$ values, with darker colors indicating lower $\chi^2$. Median walker values are shown in black, with standard deviations shown as black error bars. The $\chi^2$ best-fit values are shown in red, with the $\Delta\chi^2$-based uncertainties shown as red error bars. The median- and $\chi^2$-based best-fit values typically agree within uncertainties, but can clearly disagree by up to $\sim$1 standard deviation. The potential underestimation of uncertainties from the standard deviation method as opposed to the $\Delta\chi^2$ method can be clearly seen.}
    \label{chisq_cornerplot}
\end{figure*}

Figure~\ref{chisq_cornerplot} shows a corner plot with the distribution of walker values for each parameter for source J16082324-1930009.
The median and standard deviation-based parameter values and uncertainties are shown in black, and the $\Delta\chi^2$-based parameter values and uncertainties are shown in red.
These plots and the calculations above suggest that adopting the standard deviation (or even scaled MAD) of walker positions as the parameter uncertainty could give the false impression that a parameter is more well-constrained than it is. 
In this work, we report the parameter values from the minimum-$\chi^2$ fit, and determine upper- and lower-bound uncertainties using $\Delta\chi^2$.

\section{Results}
\label{results}
We successfully fit the CO data for 23 out of 24 sources: $^{12}$CO J$=$3$-$2 for 22 of our 24 sources, and $^{13}$CO J$=$3$-$2 for 4 out of the 5 sources with $^{13}$CO detections. 
We also obtained joint $^{12}$CO$+^{13}$CO fits for the three sources that had usable data in both lines. 
We were unable to fit the $^{12}$CO data for J16113134-1838259 B and any of the line data for J16154416-1921171. 
We discuss these sources, as well as sources which required special handling during the fitting process or returned results significantly at odds with the literature, in greater detail in Appendix~\ref{individual_notes}.

In general, parameters related to the geometry and kinematics of the source ($i$, $\theta$, RA$_{off}$, Dec$_{off}$, \vlsr, \mstar) are well-constrained by our fitting procedure. 
We note that inclination angle ($i$) has a bimodal solution for nearly all sources, with the two solution values approximately equidistant from $i$ $=$ 90$^{\circ}$.
This indicates that we are not differentiating between clockwise and counterclockwise rotation for any targets. 

We show \mstar, \vlsr, $i$, position angle, and center coordinate offsets for each $^{12}$CO fit in Table~\ref{pdspy_results}.
We show \mstar\/ and $i$ for the $^{13}$CO-only and $^{12}$CO$+^{13}$CO fits in Table~\ref{13co_results}. 
Machine-readable and ASCII tables showing the results for all free parameters for each source and line combination are available in the online material.
For the three sources with usable emission in both CO isotopologues, we find that our results for the $^{12}$CO, $^{13}$CO, and $^{12}$CO$+^{13}$CO fits all agree within uncertainties.
However, the uncertainties on the $^{13}$CO-only fit results are a factor of 2-14$\times$ larger than the others.
This is likely due to the lower signal-to-noise ratios in our $^{13}$CO detections as compared to $^{12}$CO (see Table~\ref{obstable}).
The $^{12}$CO$+^{13}$CO fits typically have uncertainties that are lower than or comparable to the $^{12}$CO-only fits. 

\begin{deluxetable*}{lrrrrrrr}
\tablecaption{$^{12}$CO Fit Results}
\tablecolumns{8}
\tablewidth{\textwidth}
\tablehead{
\colhead{Field$^a$} & \colhead{N$_{steps}^b$} & \colhead{M$_{\star}^c$} & \colhead{\vlsr} & \colhead{$i$} & \colhead{P.A.} & \colhead{RA$_{off}^d$} & \colhead{Dec$_{off}^d$}\\
& & \colhead{(\msun)} & \colhead{(\kms)} & \colhead{($^{\circ}$)} & \colhead{($^{\circ}$)} & \colhead{(mas)} & \colhead{(mas)}
}
\startdata
J15521088-2125372 & 4009 & 0.16$^{+0.04}_{-0.03}$ & 4.1$^{+0.2}_{-0.1}$ & 110.0$^{+10.0}_{-20.0}$ & 181.0$^{+7.0}_{-7.0}$ & -40$^{+10}_{-10}$ & 4$^{+8}_{-9}$ \\
J15530132-2114135 & 7343 & 0.25$^{+1.0}_{-0.09}$ & 3.9$^{+0.2}_{-0.2}$ & 140.0$^{+30.0}_{-30.0}$ & 138.0$^{+9.0}_{-20.0}$ & -80$^{+10}_{-20}$ & -50$^{+10}_{-10}$ \\
J15534211-2049282 & 15634 & 0.6$^{+0.6}_{-0.1}$ & 4.4$^{+0.6}_{-2.0}$ & 108.0$^{+5.0}_{-7.0}$ & -22.0$^{+7.0}_{-4.0}$ & -120$^{+60}_{-10}$ & -10$^{+10}_{-20}$ \\
J15562477-2225552 & 3780 & $\leq$1.9 & 3.6$^{+0.2}_{-0.2}$ & 160.0$^{+20.0}_{-20.0}$ & 250.0$^{+20.0}_{-90.0}$ & -7$^{+30}_{-10}$ & 20$^{+20}_{-10}$ \\
J16001844-2230114 & 18539 & 0.5$^{+3.0}_{-0.3}$ & 6.49$^{+0.02}_{-0.03}$ & 9.0$^{+9.0}_{-5.0}$ & 217.0$^{+5.0}_{-3.0}$ & -37$^{+4}_{-3}$ & 14$^{+4}_{-2}$ \\
J16014086-2258103 & 8015 & 0.25$^{+0.25}_{-0.05}$ & 2.45$^{+0.07}_{-0.08}$ & 135.0$^{+20.0}_{-8.0}$ & -13.0$^{+5.0}_{-7.0}$ & -47$^{+4}_{-7}$ & 18$^{+5}_{-5}$ \\
J16020757-2257467 & 3502 & 0.53$^{+0.05}_{-0.04}$ & 3.96$^{+0.06}_{-0.05}$ & 61.0$^{+3.0}_{-4.0}$ & 168.0$^{+2.0}_{-2.0}$ & -6$^{+6}_{-5}$ & -1$^{+7}_{-5}$ \\
J16035767-2031055 & 9209 & 4.0$^{+4.0}_{-2.0}$ & 3.9$^{+0.08}_{-0.06}$ & 17.0$^{+10.0}_{-5.0}$  & -16.0$^{+3.0}_{-4.0}$ & -60$^{+7}_{-5}$ & -30$^{+6}_{-6}$ \\
J16035793-1942108 & 7369 & 0.56$^{+0.05}_{-0.04}$ & 7.33$^{+0.03}_{-0.02}$ & 50.0$^{+3.0}_{-2.0}$ & 92.0$^{+2.0}_{-2.0}$ & -98$^{+4}_{-5}$ & 33$^{+4}_{-4}$ \\
J16062277-2011243 & 5505 & 0.3$^{+0.3}_{-0.1}$ & 6.3$^{+0.2}_{-0.2}$ & 60.0$^{+30.0}_{-30.0}$ & 10.0$^{+10.0}_{-20.0}$ & -150$^{+30}_{-20}$ & -20$^{+20}_{-20}$ \\
J16075796-2040087 & 6214 & 1.9$^{+0.2}_{-0.2}$ & 4.46$^{+0.03}_{-0.05}$ & 52.0$^{+1.0}_{-3.0}$ & 280.0$^{+1.0}_{-1.0}$ & -97$^{+2}_{-2}$ & 15$^{+1}_{-3}$ \\
J16081566-2222199 & 6011 & 0.5$^{+0.1}_{-0.08}$ & 3.3$^{+0.2}_{-0.2}$ & 120.0$^{+10.0}_{-10.0}$ & 3.0$^{+6.0}_{-5.0}$ & -3$^{+10}_{-20}$ & 20$^{+10}_{-10}$ \\
J16082324-1930009 & 6024 & 0.91$^{+0.04}_{-0.02}$ & 4.19$^{+0.05}_{-0.03}$ & 110.0$^{+1.0}_{-2.0}$ & 36.0$^{+1.0}_{-1.0}$ & -34$^{+5}_{-5}$ & 44$^{+5}_{-4}$ \\
J16090075-1908526 & 4004 & 0.71$^{+0.03}_{-0.03}$ & 3.69$^{+0.02}_{-0.02}$ & 125.0$^{+2.0}_{-2.0}$ & 65.3$^{+1.0}_{-0.9}$ & -43$^{+5}_{-5}$ & 2$^{+5}_{-4}$ \\
J16095933-1800090 & 2817 & 0.25$^{+0.05}_{-0.02}$ & 4.11$^{+0.04}_{-0.06}$ & 126.0$^{+9.0}_{-6.0}$ & 148.0$^{+2.0}_{-5.0}$ & -70$^{+10}_{-10}$ & 22$^{+15}_{-6}$ \\
J16104636-1840598 & 5008 & 0.13$^{+0.07}_{-0.05}$ & 4.2$^{+0.1}_{-0.1}$ & 80.0$^{+30.0}_{-30.0}$ & 124.0$^{+8.0}_{-10.0}$ & -58$^{+19}_{-8}$ & 0$^{+10}_{-10}$ \\
J16115091-2012098 & 2500 & 0.31$^{+0.05}_{-0.03}$ & 2.48$^{+0.07}_{-0.04}$ & 127.0$^{+6.0}_{-8.0}$ & 211.0$^{+3.0}_{-3.0}$ & -23$^{+10}_{-6}$ & 16$^{+9}_{-9}$ \\
J16123916-1859284 & 3133 & 0.76$^{+0.11}_{-0.07}$ & 4.79$^{+0.04}_{-0.02}$ & 51.0$^{+4.0}_{-5.0}$ & 16.0$^{+1.0}_{-1.0}$ & -21$^{+9}_{-6}$ & -32$^{+12}_{-5}$ \\
J16142029-1906481 & 10158 & 1.32$^{+0.06}_{-0.06}$ & 3.88$^{+0.02}_{-0.02}$ & 60.2$^{+0.8}_{-0.6}$ & 277.3$^{+0.5}_{-0.7}$ & -37$^{+3}_{-2}$ & 8$^{+2}_{-2}$ \\
J16143367-1900133 & 2504 & 0.23$^{+0.05}_{-0.04}$ & 3.15$^{+0.07}_{-0.07}$ & 104.0$^{+3.0}_{-2.0}$ & 227.0$^{+5.0}_{-6.0}$ & -65$^{+7}_{-6}$ & 11$^{+4}_{-7}$ \\
J16163345-2521505 & 2500 & 0.8$^{+0.1}_{-0.1}$ & 6.5$^{+0.1}_{-0.2}$ & 119.0$^{+6.0}_{-9.0}$ & -27.0$^{+5.0}_{-3.0}$ & -70$^{+20}_{-10}$ & -10$^{+10}_{-20}$ \\
J16181904-2028479 & 5214 & 0.25$^{+0.15}_{-0.05}$ & 4.6$^{+0.1}_{-0.1}$ & 130.0$^{+10.0}_{-20.0}$ & 50.0$^{+10.0}_{-6.0}$ & -10$^{+10}_{-10}$ & 0$^{+10}_{-10}$ \\
\enddata
\tablenotetext{a}{Best-fit results for selected parameters for $^{12}$CO {\tt pdspy} fits. A full table of results is available in machine-readable form in the online material. Source J16154416-1921171 could not be fit due to spatially-filtered extended emission. Source J16113134-1838259 B could not be fit in $^{12}$CO due to significant extended emission and the presence of two sources in the field.}
\tablenotetext{b}{Total number of steps (including burn-in) required for each fit to converge.}
\tablenotetext{c}{Uncertainties on all parameters come from the minimum and maximum parameter value within the $\Delta\chi^2$ contour corresponding to 1$\sigma$ uncertainty, as described in \S~\ref{chisq_uncertainties}.}
\tablenotetext{d}{The RA and Dec coordinates for which these offsets are derived are listed in Table~\ref{source_properties}.}
\label{pdspy_results}
\end{deluxetable*}

Because radiative-transfer MCMC fitting is a time- and computing resource-intensive procedure, the tradeoff between the time to complete an individual step and the total number of steps required to achieve convergence may be of interest to some readers.
We find that the $^{12}$CO$+^{13}$CO fits tend to converge in fewer steps than the single-line fits. 
This is not a universal result, however.
In the case of J16095933-1800090, the joint fit took slightly longer to converge than the $^{12}$CO fit (3013 steps versus 2817 steps). 
J16095933-1800090 is the faintest $^{13}$CO detection in the sample at just 5.1$\sigma$, and this low signal-to-noise may have contributed to the increase in convergence time. 
Our results for these three sources suggest that jointly fitting multiple lines can be an overall time-saving approach, but only if all lines are sufficiently strong (S/N $\gtrsim$ 6$\sigma$). 

\begin{deluxetable*}{l|ccc|ccc|ccc}
\tablecaption{$^{13}$CO and Joint $^{12}$CO$+$$^{13}$CO Fit Results}
\tablecolumns{10}
\tablewidth{\textwidth}
\tablehead{
\colhead{} & \multicolumn{3}{c}{$^{12}$CO} & \multicolumn{3}{c}{$^{13}$CO} &  \multicolumn{3}{c}{$^{12}$CO$+$$^{13}$CO} \\
\colhead{} & \colhead{Steps} & \colhead{M$_{\star}$} & \colhead{i} & \colhead{Steps} & \colhead{M$_{\star}$} & \colhead{i} & \colhead{Steps} & \colhead{M$_{\star}$} & \colhead{i} \\
\colhead{Field$^a$} & \colhead{} & \colhead{(\msun)} & \colhead{($^{\circ}$)} & \colhead{} & \colhead{(\msun)} & \colhead{($^{\circ}$)} & \colhead{} & \colhead{(\msun)} & \colhead{($^{\circ}$)}
}
\startdata
J16020757-2257467 & 3502 & 0.54$^{+0.05}_{-0.04}$ & 61$^{+3}_{-4}$ & 4003 & 0.52$^{+0.17}_{-0.09}$ & 60$^{10}_{-13}$ & 2949 & 0.52$^{+0.03}_{-0.03}$ & 60$^{+3}_{-3}$\\
J16035793-1942108 & 7369 & 0.56$^{+0.05}_{-0.03}$ & 50$^{+3}_{-2}$ & 6003 & 0.77$^{+0.68}_{-0.32}$ & 40$^{+19}_{-11}$ & 2520 & 0.56$^{+0.04}_{-0.04}$ & 50$^{+2}_{-3}$\\
J16095933-1800090 & 2817 & 0.23$^{+0.05}_{-0.02}$ & 126$^{+9}_{-6}$ & 3896 & 0.22$^{+0.19}_{-0.05}$ & 130$^{+19}_{-18}$ & 3013 & 0.25$^{+0.03}_{-0.03}$ & 129$^{+7}_{-7}$\\
J16113134-1838259 B & \nodata & \nodata & \nodata & 2520 & 1.5$^{+0.1}_{-0.1}$ & 65$^{+1}_{-4}$ & \nodata & \nodata & \nodata
\enddata
\tablenotetext{a}{Best-fit results for selected parameters for $^{13}$CO {\tt pdspy} fits, and for combined $^{12}$CO+$^{13}$CO fits when performed. $^{12}$CO-only results are listed for comparison for those sources with $^{12}$CO fits available. Number of steps, best-fit values, and uncertainties follow the same system as in Table~\ref{pdspy_results}. A machine-readable table showing full results for both the $^{13}$CO-only and combined $^{12}$CO$+^{13}$CO fits is available in the online material.}
\label{13co_results}
\end{deluxetable*}

Across the full sample and using the highest-confidence best-fit parameters for each source, we find minimum and maximum stellar masses of 0.13$^{+0.05}_{-0.07}$~\msun\/ (for J16104636-1840598) and 4.0$^{+2.0}_{-4.0}$~\msun\/ (for J16035767-2031055).
The sample median and mean masses are M$_{\star}$ = 0.51$\pm$0.26~\msun\/ and \mstar\/ = 0.76$\pm$0.84~\msun, respectively, where the uncertainties are the sample standard deviation and scaled MAD.
The median lower-bound uncertainty is 0.06~\msun\/ (or, 16\% of the mass of a given source), and the median upper-bound uncertainty is 0.09~\msun\/ (or, 19\% of the source mass). 

\section{Analysis}
\label{analysis}
The following analysis excludes source J16154416-1921171, which could not be fit (see \S~\ref{results}).  
This gives us a total of 23 sources with dynamically-constrained masses 
\footnote{Note that the upper-limit mass value for J15562477-2225552 is excluded from our sample statistics, but is included in the figures where possible.}.

The reliability and accuracy of pre-main sequence evolutionary tracks for deriving PMS masses and ages is a matter of great concern to the star- and planet-formation communities.
Keplerian masses are an independent measurement of stellar mass, and can serve as a powerful evaluation tool for isochronal models. 
For example, \citet{Barenfeld2016} use the PMS evolutionary models of \citet{Siess2000} to derive \mstar\/ for their sample. 
We find that our median and mean dynamical masses are 70\% and 105\% larger, respectively, than the isochronal masses derived by \citet{Barenfeld2016} for these same sources.
Our minimum dynamical mass is consistent with that of \citet{Barenfeld2016}, but our maximum dynamical mass is $\sim$4$\times$ larger.
The higher-mass outlier sources are almost certainly the reason for our higher mean mass, but cannot fully explain our higher median mass. 
Potential explanations for these trends are discussed in detail in \S~\ref{pms_masses} and Appendix~\ref{binarity}, respectively. 
We find median lower- and upper-bound mass uncertainties of 16\% and 19\%, respectively, as discussed above.
This is a moderate improvement over the results of \citet{Barenfeld2016}, which have median lower- and upper-bound uncertainties of 24\% and 26\%, respectively, for masses derived using T$_{eff}$, \lstar, and the isochrones of \citet{Siess2000}.

\subsection{Stellar Evolutionary Models Considered}
\label{evotracks_comparisons}
To more robustly compare our dynamical results with isochronal methods, we re-derive isochronal masses and ages for each source using the \lstar\/ and T$_{eff}$ in Table~\ref{source_properties} and five sets of evolutionary models: the BHAC15 tracks of \citet{Baraffe2015}, the PARSEC v1.1 and 1.2S tracks of \citet{Bressan2012} and \citet{Chen2014}, respectively, and both the non-magnetic and magnetic tracks of \citet{Feiden2016}.
We briefly describe each set of tracks in Appendix~\ref{track_descriptions}.
Figure~\ref{isochrones} shows T$_{eff}$ and \lstar\/ values for our targets overlaid on isochrone contours for each set of models. 
We compare our dynamical results to the isochrone-derived results for each model set and evaluate which, if any, agree consistently with the dynamical results.
In this paper, we focus on the dynamical masses as compared to the isochrone-inferred masses.
The stellar ages are discussed in detail in a companion paper (Towner et al. 2025b, in prep).

\begin{figure*}
    \includegraphics[width=0.33\textwidth]{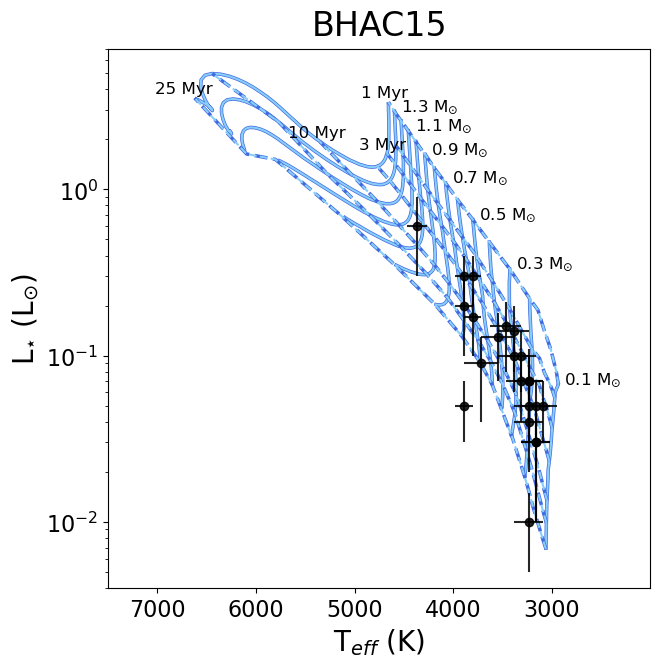}
    \includegraphics[width=0.33\textwidth]{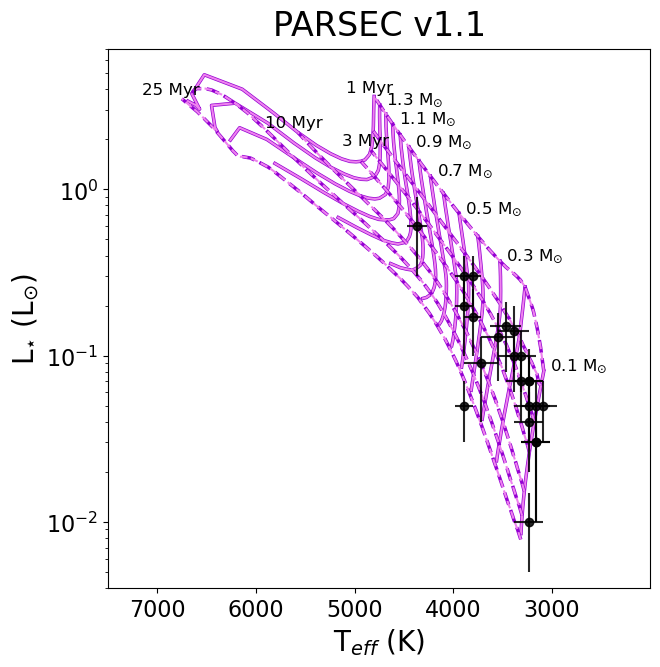}
    \includegraphics[width=0.33\textwidth]{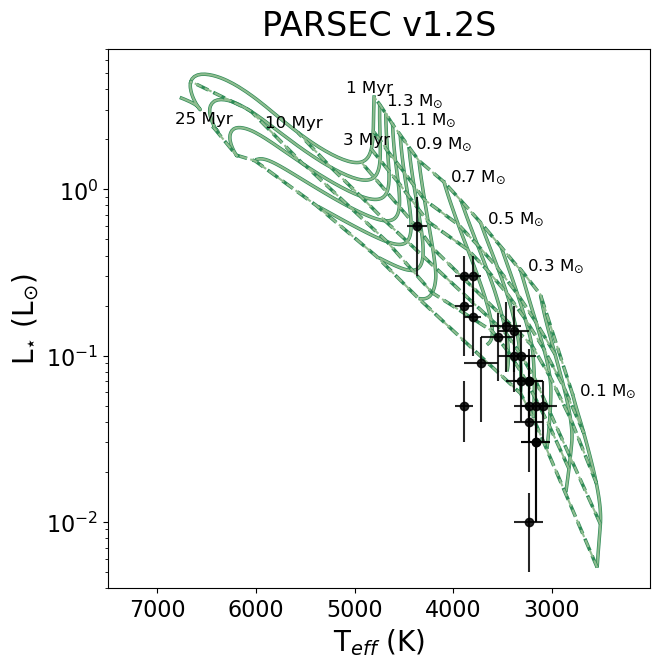}\\
    \includegraphics[width=0.33\textwidth]{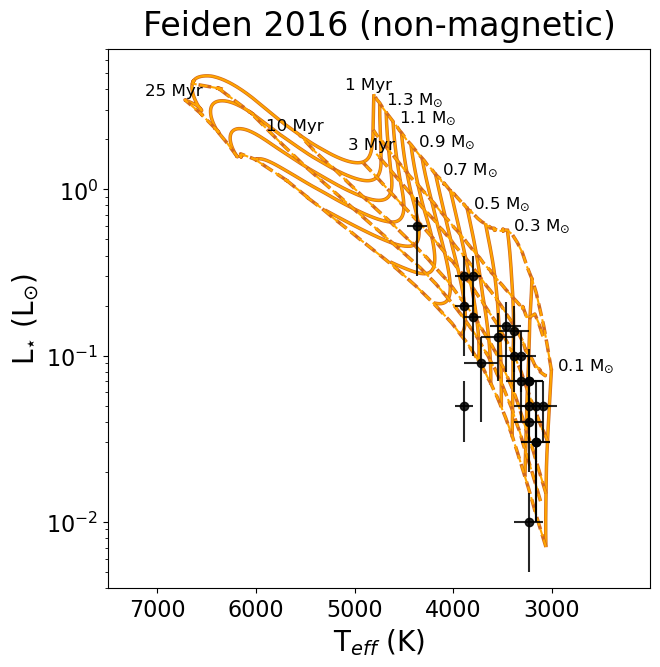}
    \includegraphics[width=0.33\textwidth]{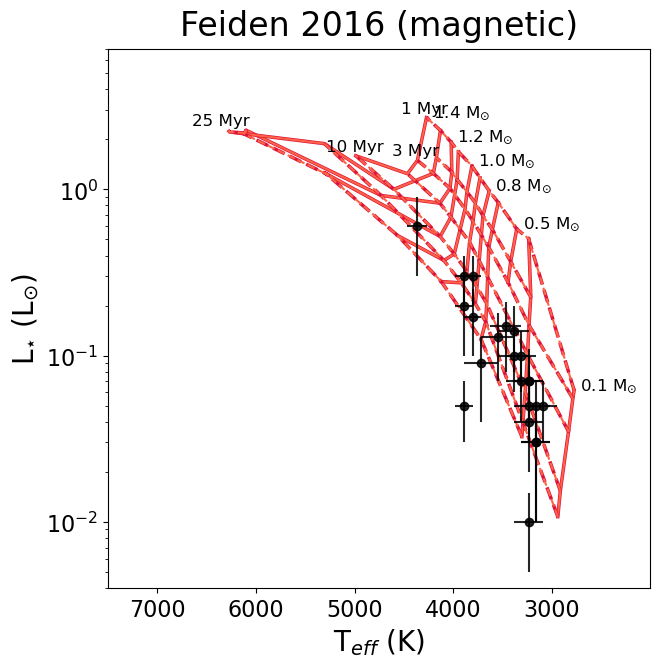}
    \caption{Isochronal age and mass contour lines for each of the five model sets considered in this paper. The mass and age ranges are 0.1 to 1.4~\msun\/ and 1 to 25~Myr in all panels. Black points show the \lstar\/ and T$_{eff}$ for our sample. Error bars come from the uncertainties listed in Table~\ref{source_properties}. The contours for each model are shown in a different color for ease of visual distinction; this color scheme is carried through all related figures for the rest of this paper. {\it Top row:} BHAC15 \citep{Baraffe2015} (left, in blue), PARSEC v1.1 \citep{Bressan2012} (center, in magenta), PARSEC v1.2S \citep{Chen2014} (right, in green). {\it Bottom row:} non-magnetic models of \citet{Feiden2016} (left, in orange), magnetic models of \citet{Feiden2016} (center, in red).}
    \label{isochrones}
\end{figure*}

To derive isochronal masses, we generate normal distributions in luminosity and temperature for each source.
The peak values and widths of each distribution are the source's L$_{\star}$ and T$_{eff}$ and their corresponding uncertainties, respectively (see Table~\ref{source_properties}).
We then feed these normal distributions into the PMS tracks of each model set. 
This step is done with the {\tt pdspy} package {\tt pdspy.stars}, 
which interpolates linearly between grid points in each isochrone.
The result is five distributions of isochronal mass for each target.
In the following analysis, we use the median and scaled MAD of each distribution as mass and uncertainty, respectively, for each model. 
These isochronal masses are listed in Table~\ref{isochronal_masses}.
Table~\ref{mass_ratios} shows minimum, maximum, median, and mean mass ratios (M$_{iso}$/M$_{dyn}$) for each set of tracks.

\begin{deluxetable*}{l|ccccc}
\tablecaption{Stellar Masses Derived from Evolutionary Tracks}
\tablecolumns{6}
\tablewidth{\textwidth}
\tablehead{
\colhead{Field} & \colhead{BHAC15$^a$} & \colhead{PARSEC v1.1} & \colhead{PARSEC v1.2S} & \colhead{Feiden (2016), n-m} & \colhead{Feiden (2016), m}\\
\colhead{} & \colhead{(\msun)} & \colhead{(\msun)} & \colhead{(\msun)} & \colhead{(\msun)} & \colhead{(\msun)}
}
\startdata
J15521088-2125372 & 0.18 (0.08) & 0.11 (0.02) & 0.23 (0.05) & 0.17 (0.06) & 0.22 (0.07) \\
J15530132-2114135 & 0.20 (0.08) & 0.13 (0.04) & 0.39 (0.10) & 0.19 (0.08) & 0.26 (0.09) \\
J15534211-2049282 & 0.26 (0.09) & 0.16 (0.05) & 0.5 (0.1)   & 0.24 (0.10) & 0.3 (0.1) \\
J15562477-2225552 & 0.21 (0.08) & 0.13 (0.04) & 0.41 (0.10) & 0.19 (0.08) & 0.27 (0.09) \\
J16001844-2230114 & 0.17 (0.07) & 0.12 (0.03) & 0.37 (0.09) & 0.15 (0.07) & 0.23 (0.08) \\
J16014086-2258103 & 0.22 (0.08) & 0.14 (0.04) & 0.4 (0.1)   & 0.20 (0.08) & 0.28 (0.10) \\
J16020757-2257467 & 0.3 (0.1)   & 0.25 (0.08) & 0.6 (0.1)   & 0.3 (0.1)   & 0.5 (0.1) \\
J16035767-2031055 & 1.0 (0.1)   & 0.9  (0.1)  & 0.9 (0.1)   & 0.9 (0.1)   & 1.1 (0.2) \\
J16035793-1942108 & 0.4 (0.1)   & 0.28 (0.08) & 0.6 (0.1)   & 0.4 (0.1)   & 0.6 (0.1)  \\
J16062277-2011243 & 0.14 (0.06) & 0.11 (0.02) & 0.33 (0.09) & 0.13 (0.05) & 0.19 (0.07) \\
J16075796-2040087 & 0.5 (0.1)   & 0.3 (0.1)   & 0.64 (0.07) & 0.5 (0.1)   & 0.6 (0.1) \\
J16081566-2222199 & 0.29 (0.09) & 0.22 (0.07) & 0.5 (0.1)   & 0.3 (0.1)   & 0.4 (0.1) \\
J16082324-1930009 & 0.65 (0.08) & 0.52 (0.08) & 0.71 (0.02) & 0.66 (0.08) & 0.8 (0.1) \\
J16090075-1908526 & 0.65 (0.09) & 0.52 (0.07) & 0.72 (0.01) & 0.65 (0.09) & 0.88 (0.08) \\
J16095933-1800090 & 0.21 (0.08) & 0.14 (0.04) & 0.4 (0.1)   & 0.20 (0.08) & 0.28 (0.10) \\
J16104636-1840598 & 0.15 (0.06) & 0.11 (0.02) & 0.33 (0.09) & 0.15 (0.06) & 0.21 (0.08) \\
J16113134-1838259 B & 0.70 (0.09) & 0.57 (0.08) & 0.72 (0.02) & 0.7 (0.1)   & 0.99 (0.08) \\
J16115091-2012098 & 0.26 (0.09) & 0.18 (0.06) & 0.5 (0.1)   & 0.24 (0.09)  & 0.3 (0.1) \\
J16123916-1859284 & 0.56 (0.08) & 0.45 (0.06) & 0.70 (0.02) & 0.57 (0.09) & 0.83 (0.08) \\
J16142029-1906481 & 0.54 (0.06) & 0.37 (0.07) & 0.57 (0.08) & 0.47 (0.08) & 0.54 (0.06) \\
J16143367-1900133 & 0.3 (0.1)   & 0.20 (0.07) & 0.5 (0.1)   & 0.3 (0.1)   & 0.4 (0.1) \\
J16163345-2521505 & 0.60 (0.07) & 0.45 (0.07) & 0.70 (0.02) & 0.61 (0.08) & 0.75 (0.09) \\
J16181904-2028479 & 0.16 (0.06) & 0.11 (0.02) & 0.33 (0.09) & 0.15 (0.06) & 0.21 (0.07) 
\enddata
\tablenotetext{a}{Isochronal masses for each of the five stellar evolutionary models considered in this work. For all models, the mass and uncertainty are, respectively, the median and scaled MAD of the distribution of masses returned by the isochrones. Column 5 shows the masses derived using the non-magnetic models of \citet{Feiden2016}, and Column 6 shows the masses derived using the magnetic models in that work.}
\label{isochronal_masses}
\end{deluxetable*}

\subsection{Comparing Dynamical and Isochronal Masses}
\label{pms_masses}
We compare isochronal versus dynamical masses for all sources in our sample for each of the five model sets.
Scatterplots of M$_{iso}$ versus M$_{dyn}$ show a clear linear relationship between M$_{iso}$ and M$_{dyn}$ up to a mass of $\sim$1~\msun, and a shallowing of that relationship above 1~\msun.
However, we also have very sparse data coverage above M$_{dyn}$ $=$ 1~\msun, and it is unclear whether this shallowing is a true physical trend or simply an artifact of our current sampling.
Additionally, a much larger fraction of our sources above 1~\msun\/ are known or candidate binary systems, which could also be contributing to this trend.

To account for these potential confounding factors, we perform a statistical analysis of dynamical versus isochronal masses for two cases: once for the full sample, and once for only those sources with M$_{dyn}$ $\leq$ 1.0~\msun, where we have much better data coverage.
This restriction excludes five sources: J15562477-2225552 (whose mass is an upper limit), J16035767-2031055 and J16142029-1906481 (discussed above), and J16075796-2040087 and J16113134-1838259 B (both of which have known or candidate companions).
Statistics for both cases are presented in Table~\ref{mass_ratios}. 
We find that, in practice, our analysis is not meaningfully impacted by using the mass-limited sample versus the full sample. 
In this subsection, we discuss results for the M$_{dyn}$ $\leq$ 1.0~\msun\/ sample only.
Figures and a brief discussion for the full range of masses can be found in Appendix~\ref{full_sample_discussion}. 

In Figure~\ref{isochrone_mass_comparison}, we show scatterplots of the isochrone-derived versus dynamical masses for each of the five sets of tracks for M$_{dyn}$ $\leq$ 1.0~\msun.
A 1:1 relationship is shown by a black solid line in all panels.
We fit a line to the data in each panel using the {\tt scipy.odr} package, which performs Orthogonal Distance Regression (ODR) fitting.
The ODR method accounts for uncertainties in both the dependent (M$_{iso}$) and independent (M$_{dyn}$) variables. 
The best fit for isochronal versus dynamical mass is shown as a dashed, colored line in each panel.

We find that the non-magnetic \citet{Feiden2016} models, BHAC15, and PARSEC v1.1 all consistently return lower masses than the dynamical mass, with slopes that do not agree with 1 within uncertainties (Fig.~\ref{isochrone_mass_comparison}, top left, top center, and bottom left panel, respectively).
The PARSEC v1.2S models, on the other hand, are about as likely to underestimate versus overestimate the dynamical masses.
However, the overall slope of the trend is still very shallow and, unlike the other model sets, does not intersect the y-axis at zero (slope = 0.6$\pm$0.1, intercept = 0.2$\pm$0.0; see 
Fig.~\ref{isochrone_mass_comparison}, top right panel).

In contrast, the magnetic models of \citet{Feiden2016} agree very consistently with the dynamical mass results for \mstar\/ $\leq$ 1~\msun\/ (Fig.~\ref{isochrone_mass_comparison}, bottom center panel).
The ODR best-fit line (slope = 1.0$\pm$0.1, intercept = 0.03$\pm$0.05) is consistent with a 1:1 relationship between the two methods within uncertainties, and the scatterplot shows no strong tendency toward over- or under-estimation by the magnetic models within this mass range.

\begin{figure*}
    \centering
    \includegraphics[width=\textwidth]{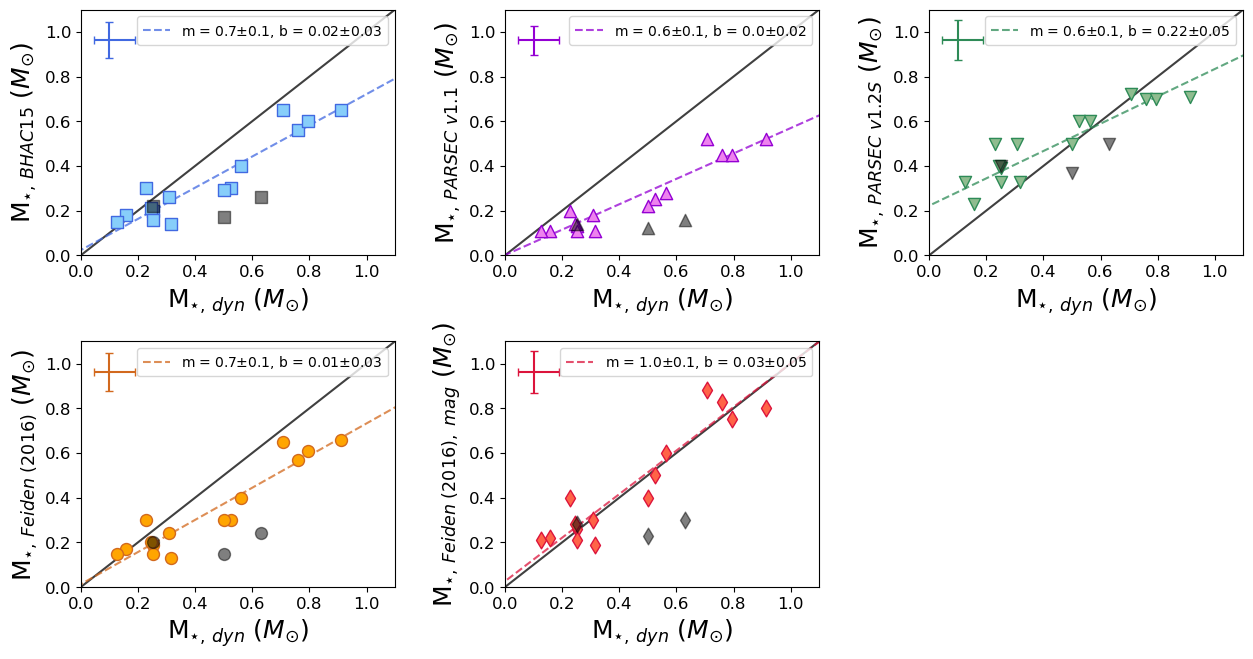}
    \caption{Comparison of stellar masses returned by each PMS evolutionary model set to the dynamical masses we derive using {\tt pdspy}. The color scheme is the same as in Figure~\ref{isochrones}. In all panels, the isochronal mass is shown on the y-axis, and the dynamical mass (same in all cases) is shown on the x-axis. The black dotted line represents a 1:1 relation in all panels. The ODR best-fit line slope (m) and intercept (b) are listed in each panel. Top row: BHAC15 (left), PARSEC v1.1 \citep{Bressan2012} (center), PARSEC v1.2S \citep{Chen2014} (right). Bottom row: non-magnetic models of \citet{Feiden2016} (left), magnetic models of \citet{Feiden2016} (center). The significant improvement in agreement of the magnetic models of \citet{Feiden2016} with the dynamical masses, as compared to the other PMS evolutionary models tested here, can be clearly seen in the bottom center panel. A figure covering the full range of masses can be found in Appendix~\ref{full_sample_discussion}.}
    \label{isochrone_mass_comparison}
\end{figure*}

These results are also consistent with the isochronal- versus dynamical-mass ratios for each model set. 
Population-level statistics for these ratios (minimum, maximum, median, and mean) are presented in Table~\ref{mass_ratios}, columns 2-5. 
The BHAC15, PARSEC v1.1, and non-magnetic \citet{Feiden2016} models all have minimum and maximum mass ratios 1.3-3.1$\times$ smaller than those of the PARSEC v1.2S and magnetic \citet{Feiden2016} models, and their median and mean mass ratios typically do not agree with unity within uncertainties.
In contrast, the magnetic \citet{Feiden2016} tracks have both median and mean mass ratios within 2\% of unity, with uncertainties comparable to those of the other models.
The PARSEC v1.2S models, meanwhile, have median and mean mass ratios larger than unity by $\geq$11\%, i.e., they are more likely to overestimate mass than to underestimate it. 
The uncertainties on the PARSEC v1.2S median and mean are also notably larger than the uncertainties for any other model set.

We also calculate Spearman $\rho$ correlation coefficients between the dynamical masses and the M$_{iso}$/M$_{dyn}$ mass ratios.
These coefficients are listed in columns 6 and 7 of Table~\ref{mass_ratios}.
We consider a correlation to be significant if it has a p-value less than 6.3$\times$10$^{-5}$ (corresponding to $\geq$ 4$\sigma$ significance assuming normally-distributed data). 
All model sets have a negative correlation between mass ratio and dynamical mass, but only the PARSEC v1.2S models have a p-value that indicates statistical significance.
Both the highly-negative correlation coefficient ($\rho$ $=$ -0.84) and small p-value ($>$4$\sigma$) indicate that the accuracy of the PARSEC v1.2S models varies with mass in a statistically-significant way.
This is consistent with Figure~\ref{isochrone_mass_comparison}, in which the data for PARSEC v1.2S have relatively little scatter but clearly do not follow a 1:1 line. 
Rather, PARSEC v1.2S is highly likely to overestimate stellar mass for M$_{dyn}$ $\lesssim$0.6~\msun, and underestimate it for M$_{dyn}$ $\gtrsim$0.6~\msun.

\begin{deluxetable*}{lcccc|cc|cc}
\tablecaption{Sample Statistics for Isochronal vs. Dynamical Masses}
\tablecolumns{9}
\tablewidth{\textwidth}
\tablehead{
\colhead{} & \multicolumn{4}{c}{Mass Ratio (M$_{iso}$/M$_{dyn}$)} & \multicolumn{2}{c}{\,M$_{iso}$/M$_{dyn}$\,\,vs\,\,M$_{dyn}$\,} & \multicolumn{2}{c}{\,M$_{iso}$\,\,vs\,\,M$_{dyn}$\,}\\
\colhead{PMS Model Set} & \colhead{Min} & \colhead{Max} & \colhead{Median} & \colhead{Mean} & \colhead{$\rho^a$} & \colhead{p} & \colhead{D$^b$} & \colhead{p} 
}
\startdata
\cutinhead{\mstar\/ $\leq$ 1.0 \msun\/ Only$^c$}
BHAC15 & 0.34 & 1.31 & 0.75 (0.22) & 0.77 (0.26) & $-$0.51 & 0.03 & 0.33 & 0.3 \\
PARSEC v1.1 & 0.24 & 0.87 & 0.56 (0.15) & 0.55 (0.17) & $-$0.28 & 0.3 & 0.56 & 7$\times$10$^{-3}$ \\
PARSEC v1.2S & 0.74 & 2.62 & 1.11 (0.47) & 1.30 (0.49) & $-$0.84 & 1$\times$10$^{-5}$ & 0.44 & 0.06 \\
Feiden (2016), non-magnetic & 0.30 & 1.31 & 0.75 (0.23) & 0.75 (0.26) & $-$0.48 & 0.04 & 0.39 & 0.1 \\
Feiden (2016), magnetic & 0.46 & 1.75 & 1.00 (0.23) & 1.02 (0.34) & $-$0.49 & 0.04 & 0.22 & 0.8 \\
\cutinhead{Full Sample$^d$}
BHAC15 & 0.25 & 1.31 & 0.71 (0.27) & 0.69 (0.29) & $-$0.69 & 4$\times$10$^{-4}$ & 0.32 & 0.2\\
PARSEC v1.1 & 0.15 & 0.87 & 0.51 (0.17) & 0.49 (0.19) & $-$0.56 & 7$\times$10$^{-3}$ & 0.45 & 0.02\\
PARSEC v1.2S & 0.23 & 2.62 & 1.03 (0.53) & 1.13 (0.58) & $-$0.91 & 4$\times$10$^{-9}$ & 0.36 & 0.1\\
Feiden (2016), non-magnetic & 0.23 & 1.31 & 0.72 (0.26) & 0.67 (0.29) & $-$0.68 & 5$\times$10$^{-4}$ & 0.32 & 0.2\\
Feiden (2016), magnetic & 0.28 & 1.75 & 0.95 (0.36) & 0.91 (0.39) & $-$0.69 & 4$\times$10$^{-4}$ & 0.18 & 0.9
\enddata
\tablenotetext{a}{Spearman $\rho$ correlation coefficients and p-values for all five model sets. 
Lower values of $\rho$ indicate that a model set more severely underestimates mass as mass increases. For normally-distributed data, p-values correspond to $\sigma$-values according to: 1$\sigma$: p $\leq$ 0.68; 2$\sigma$: p $\leq$ 0.05; 3$\sigma$: p $\leq$ 2.7$\times$10$^{-3}$; 4$\sigma$: p $\leq$ 6.3$\times$10$^{-5}$; 5$\sigma$:  p $\leq$ 5.7$\times$10$^{-7}$; 6$\sigma$: p $\leq$ 1.97$\times$10$^{-9}$.}
\tablenotetext{b}{Kolmogorov-Smirnoff (KS) test results for the comparison of M$_{dyn}$ and M$_{iso}$ for each of the five sets of isochrones. $D$ is the difference value for each KS test and p is the corresponding p-value. A higher value of $D$ indicates a greater difference between isochronal and dynamical mass.}
\tablenotetext{c}{The median differences in mass (M$_{dyn}$~-~M$_{iso}$) in the mass range $\leq$1~\msun\/ are: $+$0.13~\msun\/ for BHAC15, $+$0.2~\msun for PARSEC v1.1, $-$0.05~\msun\/ for PARSEC v1.2S, $+$0.13~\msun\/ \citet{Feiden2016} non-magnetic models, and $+$0.0001~\msun\/ for the \citet{Feiden2016} magnetic models.}
\tablenotetext{d}{The median differences in mass (M$_{dyn}$~-~M$_{iso}$) for the full sample are: $+$0.19~\msun\/ for BHAC15, $+$0.28~\msun\/ for PARSEC v1.1, $-$0.01~\msun\/ for PARSEC v1.2S, $+$0.19~\msun\/ for the \citet{Feiden2016} non-magnetic models, and $+$0.03~\msun\/ for the \citet{Feiden2016} magnetic models.}
\label{mass_ratios}
\end{deluxetable*}

\begin{figure}
    \centering
    \includegraphics[width=0.47\textwidth]{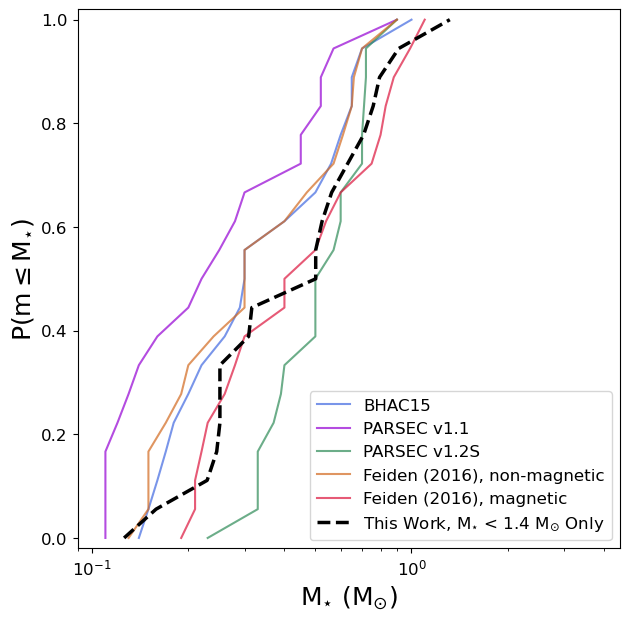}
    \caption{Cumulative distribution functions (CDFs) of stellar mass for the five PMS evolutionary models considered in this work, along with the dynamical masses we derive using {\tt pdspy}. In this Figure, we include only those sources whose dynamical mass is $\leq$1~\msun. A figure showing CDFs using the full sample can be found in Appendix~\ref{full_sample_discussion}.}
    \label{mass_cdfs}
\end{figure}

In order to more fully quantify the similarity of the isochrone-derived and dynamical masses, we employ the 2-sample Kolmogorov-Smirnov (KS) test.
The KS test quantifies how likely two sample distributions are to have been drawn from the same parent distribution.
The KS test statistic (D) is a measure of the maximum distance between the Cumulative Distribution Functions (CDFs) of the two samples, and its p-value represents the statistical likelihood of the null hypothesis (here, that the two samples were drawn from the same parent distribution). 
The KS test statistics and p-values for our sample are shown in the last two columns of Table~\ref{mass_ratios}.
Figure~\ref{mass_cdfs} shows the CDFs used in the KS tests.

Our KS test results confirm that the magnetic tracks of \citet{Feiden2016} show the smallest difference with our dynamical masses (D$=$0.22).
However, it must be noted that none of the PMS models have a p-value that explicitly rejects the null hypothesis with $\geq$4$\sigma$ confidence (p $\leq$ 6.3$\times$10$^{-5}$). 
This is consistent with the generally correlated, but rarely 1:1, isochronal versus dynamical masses shown in Figure~\ref{isochrone_mass_comparison}.

Taken together, the ODR best-fit line, mass-ratio, Spearman $\rho$, and KS-test results suggest the magnetic models of \citet{Feiden2016} are the most reliable isochronal method of determining stellar masses in the \mstar\/ $\leq$ 1~\msun\/ mass range, for our 5-11~Myr targets.

\subsection{Comparison with Literature Results}
\label{age_mass_literature}
To place our results in broader context, we compare our findings to the dynamical-isochronal mass comparisons of \citet{Rizzuto2016}, \citet{Simon2019}, and \citet{Braun2021}. 

\citet{Rizzuto2016} use astrometric measurements to derive the orbital and stellar properties of seven binary pairs, six of which are in Upper Sco: J15500499-2311537, J15573430-2321123, J16015822-2008121, J16051791-2024195, J16081474-1908327, and J16245136-2239325.
They compare these dynamical results to those of the Padova models of \citet{Girardi2002}, the Dartmouth models of \citet{Dotter2008}, and the BT-Settl models of \citet{Allard2011}.
\citet{Rizzuto2016} find that the isochronal and dynamical masses agree within uncertainties for the G-type stars in their sample. 
However, the isochronal masses for the M-type stars are highly model-dependent.
The \citet{Girardi2002} models significantly overestimate mass for the M stars in their sample, and the \citet{Dotter2008} and \citet{Allard2011} models consistently underestimate mass by 0.2$-$0.4~\msun.

\citet{Simon2019} use CO line data in disks to derive dynamical masses for 29 stars in Taurus and 3 in Ophiuchus. 
They use the DiskFit package \citep{Pietu2007} to fit a power-law disk model to the CO, assuming a vertically-isothermal temperature structure and imposing a CO depletion zone near the disk midplane.
They compare these dynamical masses to those derived from the BHAC15 and the magnetic \citet{Feiden2016} models, and find that the non-magnetic models typically underestimate stellar mass by approximately 30\%.
In contrast, the magnetic models of \citet{Feiden2016} have a typical difference of only 0.01$\pm$0.02~\msun\/ with the dynamical masses.

In a separate study, \citet{Braun2021} derive masses for 45 pre-main sequence stars in Taurus and Lupus (17 in Taurus, 28 in Lupus) using Doppler-shifted, stacked $^{12}$CO, C$^{18}$O, and CN line data.
They follow the method detailed in \citet{Yen2016,Yen2018}, which is similar to a Keplerian masking approach and assumes 1) a thin disk geometry and 2) identical excitation conditions for all line emission.
\citet{Braun2021} compare their dynamical masses to the five model sets we consider in this work, plus those of \citet{Palla1993} and \citet{Siess2000}.
Based on the mean difference between isochronal and dynamical masses in three discrete mass bins (\mstar\/ $\leq$ 0.6~\msun, 0.6~$\leq$~\mstar~$\leq$~1.3 \msun, \mstar\/ $\geq$~1.3~\msun), they conclude that the magnetic models of \citet{Feiden2016} agree well with dynamical masses for intermediate-mass sources but tend to overestimate stellar masses for the lowest-mass sources (\mstar\/ $\leq$ 0.6~\msun).

Overall, our findings agree well with the literature: the isochronal masses for PMS stars are highly model-dependent but, on the whole, non-magnetic models tend to return lower values than our dynamical masses by $\gtrsim$25\% (median absolute differences of $\gtrsim$0.13 to 0.2~\msun, see Table~\ref{mass_ratios}).
The exception is, as noted above, the PARSEC v1.2S model set, which is slightly more likely to overestimate dynamical masses than underestimate them.
In contrast, the magnetic models of \citet{Feiden2016} tend to return masses that are in good agreement with the dynamical masses (median 2\% relative difference, median 0.0001~\msun\/ absolute difference; see Table~\ref{mass_ratios}, Figure~\ref{isochrone_mass_comparison}, Figure~\ref{mass_cdfs}).
These trends suggest that the majority of non-magnetic isochronal methods may be consistently underestimating stellar mass for low-mass PMS stars by $\gtrsim$25\%.

In contrast with the results of \citet{Braun2021}, we do not find that the magnetic \citet{Feiden2016} models overestimate stellar mass for sources with M$_{dyn}$ $\leq$ 0.6~\msun. 
This is likely due to the difference in modeling methods between these two works.
Our approach includes dynamical modeling of both spatial and spectral components of the gas lines and does not assume a uniform, thin disk structure, in contrast to \citet{Braun2021}.
This possibility is supported by \citet{Braun2021} themselves, who note discrepancies between the dynamical masses they derive and those derived by \citet{Simon2019} for the 11 sources common to both samples; \citet{Simon2019} use Keplerian disk fitting, in contrast to the line-stacking approach of \citet{Braun2021}.

Finally, the work we present herein extends the evaluation of isochronal PMS masses to the 5-11~Myr age range, complementing existing work in younger star-forming regions \citep[e.g. Taurus, 1-3~Myr;][]{Simon2019}. 
The consistency of our results with those of younger regions suggests that the improved performance of magnetic over non-magnetic models is consistent for stellar ages up to $\sim$11~Myr.

\subsection{The Potential Impact of Starspots}
One potential complication in the evaluation of stellar evolutionary models' performance is the presence of starspots. 
Strong magnetic activity at or near the stellar surface can create large, cool starspots over a significant fraction ($\geq$10\%) of the stellar surface of low-mass stars \citep[e.g.][]{Rydgren1983,Bary2014,Somers2015}.
This can complicate both stellar spectral typing and inferred masses:
starspots can push a star's derived T$_{eff}$ values lower which, when combined with stellar isochrones, will yield comparatively lower stellar masses.
\citet{Somers2015}, for example, suggest that starspots can lead to isochrone-based masses being underestimated by up to a factor of two.

\citet{Flores2022} examine the impact of starspots on a star's derived T$_{eff}$ in both the optical and near-infrared regimes, and explore whether this translates to a systematic bias in isochrone-inferred masses.
They examine the spectra of 40 young stars in Taurus-Auriga and Ophiuchus, and find that starspots have a greater impact on optical temperatures than infrared temperatures.
This results in a greater discrepancy between isochronal and dynamical masses when optically-derived T$_{eff}$ are used than when infrared-derived T$_{eff}$ are used.
Similarly, \citet{PerezPaolino2024} explore the impact of starspots on the near-infrared spectra of 10 T-Tauri stars in Taurus-Auriga. 
They find that correcting for starspots results in median stellar mass increases of $+$0.24 and $+$0.44~\msun\/ (34\% and 88\%, respectively) relative to masses derived from uncorrected optical and near-infrared spectra.
\citet{Feiden2016} also consider starspots, both as a potential additional factor in isochronal fitting and as a separate, alternate explanation for the mass and age discrepancies noted in the literature \citep{Preibisch2008,Pecaut2012}.
They note that the difference in scale between the magnetic fields involved in the production of starspots (localized) versus those in their magnetic models (global) will manifest as differences in the photometric properties of each star.
In the limit that starspots cover a large fraction of the stellar surface, the two explanations should produce convergent results.

Sufficient data do not yet exist to robustly distinguish between the starspot and global-magnetic inhibition possibilities.
Likewise, a full comparison of the starspot and disk dynamical methods is beyond the scope of this paper.
However, we do compare the impact of these methods on derived stellar ages in our companion paper, Towner et al. (2025b, in prep).

\subsection{M$_{disk}$ vs. \mstar}
Given the widespread use of isochronal masses and the known dependence of isochronal mass on the specific model, it is worth considering whether known \mstar-dependent relations still hold when dynamical masses are used instead.
In this section, we consider the relationship between log(M$_{disk}$) and log(\mstar), and evaluate whether it depends on the use of isochronal versus dynamical masses. 
This comparison covers the full mass range of the sample, but excludes the sources for which we derive upper limits on disk mass and/or stellar mass.

The relationship between stellar mass and disk dust mass has been observed in numerous young star-forming regions \citet[e.g. Taurus, 1-2~Myr; Lupus, 1-3~Myr; Chameleon I, 2-3~Myr;][]{Andrews2013,Pascucci2016}, and has been found to be approximately linear: log(M$_{dust}$) $\propto$ log(\mstar). 
However, recent observations suggest that the log(M$_{disk}$)$-$log(\mstar) relation may steepen with time.
\citet{Barenfeld2016} find a slope of 1.67$\pm$0.37 for the older Upper Scorpius region assuming a dust temperature that scales with stellar luminosity, and \citet{Pascucci2016} find slopes of 1.9$\pm$0.4 (assuming L$_{\star}$-scaled T$_{dust}$) to 2.7$\pm$0.4 (assuming constant T$_{dust}$ $=$ 20~K). 
This steepening of the M$_{dust}-$\mstar\/ relation has been interpreted as a depletion of millimeter-size dust grains at larger radii as protoplanetary disks evolve \citep{Barenfeld2016,Pascucci2016}.
However, the precise mechanism of this depletion remains an open question.

Following the methods of \citet{Barenfeld2016} and \citet{Pascucci2016}, we derive disk dust masses for our targets using the equation 
\begin{equation}
    M_d = \frac{S_{\nu}d^2}{\kappa_{\nu}B_{\nu}(T_d)},
\end{equation}

where $S_{\nu}$ is our measured continuum flux density at 870~\mum, $d$ is the {\it Gaia}-derived distance to each source, $\kappa_{\nu}$ is the dust opacity, and $T_d$ is the dust temperature.
We assume a dust opacity of $\kappa_{\nu}$ $=$ 2.7 cm$^2$ g$^{-1}$ at 345~GHz, where $\kappa_{\nu}$ $=$ 2.3 cm$^2$ g$^{-1}$ at 230~GHz and scales with frequency as $\nu^{0.4}$ \citep{Barenfeld2016}.

We derive these masses under two different assumptions for dust temperature: T$_{dust}$ $=$ 20~K, and T$_{dust}$ $=$ 25~K $\times$ (\lstar/\lsun)$^{0.25}$ \citep{Andrews2013,Pascucci2016}.
Table~\ref{disk_masses} lists the integrated continuum flux densities and the disk dust masses derived for each source. 
The continuum flux densities were obtained from the deconvolved fit results of CASA's {\tt imfit} task unless otherwise noted.
The {\tt imfit} task fits a 2D Gaussian to each source in the image plane.
We verified each fit with a by-eye examination of the residual image. 
Sources with peak residuals above 3$\sigma$ were refit.
If no good fit could be obtained, we performed manual aperture photometry for the source instead.

\begin{deluxetable}{lccc}
\tablecaption{Disk Dust Masses}
\tablecolumns{4}
\tablewidth{\textwidth}
\tablehead{
\colhead{Field} & \colhead{S$_{870~\mu m}^a$} & \colhead{M$_{T_{scale}}$} & \colhead{M$_{T_{20\,K}}$} \\
& \colhead{(mJy)} & \colhead{(\mearth)} & \colhead{(\mearth)}
}
\startdata
J15521088-2125372   & 0.4 (0.2)   & 0.8 (0.3) & 0.14 (0.07) \\
J15530132-2114135   & 5.2 (0.1)   & 3.7 (0.1) & 1.40 (0.06) \\
J15534211-2049282   & 2.7 (0.2)$^{\dagger}$ & 1.3 (0.1) & 0.62 (0.06) \\
J15562477-2225552   & $<$ 0.30    & $<$ 0.18 & $<$ 0.08 \\
J16001844-2230114   & 3.6 (0.1)   & 2.0 (0.3) & 0.9 (0.1) \\
J16014086-2258103   & 3.2 (0.1)   & 1.26 (0.05) & 0.62 (0.02) \\
J16020757-2257467   & 4.4 (0.1)   & 1.60 (0.05) & 1.09 (0.03) \\
J16035767-2031055   & 5.2 (0.1)   & 1.16 (0.03) & 1.33 (0.03) \\
J16035793-1942108   & 1.4 (0.2)   & 0.7 (0.1) & 0.44 (0.06) \\
J16062277-2011243   & 0.4 (0.1)   & 0.27 (0.07) & 0.11 (0.03) \\
J16075796-2040087   & 19.1 (0.2)$^{\dagger}$ & 11.0 (0.9) & 6.0 (0.5) \\
J16081566-2222199   & 1.1 (0.1)   & 0.41 (0.04) & 0.27 (0.03) \\
J16082324-1930009   & 36.6 (0.3)$^{\dagger}$ & 11.5 (0.2) & 8.7 (0.2) \\
J16090075-1908526   & 41.4 (0.3)$^{\dagger}$ & 11.0 (0.2) & 9.8 (0.2) \\
J16095933-1800090   & 0.38 (0.08) & 0.18 (0.04) & 0.09 (0.02) \\
J16104636-1840598   & 2.0 (0.1)   & 1.6 (0.1) & 0.51 (0.03) \\
J16113134-1838259 B & 86 (2)      & 26 (2) & 27 (2) \\
J16115091-2012098   & 0.6 (0.1)   & 0.30 (0.05) & 0.17 (0.03) \\
J16123916-1859284   & 6.7 (0.2)   & 1.83 (0.07) & 1.63 (0.06) \\
J16142029-1906481   & 33.2 (0.3)$^{\dagger}$ & 20.9 (0.7) & 8.5 (0.3) \\
J16143367-1900133   & 1.3 (0.1)   & 0.57 (0.05) & 0.33 (0.03) \\
J16163345-2521505   & 2.4 (0.2)   & 1.1 (0.1) & 0.80 (0.07) \\
J16181904-2028479   & 3.5 (0.1)   & 2.5 (0.1) & 0.84 (0.04) \\
\enddata
\tablenotetext{a}{870~\mum\/ continuum flux densities and derived disk dust masses for all 23 sources with dynamical stellar masses. Flux densities were determined using the CASA task {\tt imfit}, except for sources marked with a $\dagger$, for which flux densities were determined using manual aperture photometry.}
\label{disk_masses}
\end{deluxetable}

\begin{figure*}
    \centering
    \includegraphics[width=0.9\textwidth]{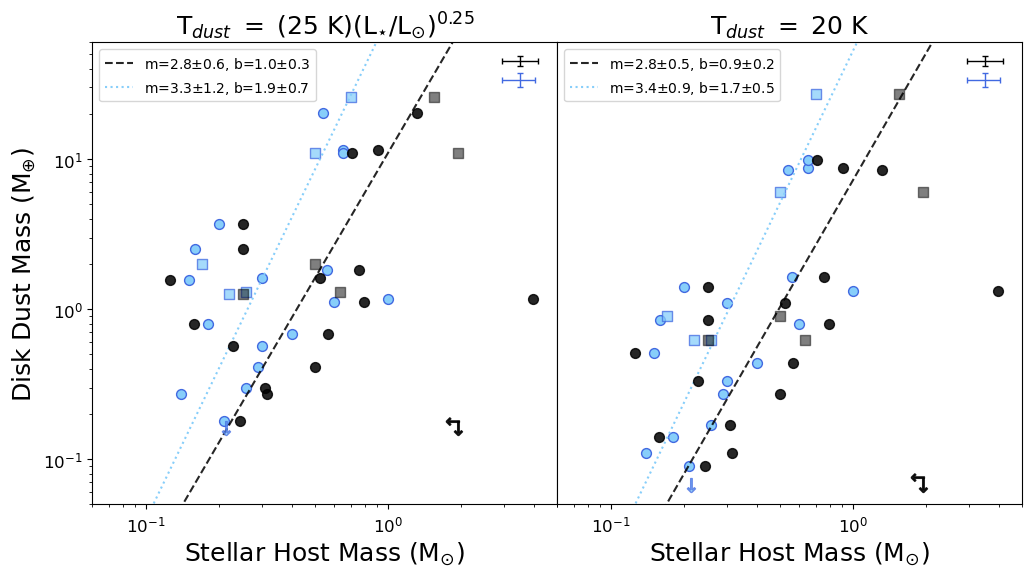}
    \caption{Disk dust mass (see Table~\ref{disk_masses}) versus stellar mass for the 23 sources in our sample. Upper limits on dust masses are marked with downward arrows. Upper limits on stellar mass are marked with left-facing arrows. In both panels, the black points denote the stellar masses we derive via Keplerian fitting, and the blue points denote stellar masses derived from the best-fit BHAC15 PMS models using \lstar\/ and T$_{eff}$. {\it Left panel:} Disk mass versus stellar mass assuming the disk dust temperature scales as 25~K (\lstar/\lsun)$^{0.25}$. The best-fit lines using the dynamical and BHAC15 stellar masses are shown in black and blue, respectively; the best-fit slope (m) and intercept (b) for each case are shown in the upper left-hand corner. Median uncertainties are shown as error bars in the upper right-hand corner. {\it Right panel:} As for the left-hand panel, except the disk dust masses were calculated assuming a constant dust temperature of 20~K.
    }
    \label{star_disk_mass}
\end{figure*}

Figure~\ref{star_disk_mass} shows scatterplots of log(M$_{disk}$) versus log(\mstar) for our data. 
The dynamical stellar masses are shown in black, and the BHAC15 stellar masses are shown in blue. 
The left-hand panel shows results under the luminosity-scaled T$_{dust}$ assumption, and the right-hand panel shows results assuming T$_{dust}$ $=$ 20~K. 
We derive a log(M$_{disk}$)$-$log(\mstar) relation using Orthogonal Distance Regression (ODR) fitting for both our dynamical masses and the BHAC15 \mstar\/ values. 
Under the luminosity-scaled temperature assumption, we derive the relation 
\begin{equation}
    log(M_{disk}) = (2.8\pm0.6)log(M_{\star}) + (1.0\pm0.3)
\end{equation}

using our dynamical masses, with Spearman $\rho$ $=$ 0.50 and a p-value of 0.02.
Using the BHAC15 masses and the scaled-temperature assumption, we derive 
\begin{equation}
    log(M_{disk}) = (3.3\pm1.2)log(M_{\star}) + (1.9\pm0.7)
\end{equation}

with Spearman $\rho$ $=$ 0.39 and a p-value of 0.08.

Under the constant-T$_{dust}$ assumption, we derive 
\begin{equation}
    log(M_{disk}) = (2.8\pm0.5)log(M_{\star}) + (0.9\pm0.2).
\end{equation}

using our dynamical masses (Spearman $\rho$ $=$ 0.71, p $=$ 2$\times$10$^{-4}$), and 
\begin{equation}
    log(M_{disk}) = (3.4\pm0.9)log(M_{\star}) + (1.7\pm0.5).
\end{equation}

using the BHAC15 masses (Spearman $\rho$ $=$ 0.65, p $=$ 1$\times$10$^{-3}$).

All Spearman $\rho$ values show a moderate correlation between \mstar\/ and M$_{disk}$, with associated p-values between corresponding to $\leq$4$\sigma$.
The p-values for the constant-T$_{dust}$ cases are lower than those for the scaled-T$_{dust}$ cases (corresponding to $\gtrsim$3$\sigma$), which would seem to indicate a more statistically-significant correlation.
However, it is possible this increased significance is actually a consequence of assuming the same T$_{dust}$ for all disks, rather than a reflection of real physics.

Regardless of the method used, we derive a relationship between log(M$_{disk}$) and log(M$_{\star}$) that is steeper than linear within the uncertainties of the ODR fit.
Our derived slopes are consistent with those of \citet{Pascucci2016} within the relevant uncertainties, and steeper than \citet{Barenfeld2016} in all cases.
The discrepancy with \citet{Barenfeld2016} is likely due to their use of a continuum-selected sample in contrast to our (more gas-rich) CO-selected sample.
The different methodologies in deriving the best-fit line and the inclusion of upper-limit disk masses in the \citet{Barenfeld2016} calculation (48 out of their 106 data points) may also be contributing.

The steeper-than-linear relationship we derive for Upper Sco is consistent with the theory that the millimeter-size dust content of protoplanetary disks evolves with time, either through inward radial drift of the millimeter dust or grain growth to larger sizes \citep{Pascucci2016}.
While we lack the spatial resolution and spectral coverage in this work to distinguish between these two possibilities, an exploration of grain growth in Upper Sco will be the subject of future work. 

Under both temperature assumptions, the slope of the log(M$_{disk}$)$-$log(\mstar) relation is steeper when derived using the BHAC15 masses than when using the dynamical masses.
This could suggest that the use of dynamical masses will shallow the slope of the log(M$_{disk}$)$-$log(\mstar) relation.
This would be consistent with our finding that dynamical methods tend to return slightly higher stellar masses ($\gtrsim$25\%) than isochronal methods.
However, the BHAC15 and dynamical-mass slopes always agree with each other within uncertainties.
Therefore, we do not consider this trend to be statistically significant at this time. 

We also find no change in slope with the T$_{dust}$ assumption.
This is in conflict with \citet{Pascucci2016}, who note a clear shallowing of the relation with luminosity-scaled T$_{dust}$ (1.9$\pm$0.4) as compared to a constant T$_{dust}$ (2.7$\pm$0.4). 
This discrepancy may be a result of our comparatively smaller sample size, though we do note that our uncertainties on the slopes derived using dynamical masses ($\pm$0.5 to 0.6) is comparable to those derived by \citet{Pascucci2016} and \citet{Barenfeld2016}. 
It is also possible that, because the disks in our sample are larger and more CO-rich than in most continuum-based samples, our disk temperatures may be naturally closer to 20~K to begin with; in that case, the use of the constant T$_{dust}$ $=$ 20~K would have less of an impact on our best-fit line.

\subsection{Potential Sources of Additional Uncertainty}
We have already examined our data for signs that the CO data are tracing outflows rather than disk rotation, and found no evidence of outflow contamination (\S~\ref{disks_vs_outflows}).
In this section, we explore two further potential sources of contamination or bias in the data: sources with known or candidate companions, and bias in stellar mass with inclination angle.
We describe the full details of this examination in Appendix~\ref{uncertainty_appendix}.

In short, we find little evidence for strong effects due to the inclusion of binary sources. 
In general, excluding the known or candidate binaries from our sample drops the correlations between mass ratio and dynamical mass to below 3$\sigma$ significance.
The one exception is the strong correlation for the PARSEC v1.2S models, which remain at $>$5$\sigma$ significance regardless of the exclusion of the binary sources.
Given the smaller number of sources in the binary-excluded sample, and the relatively low statistical significance of most of these relations to begin with, these results are not surprising.

To test for methodology-induced bias in stellar mass with disk inclination angle, we compare our \mstar\/ and $i$ to the stellar masses of \citet{Barenfeld2016} and the CO- and continuum-derived disk inclination angles of \citet{Barenfeld2017_disksize}.
All median $i$ and \mstar\/ values agree within their respective uncertainties.
We perform additional KS tests, and find that there is no statistically significant ($>$3$\sigma$) difference between the samples in either \mstar\/ or $i$. 
We therefore conclude that the {\tt pdspy}-derived dynamical masses have not introduced any additional bias in mass with inclination angle.

\section{Summary}
\label{summary}
We have derived the masses of 23 pre-main sequence K- and M-type stars with disks in Upper Scorpius using a Keplerian disk model and the open-source package {\tt pdspy}. 
We successfully fit the $^{12}$CO J$=$3$-$2 emission for 22 out of 24 sources (Table~\ref{pdspy_results}), and fit the $^{13}$CO J$=$3$-$2 emission for 4 out of the 6 sources with $>$5$\sigma$ $^{13}$CO emission (Table~\ref{13co_results}).
We also jointly fit the $^{12}$CO$+$$^{13}$CO emission for the three sources with $>$5$\sigma$ emission in both lines (Table~\ref{13co_results}).
We find a median sample mass of 0.5~\msun, and a minimum and maximum mass of 0.13$^{+0.07}_{-0.05}$~\msun\/ and 4.0$^{+4.0}_{-2.0}$~\msun, respectively. 

We have evaluated the best-fit values and uncertainties produced by {\tt pdspy} using both a $\chi^2_{min}$+$\Delta \chi^2$ method and a median plus standard deviation method.
We find that the median$+$standard deviation method is prone to underestimating parameter uncertainties.
We report as our best-fit results the values corresponding to the minimum $\chi^2$ fit, and uncertainties corresponding to the maximum and minimum values within $\Delta \chi^2$ $=$ 15.975.

We have extensively compared our results to those of five pre-main sequence evolutionary model sets: BHAC15 \citep{Baraffe2015}, PARSEC v1.1 \citep{Bressan2012}, PARSEC v1.2S \citep{Chen2014}, and both the standard and magnetic models of \citet{Feiden2016}.
We perform this comparison for both the full sample and for only those sources with 0.1~\msun\/ $\leq$ \mstar\/ $\leq$ 1~\msun.
We find the following:

\begin{enumerate}
    \item The magnetic models of \citet{Feiden2016} are in very good agreement with our dynamical results for \mstar\/ $\leq$1~\msun. 
    They have a nearly 1:1 relationship with the dynamical masses up to 1~\msun, and Kolmogorov-Smirnov tests comparing dynamical and isochronal masses for all five model sets show that the magnetic models are in the best agreement by far with our dynamical masses.
    \item The BHAC15, PARSEC v1.1, and non-magnetic \citet{Feiden2016} models all tend to underestimate dynamical mass by $\gtrsim$25\% (Figure~\ref{isochrone_mass_comparison}, Table~\ref{mass_ratios}). In contrast, the PARSEC v1.2S models tend to overestimate dynamical masses for M$_{dyn}$ $\lesssim$0.6~\msun\/ and overestimate dynamical masses for M$_{dyn}$ $\gtrsim$0.6~\msun.
    \item In most cases, model accuracy (as measured by the ratio of the PMS to dynamical-mass results) is not correlated with stellar mass at the $>$4$\sigma$ level. The exception is the PARSEC v1.2S models, which have a $\gtrsim$5$\sigma$ negative correlation between dynamical mass and the M$_{iso}$/M$_{dyn}$ ratio. This indicates that the accuracy of the PARSEC v1.2S models as compared to the dynamical masses is mass-dependent.
    \item We find no indication that the magnetic models of \citet{Feiden2016} have a tendency to overestimate mass for the lowest-mass sources (\mstar\/ $\lesssim$ 0.6~\msun). This is in contrast with the literature findings of \citet{Braun2021}. We suggest that this contrast is due to a difference in the methods used to derive dynamical mass.
    \item We find that the sample statistics and our overall interpretation are not meaningfully different when considering the full sample as opposed to the mass-limited sample, or when including versus excluding sources with known or candidate companions. The magnetic models of \citet{Feiden2016} are clearly the best match to our dynamical masses in all cases.
\end{enumerate}

This work, performed with a comparatively large sample (23 sources) on a 5-11~Myr old region, complements similar work on younger regions \citep[e.g. Taurus, 1-3~Myr;][]{Simon2019}.
Our findings extend the systematic evaluation of stellar evolutionary models' performance - and in particular, the consistently superior performance of magnetic models over non-magnetic ones - up to ages of $\sim$11~Myr.

We derive the log(M$_{disk}$)$-$log(\mstar) relationship for our data, and find a steeper-than-linear relationship regardless of the M$_{disk}$ and \mstar\/ values used. 
This is consistent with the existing literature, which finds slopes $>$ 1.0 within uncertainties for Upper Sco \citep{Barenfeld2016,Pascucci2016}. 
This steeper slope is in contrast with younger regions such as Taurus \citep{Andrews2013}, and has been interpreted as indicating either grain growth or inward radial drift of the millimeter-size grains in older disks.

We also find that using dynamical masses as opposed to isochronal masses tends to produce shallower derived slopes for the log(M$_{disk}$)$-$log(\mstar) relation, but that the values still agree with each other within uncertainties. 
We suggest that future studies with larger sample sizes will be able to explore this possibility with greater statistical significance. 
We also find no difference in slope between two different T$_{dust}$ assumptions, which contradicts the findings of \citet{Pascucci2016}. 
This may be a consequence of our comparatively gas-rich sample as compared to previous studies. 

We investigate our results for potential bias due to the presence of binary sources or other nearby companions, and find no significant impact on our results. 
We also test for correlations between stellar mass and inclination angle, and find no statistically-significant difference between our sample and those of \citet{Barenfeld2016} and \citet{Barenfeld2017_disksize}.
We conclude that our disk-modeling procedure has not introduced any mass-versus-inclination bias into our results. 

Overall, the magnetic PMS evolutionary tracks of \citet{Feiden2016} provide by far the most consistent results with our dynamical masses.
We conclude that, in the absence of direct dynamical measurements, the magnetic \citet{Feiden2016} models are the most accurate and reliable isochronal method for determining the masses of late-type (\mstar\/ $\leq$ 1~\msun) pre-main sequence stars within the 5-11~Myr age range of our sample.

\section*{acknowledgements}
We thank the anonymous referee whose thoughtful comments and suggestions have improved this paper. 
This paper makes use of the following ALMA data: ADS/JAO.ALMA\#2019.1.00493.S. ALMA is a partnership of ESO (representing its member states), NSF (USA) and NINS (Japan), together with NRC (Canada), NSTC and ASIAA (Taiwan), and KASI (Republic of Korea), in cooperation with the Republic of Chile. The Joint ALMA Observatory is operated by ESO, AUI/NRAO and NAOJ.
The National Radio Astronomy Observatory is a facility of the National Science Foundation operated under cooperative agreement by Associated Universities, Inc.
The results reported herein benefitted from collaborations and/or information exchange within NASA's Nexus for Exoplanet System Science (NExSS) research coordination network sponsored by NASA's Science Mission Directorate under Agreement No. 80NSSC21K0593 for the program ``Alien Earths''.'' 
This research has made use of the SIMBAD database, operated at CDS, Strasbourg, France.
This work has made use of data from the European Space Agency (ESA) mission {\it Gaia} (\url{https://www.cosmos.esa.int/gaia}), processed by the {\it Gaia} Data Processing and Analysis Consortium (DPAC, \url{https://www.cosmos.esa.int/web/gaia/dpac/consortium}). Funding for the DPAC has been provided by national institutions, in particular the institutions participating in the {\it Gaia} Multilateral Agreement.
We acknowledge with thanks the variable star observations from the {\it AAVSO International Database} contributed by observers worldwide and used in this research.
This paper makes use of data from the AAVSO Photometric All Sky Survey, whose funding has been provided by the Robert Martin Ayers Sciences Fund and from the NSF (AST-1412587).
The Pan-STARRS1 Surveys (PS1) and the PS1 public science archive have been made possible through contributions by the Institute for Astronomy, the University of Hawaii, the Pan-STARRS Project Office, the Max-Planck Society and its participating institutes, the Max Planck Institute for Astronomy, Heidelberg and the Max Planck Institute for Extraterrestrial Physics, Garching, The Johns Hopkins University, Durham University, the University of Edinburgh, the Queen's University Belfast, the Harvard-Smithsonian Center for Astrophysics, the Las Cumbres Observatory Global Telescope Network Incorporated, the National Central University of Taiwan, the Space Telescope Science Institute, the National Aeronautics and Space Administration under Grant No. NNX08AR22G issued through the Planetary Science Division of the NASA Science Mission Directorate, the National Science Foundation Grant No. AST-1238877, the University of Maryland, Eotvos Lorand University (ELTE), the Los Alamos National Laboratory, and the Gordon and Betty Moore Foundation.
The DENIS project has been partly funded by the SCIENCE and the HCM plans of the European Commission under grants CT920791 and CT940627. 
It is supported by INSU, MEN and CNRS in France, by the State of Baden-W\"urttemberg in Germany, by DGICYT in Spain, by CNR in Italy, by FFwFBWF in Austria, by FAPESP in Brazil, by OTKA grants F-4239 and F-013990 in Hungary, and by the ESO C\&EE grant A-04-046.
Jean Claude Renault from IAP was the Project manager. 
Observations were carried out thanks to the contribution of numerous students and young scientists from all involved institutes, under the supervision of P. Fouqu\'e, survey astronomer resident in Chile.
This publication makes use of data products from the Two Micron All Sky Survey, which is a joint project of the University of Massachusetts and the Infrared Processing and Analysis Center/California Institute of Technology, funded by the National Aeronautics and Space Administration and the National Science Foundation.

\facilities{
ALMA,
AAVSO,
PS1,
Gaia,
CTIO:2MASS,
MAST (Pan-STARRS, DENIS)
}

\software{
astropy \citep{astropy_i,astropy_ii,astropy_iii}, 
CARTA \citep{carta},
CASA \citep{CASA_2022,McMullin2007_casa},
emcee \citep{Foreman-Mackey2013_emcee},
galario \citep{Tazzari2018_galario},
pdspy \citep{Sheehan2019},
radmc3d \citep{Dullemond2012_radmc3d},
spectral-cube \citep{spectralcube}
}

\appendix

\section{Additional Figures for SED and {\tt pdspy} Fitting}
\label{additional_figs}

\subsection{BT Settl Fitting Procedure and Best-fit SEDs}
We derive \lstar\/ using BT Settl atmospheric models for two reasons.
First, the \lstar\/ derived by \citet{Barenfeld2016} were published before the first {\it Gaia} data release, and assume a single distance of 145~pc for all sources.
Given the importance of precise {\it Gaia} distances to our dynamical mass derivations, it was necessary to re-derive luminosities using the new {\it Gaia} distances.
This ensures that, when comparing dynamical with isochronal masses, we are comparing like with like.
Second, in the course of revising \lstar\/ to account for revised distances, we discovered that at least 6 sources (1/4 of the sample) appear to have excess emission in the 2MASS J-band data as compared to the optical photometry.
This meant that replicating the procedure of \citet{Barenfeld2016}, who derive \lstar\/ using J-band photometry and a bolometric correction, was likely to yield spurious results for some sources.

For each source except J16113134-1838259 A \& B (see \S~\ref{sample_description}), we fit data from each of the following, as available: {\it Gaia} Data Release 3 \citep[DR3;][]{GaiaEDR3_Overview}, the AAVSO Photometric All-Sky Survey \citep[APASS;][]{Henden2014_apass1,Henden2019_apass_dr10}, the Panoramic Survey Telescope and Rapid Response System \citep[Pan-STARRS;][]{Flewelling2020_panstarrs}, the Deep Near-Infrared Survey of the Southern Sky \citep[DENIS;][]{DENIS2005}, and the Two Micron All-Sky Survey \citep[2MASS;][]{Skrutskie2006}.
Further details of each dataset can be found at the end of this section.

Our fit parameters are as follows:
we fix the BT Settl model temperature to the T$_{eff}$ listed in Table~\ref{source_properties}, rounded to the nearest hundred (the precision of the BT Settl models).
We assume Solar metallicity values with no $\alpha$ enhancement. 
We allow log({\it g}) to vary between 3.0 and 4.5; this range was chosen based on the typical masses and radii for sources of this type and age. 
The BT Settl model grid moves in steps of 0.5 for log({\it g}).
We fit our OIR photometry data to the BT Settl models over our own grid in A$_V$ and $\theta$, where A$_{V}$ is the visual extinction, $\theta$ $=$ R$_{\star}$/$d$ is the apparent size of the source on the sky, and $d$ is the {\it Gaia} distance to the source.
The allowed A$_V$ values range from 0 to 4.5 in steps of 0.1, and the allowed $\theta$ values range from 1$\times$10$^{-6}$ arcseconds to 8$\times$10$^{-5}$ arcseconds in steps of 0.25$\times$10$^{-6}$.

For each source, we read in the OIR photometry and convert it from magnitudes to flux density in Janskys using either the zero-point values reported in the literature for each facility and filter (see below), or a zero-point flux of 3631~Jy if the reported magnitude uses the AB system.  
We de-redden our data for each A$_{V}$ grid point using the extinction laws of \citet{WangChen2019}; for filters not directly listed in \citet{WangChen2019}, we determine A$_{\lambda}$ using linear extrapolation from the nearest listed wavelengths.
To compare our model photometry to our observed photometry, we calculate the mean specific flux density of the model (S$_{\nu}$, in units of erg s$^{-1}$ cm$^{-2}$ sr$^{-1}$ Hz$^{-1}$) within the frequencies corresponding to 50\% of the maximum transmission level of each filter.
This value is then scaled according to $\theta$ for each grid point.
We calculate $\chi^2$ for each fit using the measurement and flux calibration uncertainties of the catalog photometry (see below), the uncertainties on each A$_{\lambda}$ reported by \citet{WangChen2019}, and an additional 10\% uncertainty to account for potential variability in these low-mass pre-main sequence stars. 
We determine the best-fit model by minimizing $\chi^2$.

The best-fit A$_V$ is fit directly.
The best-fit R$_{\star}$ is derived using R$_{\star}$ $=$ $\theta_{best} d$, where $\theta_{best}$ is the best-fit $\theta$ value in arcseconds and $d$ is the {\it Gaia} distance to the source.
\lstar\/ is derived by integrating the best-fit model SED over frequency space and then multiplying by 4$\pi$R$_{\star}^2$.
The uncertainties on A$_V$, R$_{\star}$ and \lstar\/ are calculated using the same $\Delta\chi^2$ method described in \S~\ref{chisq_uncertainties}.
For a 3-parameter fit, the interval corresponding to a 1$\sigma$ uncertainty is $\Delta\chi^2$ $=$ 3.52675.

\begin{figure*}
    \centering
    \includegraphics[width=\textwidth]{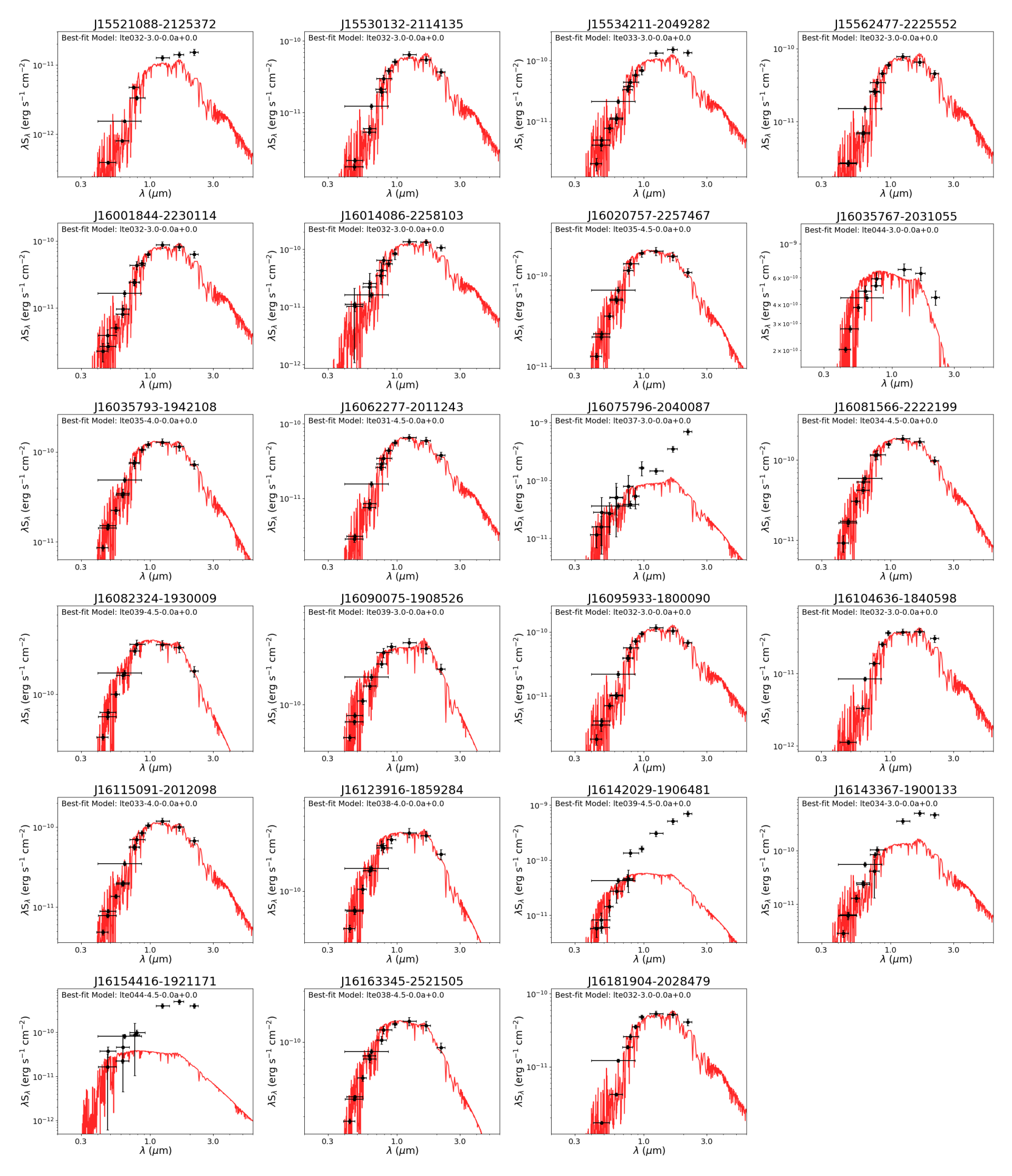}
    \caption{Optical-infrared photometry (black points) and best-fit BT Settl models (red curves) for 23 sources fit with BT Settl models (J16113134-1838259 A \& B were not excluded; see \S~\ref{sample_description}). The y-axis error bars are the uncertainties used in the fit, and the x-axis ``error'' bars show the wavelength ranges used to calculate the model S$_{\nu}$ for each observing band ($\geq$50\% of maximum transmission; see text). 
    The best-fit parameters for each source are listed in Table~\ref{btsettl_seds_parameters}.
    }
    \label{btsettl_seds}
\end{figure*}

In Figure~\ref{btsettl_seds}, we show the best-fit BT Settl model for each source in red, with the de-reddened optical and infrared photometry shown in black.
Vertical error bars show the uncertainties (including the additional 10\% to account for variability) used in the fit.
Horizontal ``error'' bars show the range of wavelengths covered by each band, as defined by the 50\%-of-maximum transmission level used for fitting (see above). 
In Table~\ref{btsettl_seds_parameters}, we report the best-fit parameters for each source, along with the specific photometric data used for each fit.
We also list the luminosities in Table~\ref{source_properties}.

Some sources exhibit notable near-infrared excess, which is not unusual for pre-main sequence, disk-bearing targets such as these.
In these cases, the fits more closely trace the optical data (dominated by the stellar photosphere) than the near-infrared data (potentially dominated by thermal emission from the disk).
This is a consequence of both the larger number of data points in the optical regime and their relative uncertainties, which are typically smaller than for the 2MASS data.
As a sanity check, we ran test fits in which we downweight the 2MASS data for sources with infrared excess.
We found that the downweighted fits produced the same \lstar\/ values as the standard fits within uncertainties.
In light of this finding and for internal consistency, we use the standard fit results for all sources.

All but three of these sources are classified as either Classical T-Tauri (CTT) or Weak T-Tauri (WTT) in the Konkoly Optical YSO Catalog (KYSO), which was compiled by cross-referencing {\it Gaia} DR3 and literature data \citep{Marton2023}.
The remaining three sources (J15521088-2125372, J15534211-2049282, and J16014086-2258103) are not listed in the KYSO catalog.
Of the 20 listed sources, 16 have specific variability types listed, which were defined and identified for {\it Gaia} data in \citet{Rimoldini2023} and \citet{Gavras2023}.
Most of our sources are classified as either ``dipper'' or ``Or*'' variables in the KYSO catalog, although some are classified as rotating spotted stars (``rot'') and one is classified as a ``burst'' variable (J16075796-2040087).
In our Figure~\ref{btsettl_seds}, the sources with the largest photometric uncertainties (J16075796-2040087, J16154416-1921171), and those with the most significant near-infrared excesses (the previous two sources, plus J16142029-1906481 and J16143367-1900133), are all classified as either ``burst'' or ``Or*'' variables.

The details of the photometric datasets and systems used in our fitting procedure are as follows: 

\begin{itemize}
    \item From {\it Gaia}, we use the G-band photometry listed in {\it Gaia} Data Release 3 (DR3). 
    All data were retrieved directly from the {\it Gaia} data archive at https://gea.esac.esa.int/archive/.
    We convert the {\it Gaia} magnitudes to flux densities using the zeropoints listed in the {\it Gaia} DR3 documentation \citet{GaiaEDR3_Overview}.
    We take the uncertainties for {\it Gaia} photometry to be the quadrature sum of the measurement uncertainty listed in the DR3 catalog and the {\it Gaia} absolute flux calibration uncertainty, which is reported as 1\% \citep{GaiaEDR3_Overview}.
    \item We use photometry from the B, V, {\it g$^{\prime}$, r$^{\prime}$, i$^{\prime}$,} and {\it z$^{\prime}$} bands of APASS. 
    Although APASS does include a {\it u$^{\prime}$} filter, none of the sources in our sample were detected in that band.
    We retrieve all APASS data from the APASS Catalog for Data Release 10, located at https://www.aavso.org/download-apass-data. 
    The APASS B and V bands use the Johnson-Cousins system, and we use the zeropoint fluxes reported by IRSA/IPAC\footnote{https://irsa.ipac.caltech.edu/data/SPITZER/docs/dataanalysistools/tools/pet/magtojy/ref.html} to convert these magnitudes to flux densities.
    The APASS {\it g$^{\prime}$, r$^{\prime}$, i$^{\prime}$,} and {\it z$^{\prime}$} bands are on the AB system, and we use a constant zeropoint of 3631~Jy to convert those data from magnitudes to flux densities.
    We take the uncertainties for all APASS photometry to be the quadrature sum of the measurement uncertainty listed in the DR10 catalog and the absolute flux calibration uncertainty, reported as $\lesssim$0.02 magnitudes in the relevant bands \citep{Henden2014_apass1,Munari2014_apass_unc,Henden2018_unc}.
    \item We use the {\it g, r, i, z,} and {\it y}-band ``Mean Object'' PSF photometry from Pan-STARRS Data Release 2 (DR2).
    All Pan-STARRS data were retrieved from the Mikulski Archive for Space Telescopes (MAST). 
    The Pan-STARRS photometric calibration procedures and sources of systematic uncertainties are described in detail in \citet{Magnier2020} and references therein, and the Pan-STARRS database and data products are described in \citet{Flewelling2020_panstarrs}.
    Pan-STARRS is a multi-epoch dataset, and the Pan-STARRS catalog reports both a measurement uncertainty from the ``Mean Object'' PSF fit (performed on the combined dataset from all epochs) and the standard deviation of a source's brightness across all epochs.
    We take the uncertainties for all Pan-STARRS photometry to be the quadrature sum of the uncertainty of the PSF fit, the standard deviation across all epochs, and the absolute flux calibration uncertainty for Pan-STARRS.
    We remove from our data any sources with a standard deviation of ``-999,'' i.e. sources for which no standard deviation could be determined.
    We use the zeropoints reported in \citet{Tonry2012_panstarrs} to convert Pan-STARRS magnitudes to flux densities.
    We assume an absolute flux calibration uncertainty of 0.03 mag in all bands \citep{Tonry2012_panstarrs}.
    \item We use DENIS I-band photometry retrieved from MAST.
    The DENIS survey uses a Gunn-$i$ filter, for which we adopt the zeropoint flux reported in \citet{Foque2000_DENIS} and an absolute flux calibration uncertainty of 0.1 mag \citep{Foque2000_DENIS,DENIS2005}.
    We take the DENIS I-band uncertainties to be the quadrature sum of the measurement and absolute flux calibration uncertainties.
    \item We retrieve all 2MASS data from the 2MASS Point-Source Catalog (PSC) on the IRSA/IPAC website. 
    We use the 2MASS J, H, and K-band zeropoint fluxes listed on that site (see previous footnote) to convert from magnitudes to flux densities.
    We calculate the uncertainty for each 2MASS data point as the quadrature sum of the ``combined photometric uncertainty'' (msigcom) for each source and the absolute flux calibration uncertainty, which we take to be 3\% in all bands \citep{Skrutskie2006}.
\end{itemize}

\begin{deluxetable}{lcccccl}
\tablecaption{BT-Settl Best-fit Parameters}
\tablecolumns{7}
\tablewidth{\textwidth}
\tablehead{
\colhead{Field} & \colhead{T$_{eff}$} & \colhead{log($g$)}& \colhead{A$_V$} & \colhead{R$_{\star}$} & \colhead{L$_{\star}$} & \colhead{Datasets$^b$} \\
 & \colhead{(K)} & \colhead{(log[cm s$^{-2}$])} & & \colhead{(R$_{\oplus}$)} & \colhead{(L$_{\odot}$)} & \colhead{Used}
}
\startdata
J15521088-2125372 & 3200 & 3.0 & 1.6$^{+0.2}_{-0.2}$ & 36$^{+3}_{-3}$ & 0.01$^{+0.005}_{-0.005}$ & {\it gri}GIJHK\\
J15530132-2114135 & 3200 & 3.0 & 0.3$^{+0.2}_{-0.3}$ & 75$^{+6}_{-8}$ & 0.04$^{+0.02}_{-0.02}$ & {\it g$^{\prime}$r$^{\prime}$i$^{\prime}$}{\it grizy}GIJHK\\
J15534211-2049282 & 3300 & 3.0 & 0.4$^{+0.3}_{-0.3}$ & 89$^{+10}_{-11}$ & 0.07$^{+0.03}_{-0.03}$ & BV{\it g$^{\prime}$r$^{\prime}$i$^{\prime}$}{\it grizy}GIJHK\\
J15562477-2225552 & 3200 & 3.0 & 0.2$^{+0.4}_{-0.2}$ & 82$^{+4}_{-8}$ & 0.05$^{+0.02}_{-0.02}$ & {\it g$^{\prime}$r$^{\prime}$i$^{\prime}$}{\it grizy}GIJHK\\
J16001844-2230114 & 3200 & 3.0 & 0.0$^{+0.1}_{-0.1}$ & 83$^{+7}_{-6}$ & 0.05$^{+0.02}_{-0.02}$ & BV{\it g$^{\prime}$r$^{\prime}$i$^{\prime}$}{\it grizy}GIJHK\\
J16014086-2258103 & 3200 & 3.0 & 0.0$^{+0.1}_{-0.1}$ & 93$^{+6}_{-5}$ & 0.07$^{+0.03}_{-0.03}$ & {\it g$^{\prime}$r$^{\prime}$i$^{\prime}$}{\it grizy}GIJHK\\
J16020757-2257467 & 3500 & 4.5 & 0.4$^{+0.1}_{-0.2}$ & 114$^{+7}_{-9}$ & 0.15$^{+0.06}_{-0.07}$ & BV{\it g$^{\prime}$r$^{\prime}$i$^{\prime}$}{\it gry}GIJHK\\
J16035767-2031055 & 4400 & 3.0 & 0.9$^{+0.1}_{-0.2}$ & 148$^{+10}_{-11}$ & 0.6$^{+0.3}_{-0.3}$ & BV{\it g$^{\prime}$r$^{\prime}$i$^{\prime}$}{\it i}GJHK\\
J16035793-1942108 & 3500 & 4.0 & 0.7$^{+0.1}_{-0.3}$ & 107$^{+6}_{-9}$ & 0.13$^{+0.05}_{-0.06}$ & BV{\it g$^{\prime}$r$^{\prime}$i$^{\prime}$}{\it grizy}GJHK\\
J16062277-2011243 & 3100 & 4.5 & 0.4$^{+0.2}_{-0.4}$ & 87$^{+7}_{-7}$ & 0.05$^{+0.02}_{-0.02}$ & {\it g$^{\prime}$r$^{\prime}$i$^{\prime}$}{\it grizy}GIJHK\\
J16075796-2040087 & 3700 & 3.0 & 0.0$^{+0.1}_{-0.1}$ & 83$^{+7}_{-7}$ & 0.09$^{+0.04}_{-0.05}$ & BV{\it g$^{\prime}$r$^{\prime}$i$^{\prime}$}{\it grzy}GIJHK\\
J16081566-2222199 & 3400 & 4.5 & 0.2$^{+0.2}_{-0.2}$ & 118$^{+10}_{-11}$ & 0.14$^{+0.06}_{-0.06}$ & BV{\it g$^{\prime}$r$^{\prime}$i$^{\prime}$}{\it gry}GIJHK\\
J16082324-1930009 & 3900 & 4.5 & 0.8$^{+0.1}_{-0.2}$ & 120$^{+7}_{-11}$ & 0.2$^{+0.1}_{-0.1}$ & BV{\it g$^{\prime}$r$^{\prime}$i$^{\prime}$}{\it g}GIJHK\\
J16090075-1908526 & 3900 & 3.0 & 0.7$^{+0.1}_{-0.2}$ & 129$^{+7}_{-12}$ & 0.3$^{+0.1}_{-0.1}$ & BV{\it g$^{\prime}$r$^{\prime}$i$^{\prime}$z$^{\prime}$}{\it g}GIJHK\\
J16095933-1800090 & 3200 & 3.0 & 0.0$^{+0.6}_{-0.1}$ & 97$^{+13}_{-4}$ & 0.07$^{+0.04}_{-0.03}$ & BV{\it g$^{\prime}$r$^{\prime}$i$^{\prime}$}{\it grizy}GIJHK\\
J16104636-1840598 & 3200 & 3.0 & 0.7$^{+0.2}_{-0.7}$ & 59$^{+5}_{-12}$ & 0.03$^{+0.01}_{-0.02}$ & {\it grizy}GJHK\\
J16115091-2012098 & 3300 & 4.0 & 0.7$^{+0.1}_{-0.2}$ & 104$^{+6}_{-6}$ & 0.10$^{+0.04}_{-0.04}$ & BV{\it g$^{\prime}$r$^{\prime}$i$^{\prime}$}{\it grizy}GIJHK\\
J16123916-1859284 & 3800 & 4.0 & 0.7$^{+0.1}_{-0.4}$ & 136$^{+8}_{-14}$ & 0.3$^{+0.1}_{-0.1}$ & BV{\it g$^{\prime}$r$^{\prime}$i$^{\prime}$z$^{\prime}$}{\it g}GIJHK\\
J16142029-1906481 & 3900 & 4.5 & 0.0$^{+0.1}_{-0.1}$ & 55$^{+5}_{-4}$ & 0.05$^{+0.02}_{-0.02}$ & BV{\it g$^{\prime}$r$^{\prime}$i$^{\prime}$}{\it giy}GIJHK\\
J16143367-1900133 & 3400 & 3.0 & 0.0$^{+0.1}_{-0.1}$ & 101$^{+5}_{-6}$ & 0.1$^{+0.04}_{-0.04}$ & BV{\it g$^{\prime}$r$^{\prime}$i$^{\prime}$}{\it gri}GIJHK\\
J16154416-1921171 & 4400 & 4.5 & 0.0$^{+0.1}_{-0.1}$ & 32$^{+2}_{-3}$ & 0.03$^{+0.01}_{-0.01}$ & {\it g$^{\prime}$r$^{\prime}$i$^{\prime}$}{\it gr}GIJHK\\
J16163345-2521505 & 3800 & 4.5 & 1.0$^{+0.1}_{-0.2}$ & 107$^{+5}_{-9}$ & 0.17$^{+0.07}_{-0.08}$ & BV{\it g$^{\prime}$r$^{\prime}$i$^{\prime}$}{\it gry}GIJHK\\
J16181904-2028479 & 3200 & 3.0 & 1.0$^{+0.2}_{-0.3}$ & 66$^{+7}_{-7}$ & 0.03$^{+0.02}_{-0.02}$ & {\it grizy}GIJHK
\enddata
\tablenotetext{a}{All fits were fixed to Solar metallicity with an $\alpha$-enhancement of zero. T$_{eff}$ for each source was fixed to the temperature value listed in Table~\ref{source_properties}, rounded to the nearest hundred (the spacing of the BT Settl model grid). Log($g$) was allowed to vary between 3.0 and 4.5 in steps of 0.5 (the spacing of the BT Settl model grid). R$_{\star}$ was derived from the fitted A$_{V}$ and the {\it Gaia} distance, A$_{V}$ was fit directly, and \lstar\/ is calculated by integrating the best-fit model SED and multiplying by 4$\pi$R$_{\star}^2$.}
\tablenotetext{b}{``BV'' refers to APASS B and V bands, which use the Johnson-Cousins system; ``{\it g$^{\prime}$, r$^{\prime}$, i$^{\prime}$,} and $z^{\prime}$'' refer to the APASS $g$, $r$, $i$, and $z$ Sloan filters, which use the AB system; ``{\it g, r, i, z, y}'' refer to the PanSTARRS $grizy$ bands, which use AB-based PanSTARRS-specific zeropoints; ``G'' refers to the {\it Gaia} G filter, which uses its own internal magnitude system; ``I'' refers to DENIS I-band data, which uses a Gunn-$i$ filter; ``J, H, K'' refer to 2MASS.}
\label{btsettl_seds_parameters}
\end{deluxetable}

In addition to removing any Pan-STARRS data with standard deviations of ``-999,'' we inspected all data for quality using the source quality flags from the above-mentioned catalogs and the measurement uncertainty for each data point.
This process resulted in the removal of data in 1-2 bands for 7 sources: APASS B- and V-band data for J16014086-2258103 and J16154416-1921171, Pan-STARRS {\it z} data for J16035767-2031055, Pan-STARRS {\it i} data for J16075796-2040087, and Pan-STARRS {\it r} data for J16142029-1906481.

\subsection{{\tt pdspy} Channel Maps and Steps Plots}
\begin{figure*}
    \centering
    \includegraphics[width=0.8\textwidth,trim={1.0cm 0cm 0.5cm 0cm},clip]{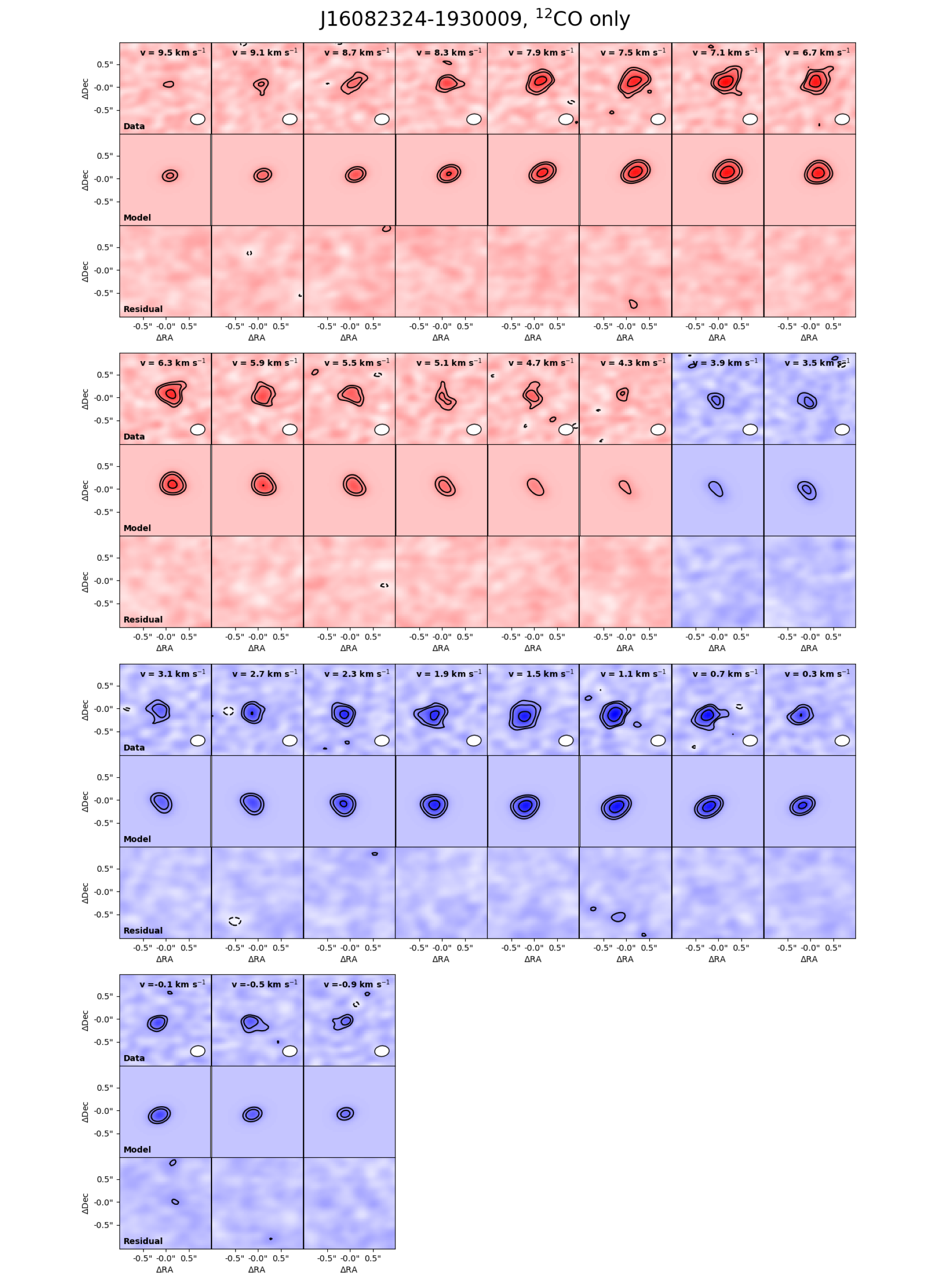}
    \caption{Channel maps showing the data, model, and residual for each channel over which the fit was performed. Panels in blue show emission that is blueshifted relative to \vlsr, and panels in red show redshifted emission. Each set of eight panels shows the data in the top row, the model in the middle row, and residuals in the bottom row. Velocity increases from left to right within each set of panels, and from top to bottom between panel sets. Contour levels are [3,5,10,15]$\times$rms, where rms is the median rms of the data cube. The complete figure set (29 items) is available in the online journal.}
    \label{residuals_plot}
\end{figure*}

In Figure~\ref{residuals_plot}, we show an example set of channel maps from the best-fit model for source J16082324-1930009 using the $^{12}$CO data.
The top row in each panel set shows the data, the middle row in each panel set shows the model, and the bottom row in each set shows the residuals (data - model). 
Channel maps for all source+line combinations are available in the online material.
For combined $^{12}$CO$+^{13}$CO fits, the ``data'' and ``residuals'' rows show the $^{12}$CO data and the ``model'' row shows the combined $^{12}$CO$+^{13}$CO model. 

\begin{figure*}
    \centering
    \includegraphics[width=0.95\textwidth]{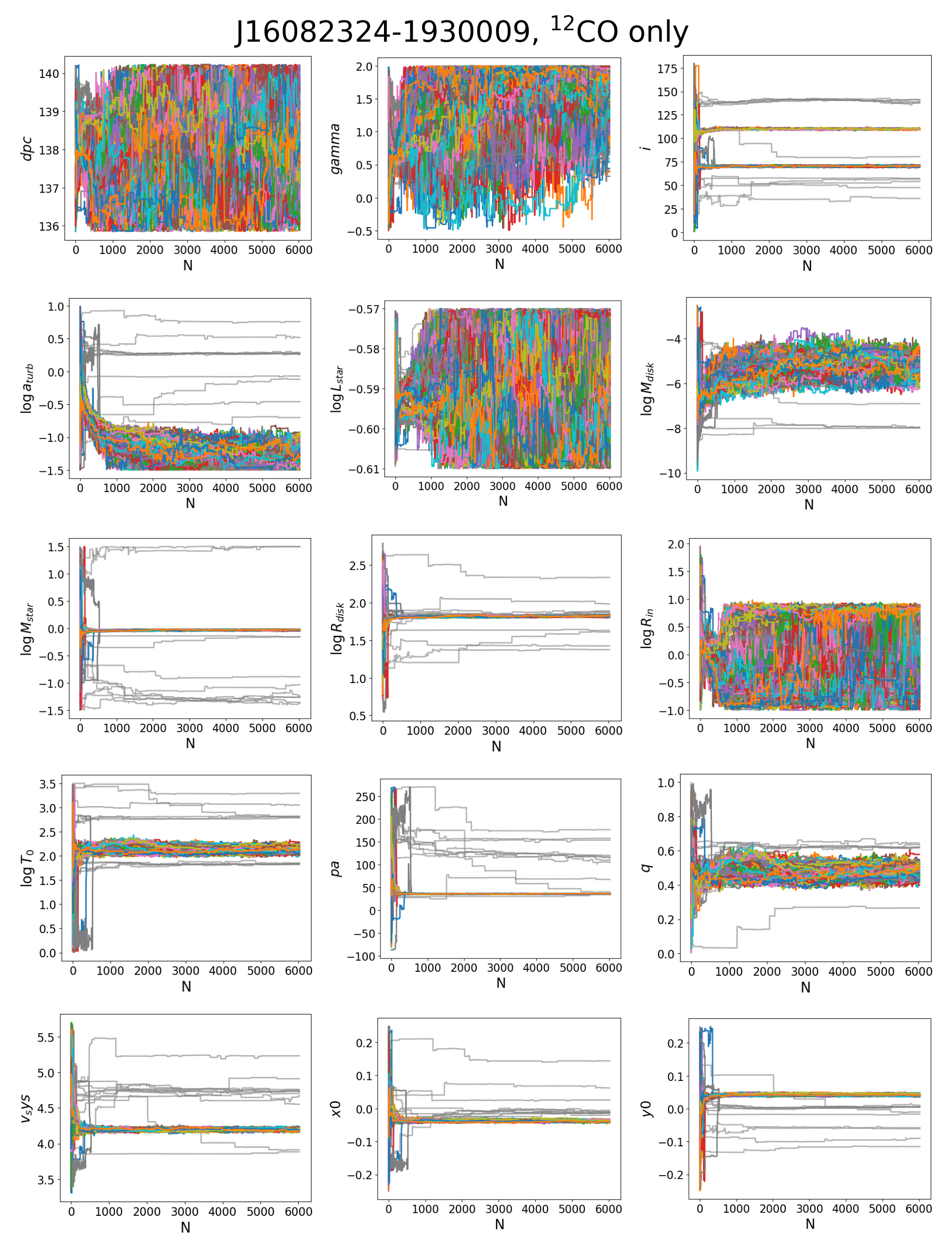}
    \caption{Steps plots (parameter value versus step number) for all 15 free parameters for target J16082324-1930009. The walkers meeting our $\chi^2$ cutoff are shown with colored lines, and the ``lost'' walkers are shown in gray. The complete figure set (29 items) is available in the online journal.}
    \label{steps_plot}
\end{figure*}

In Figure~\ref{steps_plot}, we show an example of the full set of steps plots (parameter value versus step number for all walkers, for all parameters) for J16082324-1930009.
The ``lost'' walkers are shown in gray.
Channel maps and steps plots for all source$+$line combinations are available in the online material. 

\section{Notes on Individual Sources}
\label{individual_notes}
Some sources in the sample required special handling during the fitting process, returned stellar masses that are strongly inconsistent with the current literature, or could not be fit at all.
We discuss each of these cases in greater detail below.

\subsection{J15562477-2225552}
Source J15562477-2225552 returned a distribution of \mstar\/ that is clearly clustering toward the minimum of the parameter range (0.03~\msun).
This is likely due to the low signal-to-noise of this source (5.8$\sigma$, the second-lowest in the sample) combined with its nearly face-on inclination (160$^{+20}_{-20}$), which limits the emission to a narrow (projected) velocity range.
For this source, we report the mass as an upper limit: \mstar\/ $\leq$ 1.45\msun.
This value corresponds to the 99.73\% (3$\sigma$) probability in the Cumulative Distribution Function (CDF) of walker values from the previous 1000 steps.

\subsection{J16035767-2031055 and J16142029-1906481}
The fits for both J16035767-2031055 and J16142029-1906481 returned stellar masses that were notably inconsistent with the isochronal masses derived in this work, and with the masses and spectral types listed in \citet{Barenfeld2016}.
Those authors list a stellar mass of 1.0$\pm$0.1~\msun\/ and a spectral type of K5 for J16035767-2031055, and a stellar mass of 0.56$^{+0.05}_{-0.06}$ and spectral type of M0 for J16142029-1906481.
Our Keplerian modeling instead returns stellar masses of 4.0$^{+4.0}_{-2.0}$~\msun\/ and 1.32$^{+0.06}_{-0.06}$~\msun, respectively.
These high masses persisted even when the fits were repeated with various modifications to the parameter ranges.
We could not find any evidence of known or candidate companions for either source in the literature. 
Neither source is strongly edge-on ($i$ $\sim$ 90$^{\circ}$), so they are unlikely to be ``dipper'' stars, and we were likewise unable to find clear evidence of variability in the literature. 

The most likely explanation for these sources is that the exponentially-tapered flared disk model is not capturing some additional phenomenon traced in the CO data, such as disk winds and/or outflows. 
Both sources are unresolved in the continuum in our data, so we can neither confirm nor reject this possibility based on (mis)alignment between the CO and continuum major axes. 

We could find no indications of (sub)millimeter-detected winds or outflows in the literature, but there is some useful information from shorter wavelengths. 
\citet{Fang2023_diskwinds} analyze high-resolution [OI] $\lambda$6300 and H$\alpha$ spectra from the Keck/HIRES archive for 115 T Tauri stars in Upper Sco, including both sources under consideration here.
Relative to most of our sample,  \citet{Fang2023_diskwinds} find that J16035767-2031055 has a comparatively bright low-velocity disk-wind component (log(L$_{BC}$/\lsun)$=-$4.91, FWHM $=$ 42.3~\kms), and that J16142029-1906481 has both a bright high-velocity outflow component (log(L$_{HVC}$/\lsun)$=-$4.19, FWHM $=$ 101.9~\kms) and a bright low-velocity wind component (log(L$_{SCJ}$/\lsun)$=-$5.08, FWHM $=$ 31.6~\kms). 
Both sources are also resolved in CO, as shown in Figure~\ref{mom1_12co}.
However, most sources in our sample have a low-velocity disk-wind component detected in [OI] $\lambda$6300 (typical log(L$_{LVC}$/\lsun): $-$6.6 to $-$5.4 for our sources), and some have high-velocity components as well.
Consequently, it is not clear whether the comparative brightness of these components in J16035767-2031055 and J16142029-1906481 could be solely responsible for the high dynamical masses we derive.

\subsection{J16075796-2040087}
Source J16075796-2040087 required an unusually long burn-in period ($\sim$4500 steps) due to large groups of ``lost'' walkers, i.e., walkers clumped in parameter space that did not survive our $\chi^2$ cut.
Even after the fit achieved convergence, some parameters still had distinct bimodal distributions of walkers, and the walkers are clustered at the top edge of the allowed parameter range for $\gamma$.
We suspect that, like the previous two sources, J16075796-2040087 exhibits emission that is not fully captured in our flared disk model.

J16075796-2040087 has a candidate companion identified in \citet{Barenfeld2019_binaries}.
The small projected separation between sources (6.3~au, well within their observed 1.3~mm disk radius), previous observations suggesting an accretion-powered outflow \citep{Kraus2009_J1607}, and observed periodic bursting behavior in the optical \citep{Cody2017} led those authors to suggest that this source contains a circumbinary disk.
It also has bright low-velocity (disk-wind) and high-velocity (collimated outflow) components in the [OI] $\lambda$6300 data of \citet{Fang2023_diskwinds}. 
The observed outflows and bursting behavior could both potentially be contaminating the {\tt pdspy} fit, if those processes are significantly impacting the CO emission.
This source is also resolved in our CO data (see Figure~\ref{mom1_12co}); it is possible that internal disk structures are contributing to the issues present in this fit.

\subsection{J16113134-1838259 B}
Field J16113134-1838259 (also known as AS 205) contains two disks as well as significant large-scale, extended emission near \vlsr\/ in our data.
Some of this large-scale emission takes the form of an apparent gas bridge between the two sources. 
Both disks appear to be slightly warped, and non-Keplerian rotation has been observed in the northern source, J16113134-1838259 A \citep{Kurtovic2018_AS205}. 
Though this extended emission does not appear to suffer from spatial filtering in our data, it and the presence of two sources made J16113134-1838259 B (also known as AS 205 S) impossible to fit in $^{12}$CO.
However, we did successfully fit its $^{13}$CO emission. 

Source B is compact and moderately inclined ($\sim$65$^{\circ}$).
Consequently, is exhibits line emission over a much wider velocity range than either source A or the extended emission.
We restricted the source position to $\Delta$RA, $\Delta$Dec $=$ $\pm$0$\farcs$15 as opposed to the standard $\pm$0$\farcs$25.
This stricter position range prevented the walkers from erroneously including emission from source A.
We also flagged out the channels covering \vlsr\/ $\pm$2.2~\kms\/ in the MS file.
The channel flagging ensures that the channels with the strongest emission from source A, and the strongest extended emission, are excluded from the fit.
Again, this prevented the walkers from erroneously fitting to emission from source A and the extended emission. 
Because the line emission from source B spans \vlsr\/ $\pm$~5.4~\kms\/ and we assume Keplerian rotation at all radii, the exclusion of the central channels did not strongly impact the {\tt pdspy} fit results.
Source B was fit in 2520 steps with a standard burn-in interval of 1000 steps. 

As for the field as a whole, \citet{Salyk2014} observed J16113134-1838259 in the then-highest spatial resolution (88 au at their assumed distance of 125~pc) in ALMA Cycle 0 with $^{12}$CO, $^{13}$CO, and C$^{18}$O J$=$2$-$1 line emission and 232~GHz continuum emisison.
They exclude molecular cloud emission, envelope emission, and accretion from a circumbinary disk as potential explanations for the extended emission and the non-Keplerian rotation observed in source J16113134-1838259 A (AS 205 N in their paper).
They favor a low-velocity disk wind from J16113134-1838259 A as the most likely explanation for the observed emission, but note that they cannot exclude the possibility of tidal stripping of the source A disk by source B.
\citet{Kurtovic2018_AS205}, re-observed this source with 5 au resolution as part of the Disk Substructures at High Angular Resolutions Project \citep[DSHARP,][]{Andrews2018_DSHARP}. 
These authors favor a flyby interaction as the explanation for the gas bridge between the disks, the large-scale arc-like structures, and the observed disk warping.
However, both teams note the importance of accurate stellar masses - especially for source B - in constraining the parameters of a potential flyby interaction.

A detailed kinematic and modeling analysis of the full J16113134-1838259 system is beyond the scope of this paper. 
However, we call attention to the fact that our dynamically-derived results for this source (\mstar\/ $=$ 1.53$^{+0.14}_{-0.12}$ \msun) are higher than the previous best-estimate mass (\mstar\/ $=$ 1.3~\msun) of \citet{Eisner2005} based on measurement uncertainties. 
This increase in mass ($\sim$18\%) is consistent with the typical difference between our dynamical results and the isochronal methods.
This increased, well-constrained mass will help place tighter constraints on the parameters of a potential flyby interaction in this field.

\subsection{J16154416-1921171}
Source J16154416-1921171 has strong ($\geq$6.5$\sigma$) emission in both $^{12}$CO and $^{13}$CO.
However, it also has significant, large angular-scale line emission.
This emission is partially resolved-out at \vlsr\/ in both isotopologues.
This spatial filtering made the source impossible to fit, even with channel and {\it (u,v)}-range flagging.
It is therefore excluded from our results and analysis.

\section{Full Sample versus Mass-limited Sample}
\label{full_sample_discussion}
Table~\ref{mass_ratios} showed statistics for both the full sample and the mass-limited sample.
In \S~\ref{pms_masses}, we examined sample statistics and figures for only those sources with \mstar\/ $\leq$ 1~\msun, which is the mass range common to all PMS model sets. 
Here, we show additional figures using the full sample and briefly discuss the impact of using the full sample on the statistics in Table~\ref{mass_ratios}.

\begin{figure*}
    \centering
    \includegraphics[width=\textwidth]{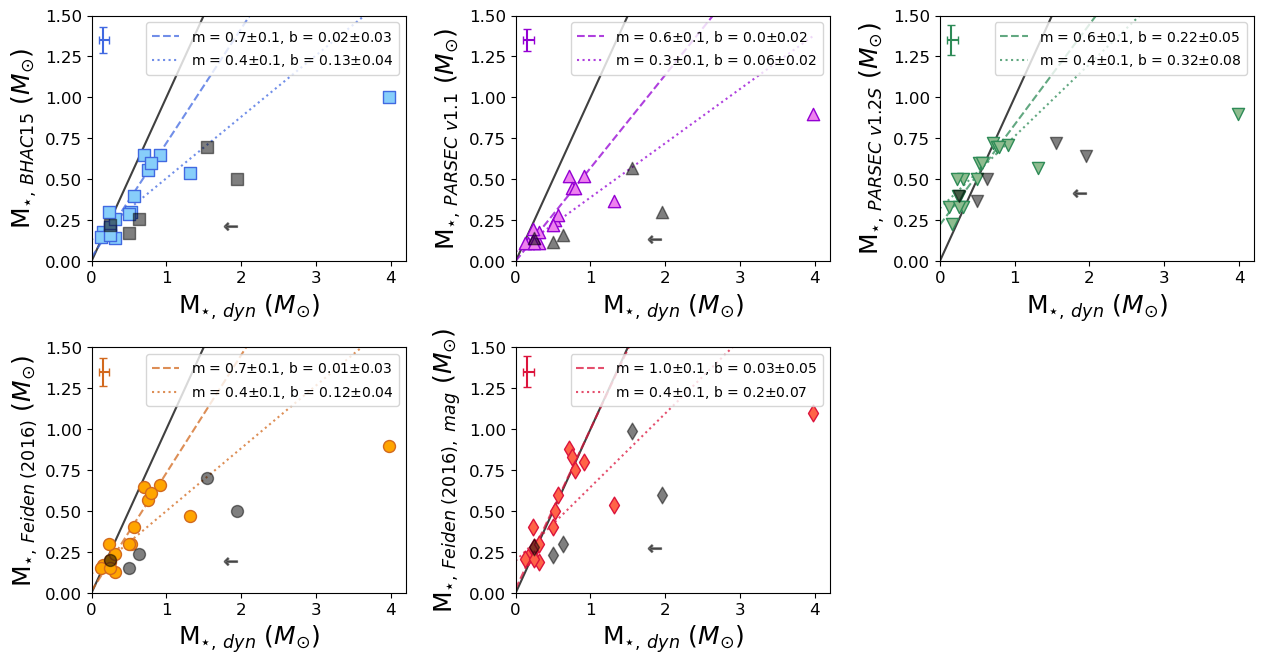}
    \caption{As for Figure~\ref{isochrone_mass_comparison}, but showing the full range of dynamical masses for our data. The best-fit line for the full sample is shown as a dotted colored line in each panel, and the best fit for the mass-limited sample is shown as a dashed colored line. As in Figure~\ref{isochrone_mass_comparison}, known or candidate binaries are shown with gray symbols, and the 1:1 relation is shown as a black solid line. Although no PMS model sets show a 1:1 relationship with dynamical mass when using the full sample, the mass values for \citet{Feiden2016} magnetic models are still the least discrepant.}
    \label{mass_isochrone_full}
\end{figure*}

In Figure~\ref{mass_isochrone_full}, we show an alternate version of Figure~\ref{isochrone_mass_comparison} in which the x-axis extends up to 4.2~\msun.
We show the best-fit lines for both the full-sample and mass-limited sample.
The impact of the \mstar\/ $\geq$ 1~\msun\/ sources, and in particular the source with M$_{dyn}$ = 4.0~\msun, on the best-fit line can be clearly seen.
Although none of the PMS model sets agree with the 1:1 line within uncertainties when the full sample is considered, the masses derived using the magnetic models of \citet{Feiden2016} are still the least discrepant out of all model sets.

\begin{figure}
    \centering
    \includegraphics[width=0.5\textwidth]{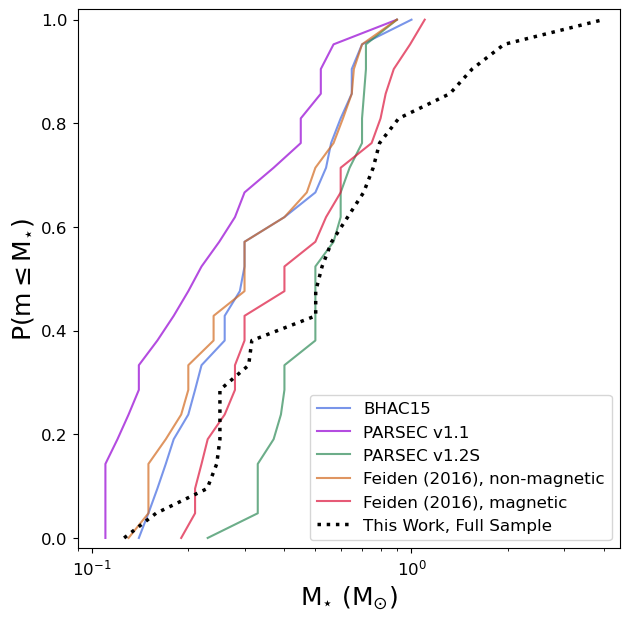}
    \caption{As for Figure~\ref{mass_cdfs}, but showing the full range of dynamical masses for our data. The Cumulative Distribution Function (CDF) using our full sample is shown as a black dotted line. Compare to Figure~\ref{mass_cdfs}.}
    \label{cdfs_full}
\end{figure}

Figure~\ref{cdfs_full} shows the Cumulative Distribution Functions (CDFs) of the isochronal masses and our dynamical masses using the full sample.
Although the CDFs of the isochronal masses are not significantly impacted by including the M$_{dyn}$ $>$ 1~\msun\/ sources (compare to Figure~\ref{mass_cdfs}), the CDF for the dynamical masses clearly is.
However, as demonstrated in Table~\ref{mass_ratios}, the impact of this inclusion on the KS test statistics is not large: $D$ changes by $\leq$0.1, with only nominal changes in corresponding p-values.
No model sets are rejected with $\geq$4$\sigma$ confidence. 
The magnetic models of \citet{Feiden2016} are still clearly the most similar to the dynamical masses ($D$ $=$ 0.18) compared to the other four model sets ($D$ $\geq$ 0.32). 

Finally, for the remaining statistics listed in Table~\ref{mass_ratios}, excluding sources with \mstar\/ $>$ 1~\msun\/ primarily impacts the mass ratios.
The mass ratio minima decrease by $\sim$23-69\% when the full sample is used.
This is to be expected, as the most massive sources often show the greatest discrepancies between dynamical and isochronal mass.
The mass ratio maxima are unchanged, and the median and mean mass ratios are unchanged within uncertainties.
The PARSEC v1.2S median and mean ratios still have the largest uncertainties out of all the model sets, and the magnetic \citet{Feiden2016} mean and mean mass ratios are both still with $\leq$10\% of unity.
The Spearman $\rho$ correlation coefficients become more negative in all cases, which is expected considering the positions of the excluded sources on the plots in Figure~\ref{mass_isochrone_full}.
The p-values of the coefficients all decrease by $>$2 orders of magnitude.
However, the p-values for the BHAC15, PARSEC v1.1, and both \citet{Feiden2016} models all remain $<$4$\sigma$, i.e. they still do not meet our threshold for statistical significance.
The p-value for the PARSEC v1.2S models decreases to 4$\times$10$^{-9}$, and remains at $\geq$4$\sigma$ significance.

Taken together, these changes do not meaningfully alter our interpretations.
We still conclude that the magnetic models of \citet{Feiden2016} are more consistent with our dynamical masses than the other four model sets tested.
We likewise still conclude that only the PARSEC v1.2S models show a statistically-significant correlation between mass and mass ratio (i.e. potential that the accuracy of the model set is mass-dependent).
In short, we find that our results are robust against the inclusion or exclusion of sources with \mstar\/ $>$ 1~\msun\/ from our sample.

\section{Impacts of Binarity and Methodology on the Observed Trends}
\label{uncertainty_appendix}
\subsection{Confirmed or Candidate Companions}
\label{binarity}
The presence of binary or other close companions has the potential to impact the dynamical fitting results.
In the case of bound companions outside the disk, the companion star has the potential distort the disk, depending on the companion's current distance and/or any recent flyby encounters \citep[e.g.][]{Kurtovic2018_AS205}.
In the case of a circumbinary disk, the dynamical mass may be more discrepant with the isochronal mass if the companion is dim and undetected; likewise, circumbinary disks are also both theorized and observed to exhibit truncation, spirals, larger inner cavities and accretion rates than in single systems, and disk misalignment \citep{Rosotti2018,Penzlin2024,Cuello2025}. 
In both cases, the comparison between isochronal and dynamical results could be strongly affected.

Five targets in our sample have confirmed or candidate companions: 
J15534211$-$2114135, J16001844$-$2230114, J16014086$-$2258103, J16075796$-$2040087, and J16113134-183259 B.
\citet{Bouy2006} detect a companion for J16014086$-$2258103 in $Ks$-band (2.18~\mum) using the Nasmyth Adaptive Optics System Near-Infrared Imager and Spectrograph \citep[NACO;][]{Rousset2003,Lenzen2003} on the Very Large Telescope (VLT).
The separation between this source and its companion put the companion outside the disk radius, at a projected linear separation of 706$\pm$1~mas ($\sim$~88~au based on the {\it Gaia} distance).
\citet{Barenfeld2019_binaries} search for candidate companions for 113 G2$-$M5 stars in Upper Sco, including all targets in our sample.
They use the Near-Infrared Camera (NIRC2; instrument PI: Keith Matthews) on the Keck II telescope to obtain both AO and aperture-masked $K$-band (2.196~\mum) images for their targets. 
\citet{Barenfeld2019_binaries} find new candidate companions for J15534211$-$2114135, J16001844$-$2230114, and J16075796$-$2040087.
In all three cases, the candidate companion lies within the radius of the continuum disk. 
J15534211$-$2114135 has two candidate companions: one at a projected linear separation of 321.8$\pm$1~mas ($\sim$11.5~au at the distance to this source), and the other at 1097.1$\pm$0.1~mas ($\sim$38~au). 
The location of the more distant companion puts it well outside the observed disk radius. 
\citep{Eisner2005} observe source J16113134-183259 B (designated AS205B in that work) using the Keck I High-Resolution Echelle Spectrometer \citep[HIRES,][]{Vogt1994}.
They classify this source as a spectroscopic binary.

To investigate whether binary systems may be skewing our results, we re-calculate all statistics excluding these five sources.
We find that the majority of our results do not change.
The main exception is the relationship between source mass ratio and dynamical mass, which is moderately sensitive to the inclusion of the binary candidates.
We describe the impact of binary exclusion on each statistical value below:

\begin{enumerate}
    \item The mass ratios themselves exhibit three consistent changes: 1) no change or a slight increase in the minimum mass ratio (never a decrease), 2) a slight increase in the mean and median mass ratios, and 3) a slight decrease in the scaled MAD and standard deviation of the mass ratios. 
    This is true regardless of whether we consider the full sample or only the mass-limited sample. 
    The increases in minimum, median, and mean mass ratio are expected, because two of the three most massive sources in the sample (by M$_{dyn}$) are binaries.
    The corresponding decrease in scaled MAD and standard deviation are also consistent with this explanation.
    However, all changes are within the uncertainties of both the new and original values, i.e., there are no statistically-significant changes in mass ratio. 
    \item The Spearman $\rho$ coefficients for the full sample show the largest change. The coefficients themselves (Table~\ref{mass_ratios}, column 6) all remain negative, but their p-values (column 7) increase by an order of magnitude in all cases. This drops the p-values below a 3$\sigma$ significance level for all model sets except the PARSEC v1.2S models, which remain at $>$5$\sigma$ significance.
    \item The Spearman $\rho$ coefficients for the mass-limited sample show almost no change. As with the mass ratios, this is consistent with the fact that two of the four most massive sources are binaries - applying a mass limit of \mstar\/ $\leq$ 1~\msun\/ effectively removes them from the sample before binarity is even considered. The only exception is, again, the PARSEC v1.2S models, for which the p-value decreases from 1$\times$10$^{-5}$ to 3$\times$10$^{-6}$.
    \item The KS test p-values increase in all cases. However, because all KS-test p-values were high to begin with, this does not change our interpretations. The D-values decrease slightly in all cases except for the PARSEC v1.2S models, for which the D-values increase slightly in both the full-sample and mass-limited cases. The smaller D-values are consistent with decreased scatter in our data. As Figure~\ref{cdfs_full} shows, most of the deviation between the dynamical mass CDF and those of the isochronal masses comes at the high-mass end. Removing most of the highest-mass sources (50\% of which are known binaries) brings our CDF into better alignment with all of the isochrones except for the PARSEC v1.2S models. 
\end{enumerate}

Overall, we find that excluding the binary sources from our sample has little impact on the sample statistics for either the full or the mass-limited sample case.
Our interpretations and conclusions remain unchanged.

Finally, we find that excluding our known or candidate binary systems has no significant impact on the derived relationship between disk dust mass and stellar dynamical mass.
For the assumption that dust temperature scales with \lstar, the best-fit line has an initial slope of 2.8$\pm$0.6 and intercept of 1.0$\pm$0.3.
After removing the binary systems, the best-fit line has a slope of 2.8$\pm$0.7 and intercept of 1.1$\pm$0.3, i.e. statistically identical but with a slight increase in uncertainty.
Assuming a constant T$_{dust}$ $=$ 20~K, our best-fit line using all sources has a slope of 2.8$\pm$0.5 and intercept of 0.9$\pm$0.2; removing the binary systems changes this to a slope of 2.8$\pm$0.6 and intercept of 0.9$\pm$0.3, i.e. again identical within uncertainties.

\subsection{Disk Inclination, Stellar Mass, and the Keplerian-fitting Method}
The inclination of a disk can have a strong impact on our ability to accurately model the stellar mass. 
For a star of any given mass $M$ and inclination $i$, the true Keplerian velocity at radius $r$ in the disk will be $v(r)_{true}$ $=$ $\sqrt{GM/r}$, but the corresponding observed velocity will be $v(r)_{obs}$ $=$ $(sin\,\,i)\sqrt{GM/r}$.
As inclination decreases, the observed range of velocities $v_{max,\,obs}$ $-$ $v_{min,\,obs}$ will also decrease.
This makes a disk of mass $M$ progressively more difficult to detect at lower inclinations, and is especially detrimental for lower-mass sources, whose maximum Keplerian velocities will be lower to begin with.

In principle, the {\tt pdspy} modeling procedure should be agnostic to both stellar mass and inclination angle.
The physics of the disk model itself, the specifics of the walker step selection, and the $\chi^2$ minimization routine neither assume nor require a particular stellar mass or inclination value. 
However, as a practical matter, the comparatively smaller velocity range observed for a low-inclination disk means that the model will have fewer data points to fit to overall.
This could result in high uncertainties, an inability to distinguish between degenerate possibilities (e.g. a low-mass, high-inclination source and a high-inclination, low-mass source), or an inability to achieve a successful fit at all.
This is an even more significant concern for small or low signal-to-noise disks, which will further narrow the range of ($u,v$) space over which {\tt pdspy} can fit to the data.

Conversely, the high extinction caused by edge-on disks can bias inferred stellar spectral types in the opposite direction. 
High extinction will decrease the overall stellar luminosity, and will redden the observed colors of the star.
Both factors can result in a later inferred spectral type than is actually true, and can potentially bias the spectral-type analysis toward lower masses and/or higher ages. 

\begin{figure}
    \centering
    \includegraphics[width=0.47\textwidth]{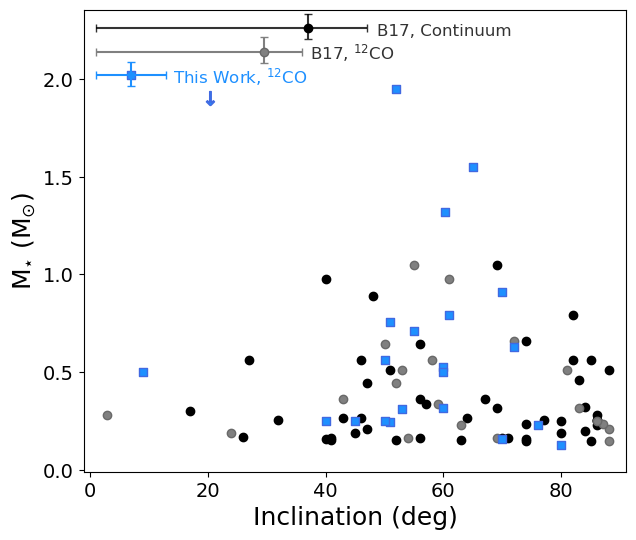}
    \caption{Stellar mass versus inclination angle for the CO-derived \mstar\/ and $i$ in this work (blue), the CO-derived \mstar\/ and $i$ in \citet{Barenfeld2017_disksize} (grey), and the continuum-derived \mstar\/ and $i$ in \citet{Barenfeld2017_disksize} (black). We find that our derived masses and inclination angles show no bias compared to the \citet{Barenfeld2017_disksize} data, but that all samples suffer from small-number statistics at the lower ($<$30$^{\circ}$) and highest ($>$85$^{\circ}$) inclinations.}
    \label{Barenfeld_sample_bias}
\end{figure}

We have examined whether there is any trend in stellar mass with derived inclination angle for our sample. 
We compare stellar mass versus inclination for our Upper Sco sources to the trends detailed in \citet{Barenfeld2017_disksize}, which examines a sample of 55 continuum-detected disks in Upper Sco that includes all sources in our sample.
These authors use the stellar masses of \citet{Barenfeld2016}, and derive inclination angle(s) for each disk using dust-continuum emission (55 disks) and, where possible, $^{12}$CO emission (23 disks).
Figure~\ref{Barenfeld_sample_bias} shows mass versus inclination for our data compared to both the continuum- and CO-derived inclinations from \citet{Barenfeld2017_disksize}.

The three samples appear to occupy the same parameter space in \mstar\/ versus $i$ in Figure~\ref{Barenfeld_sample_bias}.
To quantify the similarity between samples, we calculate Spearman $\rho$ coefficients and p-values between $i$ and \mstar\/ for all three data sets.
The data obtained from line emission (our data and the \citet{Barenfeld2017_disksize} CO data) have slight negative correlations between mass and inclination angle, but the correlation does not rise above 2$\sigma$ in either case (p $=$ 0.68 and 0.25, respectively).
The mass$-$inclination data obtained from the continuum observations have a slight positive correlation, but the p-value again indicates no statistical significance (p $=$ 0.9).
We additionally examine the median and scaled MAD of \mstar\/ and $i$ for each of the three data sets.
We find that the medians agree within uncertainties for both properties for all three data sets ($i$ $=$ 58$\pm$11, 60$\pm$16, 68$\pm$24, and \mstar\/ $=$ 0.5$\pm$0.4, 0.3$\pm$0.2, 0.3$\pm$0.1).
Finally, we perform a KS-test analysis for $i$ and \mstar\/ separately, and compare each data set to the other two.
We find that no KS-test p-values are consistent with a $\geq$4$\sigma$ significance (p-values ranged from 0.04 to 0.8), indicating that the three data sets are not significantly different in either $i$ or \mstar.
We conclude from these results that our method of deriving stellar mass $-$ Keplerian fitting using the {\tt pdspy} package $-$ has not introduced a bias in derived mass with inclination angle.

\section{Description of the Pre-Main Sequence Evolutionary Models}
\label{track_descriptions}
The five stellar evolutionary models we evaluate in this work represent three of the most widely-used schools of models applicable to low-mass stars: the Baraffe models (BHAC15), the Padova stellar evolution code (PARSEC v1.1 and 1.2S), and the Dartmouth models \citep{Feiden2016}. 

\subsection{BHAC15}
The BHAC15 stellar evolutionary models \citep{Baraffe2015} are a set of models developed primarily for low-mass stars, and cover the mass range 0.01-1.4~\msun.
They are an update to the widely-used BCAH98 models \citep{Baraffe1998}, which cover 0.075-1.0~\msun.
Both BCAH98 and BHAC15 model the structure and evolution of stellar interiors following the physics of \citet{Chabrier1997}. 
The most significant change in BHAC15 is a replacement of the NextGen stellar atmospheric models \citep{Hauschildt1999} with the BT-Settl atmospheric models \citep{Allard2012}.
Additionally, BHAC15 includes an expanded molecular linelist compared to BCAH98, which used only solar-type metallicities. 
The revised models of \citet{Baraffe2015} solve the known issue of BCAH98 predicting optical colors (V-I) that were too blue for a given magnitude.

\subsection{PARSEC v1.1}
The {\it PA}dova and T{\it R}ieste {\it S}tellar {\it E}volution {\it C}ode \citep[PARSEC,][]{Bressan2012} is a revision and expansion of the Padova stellar evolution code \citep{Bressan1993_padova}.
Most notably, the PARSEC models expand Padova to include the pre-main sequence (PMS) phase of stellar evolution, using the physics and boundary conditions described below.
They span a mass range of \mstar\/ $=$ 0.1-12~\msun\/ and a metallicity range of $Z$ $=$ 0.0005 to 0.07.
The PARSEC models calculate the equation of state using the FreeEOS library in FORTRAN (Alan W. Irwin, copyright 2008 under GPL license, \href{https://freeeos.sourceforge.net/}{https://freeeos.sourceforge.net/}).
They implement a temperature gradient described by the mixing-length theory of \citet{Bohm-Vitense1958}, include a revised nuclear network and reaction rates \citep{Cyburt2010}, and add microscopic diffusion to the model physics.
They also implement an opacity-table generation tool that allows the user to create models for any initial chemical composition using opacities from AESOPUS \citep{Marigo2009} and OPAL \citep{Iglesias1996}.
As the PARSEC code later underwent several revisions, this original version is referred to as PARSEC v1.1.

\subsection{PARSEC v1.2S}
The PARSEC v1.2S \citep{Chen2014} models are an update to the PARSEC v1.1 code of \citet{Bressan2012}, described above.
The v1.2S models aim to address known issues with modeling low-mass stars specifically ($\leq$0.6~\msun).
In particular, the PARSEC v1.1 models show a consistent discrepancy between observed and model radii, which leads to a $\sim$8\% discrepancy in the mass-radius relation in the models.
They also tend to produce too-blue optical colors for low-mass main-sequence stars. 
PARSEC v1.2S applies the $T-\tau$ relation of the BT-Settl atmospheric models as an outer boundary condition in the PARSEC code.
This change reduces the mass-radius discrepancy to 5\%.
It also reduces, though does not eliminate, the discrepancies in the optical colors.

\subsection{Feiden (2016) standard and magnetic models}
\citet{Feiden2016} seeks to address the known tendency of PMS stellar evolutionary models of late-type (K, M) stars to return much younger ages than for more massive stars (A, F, G) in the same group or cluster.
The underlying premise of the \citet{Feiden2016} magnetic models is that magnetic fields in these cool dwarfs inhibit convection, which slows the star's contraction and thus extends the amount of time they spend on the Hayashi track.
The result is that the age inferred from a given mass, radius, and isochrone will be too low for these late-type stars, if the stellar evolutionary model used does not account for this magnetic inhibition.

The magnetic models of \citet{Feiden2016} are based on the magnetic Dartmouth stellar evolution models of \citet{Feiden2012}.
Those models are, in turn, based on the original standard models produced by the Dartmouth Stellar Evolution Program \citep[DSEP,][]{Dotter2008}. 
In addition to magnetic fields, the \citet{Feiden2016} models incorporate model atmosphere structures generated using a custom grid that combines the solar composition of \citet{Grevesse1998} with the opacities of \citet{Ferguson2005} for low temperatures.
The models also include explicit treatment of deuterium burning, and the implementation of a boundary condition at the $\tau_{5000}$ $=$ 10 surface.

In the \citet{Feiden2016} models, the magnetic field strength is Gaussian with radius, centered at the location of the inner boundary of the stars' convective layer. The interaction between the magnetic field, convection, and the plasma equation of state follow the physics described in \citet{Feiden2012,Feiden2013}.
The model assumes equipartition between the gas pressure and the magnetic field pressure at the star's surface, with typical surface magnetic field strengths of 2.5~kG (corresponding to typical peak  magnetic field strengths of 50~kG in the star's interior).
The magnetic field is ``switched on'' at a stellar age of 0.1 to 0.5~Myr.
It is switched off if and when the star's outer convective envelope disappears, which occurs only for \mstar\/ $\gtrsim$ 1.15~\msun.




\begin{thebibliography}{}
\bibitem[Allard et al.(2011)]{Allard2011} Allard, F., Homeier, D., \& Freytag, B.\ 2011, 16th Cambridge Workshop on Cool Stars, Stellar Systems, and the Sun, 448, 91
\bibitem[Allard et al.(2012)]{Allard2012} Allard, F., Homeier, D., \& Freytag, B.\ 2012, Philosophical Transactions of the Royal Society of London Series A, 370, 2765
\bibitem[Allen(1973)]{Allen1973} Allen, C.~W.\ 1973, London: University of London, Athlone Press, |c1973, 3rd ed.
\bibitem[Andrews et al.(2013)]{Andrews2013} Andrews, S.~M., Rosenfeld, K.~A., Kraus, A.~L., et al.\ 2013, \apj, 771, 129
\bibitem[Andrews et al.(2018)]{Andrews2018_DSHARP} Andrews, S.~M., Huang, J., P{\'e}rez, L.~M., et al.\ 2018, \apjl, 869, L41
\bibitem[Astropy Collaboration et al.(2013)]{astropy_i} Astropy Collaboration, Robitaille, T.~P., Tollerud, E.~J., et al.\ 2013, \aap, 558, A33
\bibitem[Astropy Collaboration et al.(2018)]{astropy_ii} Astropy Collaboration, Price-Whelan, A.~M., Sip{\H{o}}cz, B.~M., et al.\ 2018, \aj, 156, 123
\bibitem[Astropy Collaboration et al.(2022)]{astropy_iii} Astropy Collaboration, Price-Whelan, A.~M., Lim, P.~L., et al.\ 2022, \apj, 935, 167
\bibitem[Baraffe et al.(1998)]{Baraffe1998} Baraffe, I., Chabrier, G., Allard, F., et al.\ 1998, \aap, 337, 403
\bibitem[Baraffe et al.(2015)]{Baraffe2015} Baraffe, I., Homeier, D., Allard, F., et al.\ 2015, \aap, 577, A42 
\bibitem[Barenfeld et al. (2016)]{Barenfeld2016} Barenfeld, S.~A., Carpenter, J.~M., Ricci, L., et al.\ 2016, \apj, 827, 142
\bibitem[Barenfeld et al.(2017)]{Barenfeld2017_disksize} Barenfeld, S.~A., Carpenter, J.~M., Sargent, A.~I., et al.\ 2017, \apj, 851, 85
\bibitem[Barenfeld et al.(2019)]{Barenfeld2019_binaries} Barenfeld, S.~A., Carpenter, J.~M., Sargent, A.~I., et al.\ 2019, \apj, 878, 45
\bibitem[Bary \& Petersen(2014)]{Bary2014} Bary, J.~S. \& Petersen, M.~S.\ 2014, \apj, 792, 64
\bibitem[Bell et al.(2012)]{Bell2012} Bell, C.~P.~M., Naylor, T., Mayne, N.~J., et al.\ 2012, \mnras, 424, 3178
\bibitem[Bergin et al.(2024)]{Bergin2024} Bergin, E.~A., Bosman, A., Teague, R., et al.\ 2024, \apj, 965, 147
\bibitem[Blaauw(1946)]{Blaauw1946} Blaauw, A.\ 1946, Publications of the Kapteyn Astronomical Laboratory Groningen, 52, 1
\bibitem[Blaauw(1964)]{Blaauw1964} Blaauw, A.\ 1964, \araa, 2, 213
\bibitem[B{\"o}hm-Vitense(1958)]{Bohm-Vitense1958} B{\"o}hm-Vitense, E.\ 1958, \zap, 46, 108
\bibitem[Bouy et al.(2006)]{Bouy2006} Bouy, H., Mart{\'\i}n, E.~L., Brandner, W., et al.\ 2006, \aap, 451, 177
\bibitem[Braun et al.(2021)]{Braun2021} Braun, T.~A.~M., Yen, H.-W., Koch, P.~M., et al.\ 2021, \apj, 908, 46
\bibitem[Bressan et al.(1993)]{Bressan1993_padova} Bressan, A., Fagotto, F., Bertelli, G., et al.\ 1993, \aaps, 100, 647
\bibitem[Bressan et al.(2012)]{Bressan2012} Bressan, A., Marigo, P., Girardi, L., et al.\ 2012, \mnras, 427, 127
\bibitem[Cardelli et al.(1989)]{Cardelli1989} Cardelli, J.~A., Clayton, G.~C., \& Mathis, J.~S.\ 1989, \apj, The Relationship between Infrared, Optical, and Ultraviolet Extinction, 345, 245.
\bibitem[Carpenter et al.(2025)]{Carpenter2025} Carpenter, J.~M., Esplin, T.~L., Luhman, K.~L., et al.\ 2025, \apj, 978, 117
\bibitem[CASA Team et al.(2022)]{CASA_2022} CASA Team, Bean, B., Bhatnagar, S., et al.\ 2022, \pasp, 134, 114501
\bibitem[Chabrier \& Baraffe(1997)]{Chabrier1997} Chabrier, G. \& Baraffe, I.\ 1997, \aap, 327, 1039
\bibitem[Chen et al.(2014)]{Chen2014} Chen, Y., Girardi, L., Bressan, A., et al.\ 2014, \mnras, 444, 2525
\bibitem[Cody et al.(2017)]{Cody2017} Cody, A.~M., Hillenbrand, L.~A., David, T.~J., et al.\ 2017, \apj, 836, 41
\bibitem[Cohen \& Kuhi(1979)]{Cohen1979} Cohen, M. \& Kuhi, L.~V.\ 1979, \apjs, 41, 743
\bibitem[Comrie et al.(2021)]{carta} Comrie, A., Wang, K.-S., Hsu, S.-C., et al.\ 2021, Zenodo
\bibitem[Cuello et al.(2025)]{Cuello2025} Cuello, N., Alaguero, A., \& Poblete, P.~P.\ 2025, arXiv:2501.19249
\bibitem[Cyburt et al.(2010)]{Cyburt2010} Cyburt, R.~H., Amthor, A.~M., Ferguson, R., et al.\ 2010, \apjs, 189, 240
\bibitem[Czekala et al.(2015)]{Czekala2015} Czekala, I., Andrews, S.~M., Jensen, E.~L.~N., et al.\ 2015, \apj, 806, 154
\bibitem[David et al.(2019)]{David2019} David, T.~J., Hillenbrand, L.~A., Gillen, E., et al.\ 2019, \apj, 872, 161
\bibitem[DENIS Consortium(2005)]{DENIS2005} DENIS Consortium\ 2005, VizieR Online Data Catalog, VizieR Online Data Catalog: The DENIS database (DENIS Consortium, 2005), 1. B/denis
\bibitem[Dickman(1978)]{Dickman1978} Dickman, R.~L.\ 1978, \apjs, 37, 407
\bibitem[Dotter et al.(2008)]{Dotter2008} Dotter, A., Chaboyer, B., Jevremovi{\'c}, D., et al.\ 2008, \apjs, 178, 89
\bibitem[Dullemond et al.(2012)]{Dullemond2012_radmc3d} Dullemond, C.~P., Juhasz, A., Pohl, A., et al.\ 2012, Astrophysics Source Code Library
\bibitem[Eisner et al.(2005)]{Eisner2005} Eisner, J.~A., Hillenbrand, L.~A., White, R.~J., et al.\ 2005, \apj, 623, 952
\bibitem[Encrenaz et al.(1975)]{Encrenaz1975} Encrenaz, P.~J., Falgarone, E., \& Lucas, R.\ 1975, \aap, 44, 73
\bibitem[Fang et al.(2023)]{Fang2023_diskwinds} Fang, M., Pascucci, I., Edwards, S., et al.\ 2023, \apj, 945, 112
\bibitem[Feiden \& Chaboyer(2012)]{Feiden2012} Feiden, G.~A. \& Chaboyer, B.\ 2012, \apj, 761, 30
\bibitem[Feiden \& Chaboyer(2013)]{Feiden2013} Feiden, G.~A. \& Chaboyer, B.\ 2013, \apj, 779, 183
\bibitem[Feiden(2016)]{Feiden2016} Feiden, G.~A.\ 2016, \aap, 593, A99
\bibitem[Ferguson et al.(2005)]{Ferguson2005} Ferguson, J.~W., Alexander, D.~R., Allard, F., et al.\ 2005, \apj, 623, 585
\bibitem[Flores et al.(2022)]{Flores2022} Flores, C., Connelley, M.~S., Reipurth, B., et al.\ 2022, \apj, 925, 21
\bibitem[Flewelling et al.(2020)]{Flewelling2020_panstarrs} Flewelling, H.~A., Magnier, E.~A., Chambers, K.~C., et al.\ 2020, \apjs, 251, 1, 7
\bibitem[Fouqu{\'e} et al.(2000)]{Foque2000_DENIS} Fouqu{\'e}, P., Chevallier, L., Cohen, M., et al.\ 2000, \aaps, 141, 313
\bibitem[Foreman-Mackey et al.(2013)]{Foreman-Mackey2013_emcee} Foreman-Mackey, D., Hogg, D.~W., Lang, D., et al.\ 2013, \pasp, 125, 306
\bibitem[Frank et al.(2014)]{Frank2014} Frank, A., Ray, T.~P., Cabrit, S., et al.\ 2014, Protostars and Planets VI, 451
\bibitem[Gaia Collaboration et al.(2016)]{Gaia_i} Gaia Collaboration, Prusti, T., de Bruijne, J.~H.~J., et al.\ 2016, \aap, 595, A1
\bibitem[Gaia Collaboration et al.(2018)]{Gaia_DR2} Gaia Collaboration, Brown, A.~G.~A., Vallenari, A., et al.\ 2018, \aap, 616, A1
\bibitem[Gaia Collaboration et al.(2021)]{GaiaEDR3_Overview} Gaia Collaboration, Brown, A.~G.~A., Vallenari, A., et al.\ 2021, \aap, 649, A1
\bibitem[Gavras et al.(2023)]{Gavras2023} Gavras, P., Rimoldini, L., Nienartowicz, K., et al.\ 2023, \aap, 674, A22
\bibitem[Ginsburg et al.(2014)]{spectralcube} Ginsburg, A., Robitaille, T., Beaumont, C., \& ZuHone, J. (2014). Release Candidate 2 - includes yt interop (v0.2-rc2). Zenodo. \href{https://doi.org/10.5281/zenodo.11485}{https://doi.org/10.5281/zenodo.11485}
\bibitem[Girardi et al.(2002)]{Girardi2002} Girardi, L., Bertelli, G., Bressan, A., et al.\ 2002, \aap, 391, 195
\bibitem[Grevesse \& Sauval(1998)]{Grevesse1998} Grevesse, N. \& Sauval, A.~J.\ 1998, \ssr, 85, 161
\bibitem[Hauschildt et al.(1999)]{Hauschildt1999} Hauschildt, P.~H., Allard, F., \& Baron, E.\ 1999, \apj, 512, 377
\bibitem[Haworth et al.(2018)]{Haworth2018} Haworth, T.~J., Booth, R.~A., Homan, W., et al.\ 2018, \mnras, 473, 317
\bibitem[Henden \& Munari(2014)]{Henden2014_apass1} Henden, A. \& Munari, U.\ 2014, Contributions of the Astronomical Observatory Skalnat\'{e} Pleso, 43, 3, 518
\bibitem[Henden et al.(2018)]{Henden2018_unc} Henden, A.~A., Levine, S., Terrell, D., et al.\ 2018, \aas, 232, 223.06
\bibitem[Henden(2019)]{Henden2019_apass_dr10} Henden, A.~A.\ 2019, \jaavso, 47, 1, 130
\bibitem[Herczeg \& Hillenbrand(2015)]{Herczeg2015} Herczeg, G.~J. \& Hillenbrand, L.~A.\ 2015, \apj, 808, 23
\bibitem[Hillenbrand \& White(2004)]{Hillenbrand2004} Hillenbrand, L.~A. \& White, R.~J.\ 2004, \apj, 604, 741
\bibitem[Iglesias \& Rogers(1996)]{Iglesias1996} Iglesias, C.~A. \& Rogers, F.~J.\ 1996, \apj, 464, 943
\bibitem[Kraus et al.(2008)]{Kraus2008} Kraus, A.~L., Ireland, M.~J., Martinache, F., et al.\ 2008, \apj, 679, 762
\bibitem[Kraus \& Hillenbrand(2009)]{Kraus2009_J1607} Kraus, A.~L. \& Hillenbrand, L.~A.\ 2009, \apj, 703, 1511
\bibitem[Kurtovic et al.(2018)]{Kurtovic2018_AS205} Kurtovic, N.~T., P{\'e}rez, L.~M., Benisty, M., et al.\ 2018, \apjl, 869, L44
\bibitem[Lenzen et al.(2003)]{Lenzen2003} Lenzen, R., Hartung, M., Brandner, W., et al.\ 2003, \procspie, 4841, 944
\bibitem[Luhman(1999)]{Luhman1999} Luhman, K.~L.\ 1999, \apj, Young Low-Mass Stars and Brown Dwarfs in IC 348, 525, 1, 466
\bibitem[Luhman \& Mamajek(2012)]{Luhman2012} Luhman, K.~L. \& Mamajek, E.~E.\ 2012, \apj, The Disk Population of the Upper Scorpius Association, 758, 1, 31
\bibitem[Luhman \& Esplin(2020)]{Luhman2020} Luhman, K.~L. \& Esplin, T.~L.\ 2020, \aj, 160, 44
\bibitem[Luhman(2022a)]{Luhman2022} Luhman, K.~L.\ 2022, \aj, 163, 24
\bibitem[Lynden-Bell \& Pringle(1974)]{Lynden-Bell1974} Lynden-Bell, D. \& Pringle, J.~E.\ 1974, \mnras, 168, 603
\bibitem[Magnier et al.(2020)]{Magnier2020} Magnier, E.~A., Schlafly, E.~F., Finkbeiner, D.~P., et al.\ 2020, \apjs, 251, 1, 6
\bibitem[Manara et al.(2023)]{Manara2023_review} Manara, C.~F., Ansdell, M., Rosotti, G.~P., et al.\ 2023, Protostars and Planets VII, 534, 539
\bibitem[Marigo \& Aringer(2009)]{Marigo2009} Marigo, P. \& Aringer, B.\ 2009, \aap, 508, 1539
\bibitem[Marigo et al.(2017)]{Marigo2017} Marigo, P., Girardi, L., Bressan, A., et al.\ 2017, \apj, 835, 77
\bibitem[Marton et al.(2023)]{Marton2023} Marton, G., {\'A}brah{\'a}m, P., Rimoldini, L., et al.\ 2023, \aap, 674, A21
\bibitem[McMullin et al.(2007)]{McMullin2007_casa} McMullin, J.~P., Waters, B., Schiebel, D., et al.\ 2007, Astronomical Data Analysis Software and Systems XVI, 376, 127
\bibitem[Miret-Roig et al.(2022)]{Miret-Roig2022} Miret-Roig, N., Galli, P.~A.~B., Olivares, J., et al.\ 2022, \aap, 667, A163
\bibitem[Munari et al.(2014)]{Munari2014_apass_unc} Munari, U., Henden, A., Frigo, A., et al.\ 2014, \aj, 148, 5, 81
\bibitem[Palla \& Stahler(1993)]{Palla1993} Palla, F. \& Stahler, S.~W.\ 1993, \apj, 418, 414
\bibitem[Pascucci et al.(2016)]{Pascucci2016} Pascucci, I., Testi, L., Herczeg, G.~J., et al.\ 2016, \apj, 831, 125
\bibitem[Pecaut et al.(2012)]{Pecaut2012} Pecaut, M.~J., Mamajek, E.~E., \& Bubar, E.~J.\ 2012, \apj, 746, 154
\bibitem[Pecaut \& Mamajek(2013)]{Pecaut2013} Pecaut, M.~J. \& Mamajek, E.~E.\ 2013, \apjs, Intrinsic Colors, Temperatures, and Bolometric Corrections of Pre-main-sequence Stars, 208, 1, 9
\bibitem[Pegues et al.(2021)]{Pegues2021} Pegues, J., Czekala, I., Andrews, S.~M., et al.\ 2021, \apj, 908, 42
\bibitem[Penzlin et al.(2024)]{Penzlin2024} Penzlin, A.~B.~T., Booth, R.~A., Nelson, R.~P., et al.\ 2024, \mnras, 532, 3166
\bibitem[P{\'e}rez Paolino et al.(2024)]{PerezPaolino2024} P{\'e}rez Paolino, F., Bary, J.~S., Hillenbrand, L.~A., et al.\ 2024, \apj, 967, 45
\bibitem[Pi{\'e}tu et al.(2007)]{Pietu2007} Pi{\'e}tu, V., Dutrey, A., \& Guilloteau, S.\ 2007, \aap, 467, 163
\bibitem[Preibisch et al.(2002)]{Preibisch2002} Preibisch, T., Brown, A.~G.~A., Bridges, T., et al.\ 2002, \aj, 124, 404
\bibitem[Preibisch \& Mamajek(2008)]{Preibisch2008} Preibisch, T. \& Mamajek, E.\ 2008, Handbook of Star Forming Regions, Volume II, 235
\bibitem[Press et al.(1992)]{NumericalRecipesBook} Press, W.~H., Teukolsky, S.~A., Vetterling, W.~T., et al.\ 1992, Cambridge: University Press, |c1992, 2nd ed.
\bibitem[Ratzenb{\"o}ck et al.(2023a)]{Ratzenbock2023a} Ratzenb{\"o}ck, S., Gro{\ss}schedl, J.~E., M{\"o}ller, T., et al.\ 2023, \aap, 677, A59
\bibitem[Ratzenb{\"o}ck et al.(2023b)]{Ratzenbock2023b} Ratzenb{\"o}ck, S., Gro{\ss}schedl, J.~E., Alves, J., et al.\ 2023, \aap, 678, A71
\bibitem[Rimoldini et al.(2023)]{Rimoldini2023} Rimoldini, L., Holl, B., Gavras, P., et al.\ 2023, \aap, 674, A14
\bibitem[Rizzuto et al.(2016)]{Rizzuto2016} Rizzuto, A.~C., Ireland, M.~J., Dupuy, T.~J., et al.\ 2016, \apj, 817, 164
\bibitem[Rosotti \& Clarke(2018)]{Rosotti2018} Rosotti, G.~P. \& Clarke, C.~J.\ 2018, \mnras, 473, 5630
\bibitem[Rousset et al.(2003)]{Rousset2003} Rousset, G., Lacombe, F., Puget, P., et al.\ 2003, \procspie, 4839, 140
\bibitem[Roy(2020)]{Roy2020_MCMCstop} Roy, V.\ 2020, Annual Review of Statistics and Its Application, 7, 387
\bibitem[Rydgren \& Vrba(1983)]{Rydgren1983} Rydgren, A.~E. \& Vrba, F.~J.\ 1983, \aj, 88, 1017
\bibitem[Salyk et al.(2014)]{Salyk2014} Salyk, C., Pontoppidan, K., Corder, S., et al.\ 2014, \apj, 792, 68
\bibitem[Sanna et al.(2019)]{Sanna2019} Sanna, A., K{\"o}lligan, A., Moscadelli, L., et al.\ 2019, \aap, 623, A77
\bibitem[Sartori et al.(2003)]{Sartori2003} Sartori, M.~J., L{\'e}pine, J.~R.~D., \& Dias, W.~S.\ 2003, \aap, 404, 913
\bibitem[Schmidt-Kaler(1982)]{Schmidt-Kaler1982} Schmidt-Kaler, T.\ 1982, Bulletin d'Information du Centre de Donnees Stellaires, Automated spectral classification. A survey, 23, 2. 
\bibitem[Semenov et al.(2024)]{Semenov2024} Semenov, D., Henning, T., Guilloteau, S., et al.\ 2024, \aap, 685, A126
\bibitem[Sheehan(2018)]{Sheehan2018_zenodo} Sheehan, P.\ 2018, Zenodo
\bibitem[Sheehan et al. (2019)]{Sheehan2019} Sheehan, P.~D., Wu, Y.-L., Eisner, J.~A., et al.\ 2019, \apj, 874, 136
\bibitem[Skrutskie et al.(2006)]{Skrutskie2006} Skrutskie, M.~F., Cutri, R.~M., Stiening, R., et al.\ 2006, \aj, The Two Micron All Sky Survey (2MASS), 131, 2, 1163
\bibitem[Siess et al.(2000)]{Siess2000} Siess, L., Dufour, E., \& Forestini, M.\ 2000, \aap, 358, 593
\bibitem[Simon et al.(2019)]{Simon2019} Simon, M., Guilloteau, S., Beck, T.~L., et al.\ 2019, \apj, 884, 42
\bibitem[Soderblom et al.(2014)]{Soderblom2014} Soderblom, D.~R., Hillenbrand, L.~A., Jeffries, R.~D., et al.\ 2014, Protostars and Planets VI, 219
\bibitem[Somers \& Pinsonneault(2015)]{Somers2015} Somers, G. \& Pinsonneault, M.~H.\ 2015, \apj, 807, 174
\bibitem[Somers et al.(2020)]{Somers2020} Somers, G., Cao, L., \& Pinsonneault, M.~H.\ 2020, \apj, 891, 29
\bibitem[Strai{\v{z}}ys(1992)]{Straizys1992} Strai{\v{z}}ys, V.\ 1992, , Multicolor stellar photometry. 
\bibitem[Steenman \& The(1991)]{Steenman1991} Steenman, H. \& The, P.~S.\ 1991, \apss, The Anomalous Extinction Law - Part Two - the Effect of Changing the Lower Size Cutoff of the Particle Size Distribution, 184, 1, 9
\bibitem[Tazzari et al.(2018)]{Tazzari2018_galario} Tazzari, M., Beaujean, F., \& Testi, L.\ 2018, \mnras, 476, 4527
\bibitem[Tonry et al.(2012)]{Tonry2012_panstarrs} Tonry, J.~L., Stubbs, C.~W., Lykke, K.~R., et al.\ 2012, \apj, 750, 2, 99
\bibitem[Vogt et al.(1994)]{Vogt1994} Vogt, S.~S., Allen, S.~L., Bigelow, B.~C., et al.\ 1994, \procspie, 2198, 362
\bibitem[Wang \& Chen(2019)]{WangChen2019} Wang, S. \& Chen, X.\ 2019, \apj, 877, 2, 116
\bibitem[Wenger et al.(2000)]{Wenger2000_SIMBAD} Wenger, M., Ochsenbein, F., Egret, D., et al.\ 2000, \aaps, The SIMBAD astronomical database. The CDS reference database for astronomical objects, 143, 9
\bibitem[Yen et al.(2016)]{Yen2016} Yen, H.-W., Koch, P.~M., Liu, H.~B., et al.\ 2016, \apj, 832, 204
\bibitem[Yen et al.(2018)]{Yen2018} Yen, H.-W., Koch, P.~M., Manara, C.~F., et al.\ 2018, \aap, 616, A100
\bibitem[Zacharias et al.(2013)]{Zacharias2013_ucac4} Zacharias, N., Finch, C.~T., Girard, T.~M., et al.\ 2013, \aj, 145, 2, 44

\end{thebibliography}
\end{document}